\documentclass[twocolumn,tighten]{aastex701} 
\hypersetup{linkcolor=red,citecolor=blue,filecolor=cyan,urlcolor=magenta}
\usepackage{amsmath}
\usepackage{natbib}
\usepackage{todonotes}
\usepackage{CJK}
\usepackage{xspace}
\usepackage[shortlabels]{enumitem}

\newif\ifextrastamps
\extrastampstrue

\newcommand{\Latte}{\textit{Latte}\xspace}
\newcommand{\Fire}{\textsc{FIRE}\xspace}

\newcommand{\msun}{\mbox{$M_{\odot}$}}

\newcommand{\RN}[1]{%
  \textup{\uppercase\expandafter{\romannumeral#1}}%
}

\shorttitle{From starlight to dark matter}
\shortauthors{Ou et al.}
\begin{document}
\begin{CJK*}{UTF8}{gbsn}

\title{From starlight to dark matter: a stochastic interpolation approach to map dark matter from stellar density}

\author[0000-0002-4669-9967]{Xiaowei~Ou (欧筱葳)} 
\altaffiliation{Galaxy Evolution and Cosmology (GECO) Fellow}
\altaffiliation{CosmicAI Fellow}
\affiliation{%
Department of Astronomy, University of Virginia,
530 McCormick Rd, Charlottesville, VA 22904, USA}
\affiliation{The NSF-Simons AI Institute for Cosmic Origins, USA}
\email{Email:\ xwou@virginia.edu}

\author[0000-0003-2806-1414]{Lina~Necib}
\affiliation{Department of Physics and MIT Kavli Institute for Astrophysics and Space Research, \\
Massachusetts Institute of Technology,
77 Massachusetts Avenue, Cambridge, MA 02139, USA}
\affiliation{The NSF AI Institute for Artificial Intelligence and Fundamental Interactions, \\
Massachusetts Institute of Technology,
77 Massachusetts Avenue, Cambridge, MA 02139, USA}
\email{Email:\ lnecib@mit.edu}

\author[]{Carolina~Cuesta-Lazaro}
\affiliation{Center for Cosmology and Particle Physics, 
Department of Physics, New York University, \\
4 Washington Place, New York, NY, 10003, USA}
\affiliation{The NSF AI Institute for Artificial Intelligence and Fundamental Interactions, \\
Massachusetts Institute of Technology,
77 Massachusetts Avenue, Cambridge, MA 02139, USA}
\affiliation{Center for Computational Astrophysics, 
Flatiron Institute, 
162 Fifth Avenue, New York, NY 10010, USA}
\email{Email:\ cuestalz@mit.edu}

\author[0000-0002-5653-0786]{Paul~Torrey}
\affiliation{%
Department of Astronomy, University of Virginia,
530 McCormick Rd, Charlottesville, VA 22904, USA}
\affiliation{The NSF-Simons AI Institute for Cosmic Origins, USA}
\email{Email:\ paul.torrey@virginia.edu}

\author[0009-0002-1233-2013]{Niusha~Ahvazi}
\altaffiliation{Galaxy Evolution and Cosmology (GECO) Fellow}
\affiliation{%
Department of Astronomy, 
University of Virginia,
530 McCormick Rd, Charlottesville, VA 22904, USA}
\affiliation{The NSF-Simons AI Institute for Cosmic Origins, USA}
\email{Email:\ dkk9en@virginia.edu}

\author[0000-0002-0372-3736]{Alyson~M.~Brooks}
\affiliation{
Department of Physics \& Astronomy, Rutgers, the State University of New Jersey, \\
Piscataway, NJ 08854, USA
}
\email{Email:\ abrooks@physics.rutgers.edu}

\author[]{Berthy~T.~Feng}
\affiliation{The NSF AI Institute for Artificial Intelligence and Fundamental Interactions, \\
Massachusetts Institute of Technology,
77 Massachusetts Avenue, Cambridge, MA 02139, USA}
\email{Email:\ berthy@mit.edu}

\author[0000-0002-8111-9884]{Alex~M.~Garcia}
\affiliation{The Center for Astrophysics, Harvard University, Cambridge, MA 02138, USA}
\affiliation{The NSF-Simons AI Institute for Cosmic Origins, USA}
\email{Email:\ alexgarcia@virginia.edu}

\author[0000-0001-9592-4190]{Jiaxuan~Li (李嘉轩)}
\email{jiaxuanl@stanford.edu}
\affiliation{Department of Astrophysical Sciences, 4 Ivy Lane, Princeton University, Princeton, NJ 08540, USA}
\affiliation{Kavli Institute for Particle Astrophysics and Cosmology, Stanford University, Stanford, CA 94305, USA}

\author[0000-0002-2628-0237]{Jonah~C.~Rose}
\affiliation{%
Department of Physics, Princeton University, 
Princeton, NJ 08544, USA}
\affiliation{
Center for Computational Astrophysics, Flatiron Institute, 
162 5th Avenue, New York, NY 10010, USA}
\email{Email:\ jr8952@princeton.edu}

\author[0000-0002-6196-823X]{Xuejian~Shen}
\affiliation{The Center for Astrophysics, Harvard University, Cambridge, MA 02138, USA}
\email{Email:\ xuejianshen@fas.harvard.edu}

\author[0000-0001-8593-7692]{Mark~Vogelsberger}
\affiliation{Department of Physics and MIT Kavli Institute for Astrophysics and Space Research, \\
Massachusetts Institute of Technology,
77 Massachusetts Avenue, Cambridge, MA 02139, USA}
\affiliation{Fachbereich Physik, Philipps Universit\"at Marburg, D-35032 Marburg, Germany}
\email{Email:\ mvogelsb@mit.edu}


\begin{abstract}

The dark matter halo profile in galaxies holds key information about the nature of dark matter and galaxy formation.
Constraining the dark matter profile of galaxies beyond the Milky Way traditionally requires expensive spectroscopic observations for kinematic information.
In this paper, we explore a conditional generative model framework to map the dark matter profile of Milky Way-mass galaxies from stellar density profiles, more easily obtainable through large photometric imaging surveys.
As a proof of concept, we train the model to learn a stochastic bridge between instrument systematics-free baryonic stellar distributions and underlying dark matter density maps from the DREAMS hydrodynamics simulation suite.
We recover 2D dark matter density profiles with a typical accuracy of $\sim0.1$\,dex ($\sim1.5$\% of the truth in log scale).
The stochastic sampling procedure provides uncertainty estimates of the predicted dark matter map, with typical values $\sim0.1$\,dex.
Out-of-domain tests with Milky Way-mass galaxies from IllustrisTNG and \Fire simulations show that, while the model can qualitatively be generalized to TNG50 galaxies from IllustrisTNG, the model is sensitive to the galaxy formation model, with $\sim0.2$-$0.4$\,dex over-prediction for the inner profiles ($r\lesssim5$\,kpc) of the \Fire test galaxies.
Future work will explore training with additional suites of simulations and/or conditioning on additional information, such as multi-band images.
Our results are a first step towards using generative models as a flexible, uncertainty-aware framework for turning forthcoming data from large imaging surveys into spatially resolved dark matter maps.

\end{abstract}


\section{Introduction}

The spatial distribution of dark matter in galaxies encodes information about both the particle nature of dark matter and the baryonic processes that regulate galaxy formation. 
In collisionless cold dark matter simulations, halos develop approximately universal density profiles whose normalization and concentration reflect halo mass and assembly history \citep{navarro96,navarro97}. 
On galactic scales, however, the connection between the predicted halo profile and the observed mass distribution is modified by baryonic physics. 
Gas cooling, star formation, stellar feedback, black-hole feedback, and mergers can contract, heat, or reshape the inner dark matter distribution, producing galaxy-to-galaxy diversity even at fixed halo mass \citep[e.g.,][]{pontzen12,dicintio14,chan15,bose19,lazar20,mostow25}. 
Measuring dark matter profiles across a representative population of galaxies is therefore essential for separating signatures of dark matter physics from the astrophysical response of halos to galaxy formation.

A range of observational probes has been developed to infer galaxy dark matter distributions. 
For rotationally supported galaxies, gas and stellar rotation curves provide one of the most direct constraints on the enclosed mass profile, and large compilations such as the SPARC survey have enabled detailed comparisons between baryonic structure and inferred dynamical mass \citep{lelli16,mcgaugh16}. 
For pressure-supported systems, stellar kinematics combined with Jeans modeling, distribution-function modeling, or orbit-based modeling can constrain the total gravitational potential, although degeneracies with velocity anisotropy, geometry, and stellar mass-to-light ratio remain significant \citep[e.g.,][]{Merrifield1990, walker09,wolf10}. 
Strong gravitational lensing provides precise projected mass constraints in the central regions of massive galaxies, especially when combined with stellar dynamics \citep[e.g.,][]{gavazzi07,auger10}. 
Weak gravitational lensing and satellite kinematics extend these measurements statistically to larger radii and larger samples, but generally constrain ensemble-averaged halo properties rather than detailed dark matter maps for individual galaxies \citep[e.g.,][]{mandelbaum06,mandelbaum08}. 
These methods have been highly successful, but they usually require spectroscopy, well-resolved kinematics, lensing configurations, or large statistical samples. 
As a result, spatially resolved dark matter constraints remain expensive compared with the photometric imaging that is available for orders of magnitude larger galaxy samples.

The next generation of wide-field imaging surveys will dramatically expand the available photometric view of galaxy structure. 
Euclid \citep{mellier25}, the Vera C. Rubin Observatory Legacy Survey of Space and Time \citep{ivezic19}, and the Nancy Grace Roman Space Telescope \citep{akeson19,eifler21} will deliver deep, wide, multi-band imaging for enormous galaxy samples, with Euclid alone designed to provide high-resolution optical imaging and near-infrared imaging/spectroscopy over $\sim 14,000\,{\rm deg}^2$ of extragalactic sky. 
These surveys are primarily designed to constrain cosmology through weak lensing, galaxy clustering, and related large-scale probes. 
However, the same images also contain detailed information about galaxy morphology such as stellar mass distributions. 
If these photometric observables can be connected to the underlying dark matter distribution, upcoming imaging surveys could provide a complementary route to dark matter constraints.

Such a connection is plausible because the stellar distribution is shaped by the same gravitational potential and assembly history that determine the dark matter halo (see e.g., \citealt{sanchez-almeida23,riggs24,keith25,sanchez-almeida25,hakkinen26}). 
The correlation is not expected to be deterministic. 
For dwarf galaxies, as an example, ongoing debates on whether a cuspy dark matter halo can host a cored stellar mass distribution argue for both cases based on analytical models and idealized simulations \citep{sanchez-almeida23,hakkinen26}.
Beyond the gravitation potential from dark matter, stellar formation and feedback history also shape the present-day stellar mass distribution \citep{riggs24,keith25}. 
Additionally, observational systematics such as instrument passbands also make the same physical galaxy map to different observations.
A useful model must therefore learn not only a mean mapping from baryonic structure to dark matter structure, but also the intrinsic scatter and non-uniqueness of that mapping. 
This makes the problem naturally suited to probabilistic generative modeling rather than deterministic regression.

Diffusion and generative models provide a flexible framework for learning such probabilistic mappings. 
Originally developed as powerful generative models for high-dimensional data, diffusion and score-based models learn to transform simple probability distributions into complex target distributions through a sequence of stochastic denoising or transport steps \citep{ho20,song20}. 
More recently, stochastic interpolant formulations have provided a unified view of flow- and diffusion-based generative modeling by learning a continuous bridge between two distributions \citep{albergo23}. 
In the conditional setting, the model can generate samples from a target field given an observed input field. 

For astrophysical applications, this is especially attractive because many inference problems are naturally field-to-field: galaxies trace dark matter fields, gas traces gravitational potentials, stellar populations trace formation histories, and mock observations trace intrinsic physical quantities through an observational forward model. 
A conditional generative model can therefore be used not only to predict a single best-fit map, but to generate an ensemble of physically plausible target maps conditioned on the same observed baryonic structure.
Recent studies have applied such generative models on generating realistic mock galaxy images \citep{smith22} and constraining large scale structures and cosmology with promising results \citep{mudur23,ono24,mudur25,mishra26}.

Generative field-level models are also increasingly timely because cosmological simulations now provide large, diverse training sets. 
Hydrodynamical simulation suites such as IllustrisTNG \citep{nelson19}, CAMELS \citep{villaescusanavarro21}, \Fire \citep{hopkins18}, and DREAMS \citep{rose25a} probe galaxy formation over different volumes, resolutions, feedback models, and cosmological parameters, with CAMELS and DREAMS specifically sampling that parameter space systematically (see Section~\ref{sec:methods}). 
These simulations do not remove the need for observational validation, but they provide controlled paired data: the observable baryonic fields and the underlying dark matter fields are known simultaneously. 
Such paired datasets make it possible to train models that learn the statistical relationship between light, stellar mass, gas, and dark matter, while explicitly testing how the learned relation changes under halo-to-halo variance, feedback variation, resolution changes, and simulation-domain shifts.

In this work, as a proof of concept, we take a first step toward using deep-learning-based imaging as a dark matter probe by asking whether the stellar density field contains enough information to reconstruct the dark matter density field of Milky Way-mass galaxies. 
We train a conditional generative model, BaryonBridge \citep{horowitz25}, on the Milky Way-mass halo suite from the DREAMS project. 
The model learns a stochastic bridge from two-dimensional stellar density maps to two-dimensional dark matter density maps, producing an ensemble of dark matter realizations for each input galaxy. 
We focus in this study on the idealized stellar-density-to-dark-matter-density problem, rather than on fully forward-modeled survey images. 
This isolates the physical baryon-dark matter mapping from observational effects such as sky background, point-spread functions, surface-brightness limits, bandpass dependence, and stellar-population modeling. 
Since simulation-trained models can learn features specific to the numerical method, resolution, subgrid feedback implementation, and sample selection of their training suite, validation on independent simulations is necessary to distinguish a genuinely physical stellar-dark-matter mapping from a simulation-domain-specific correlation.

The paper is organized as follows. 
We describe the conditional generative model and the construction of the stellar and dark matter density maps from the DREAMS training suite in Section~\ref{sec:methods}. 
We also introduce the out-of-domain validation samples from TNG50 and \Fire. 
In Section~\ref{sec:results}, we quantify performance on held-out DREAMS galaxies and out-of-domain simulations, with discussion on the implications for photometric dark matter inference in Section~\ref{sec:discussion}.
The main results are summarized in Section~\ref{sec:conclusion}.

\section{Methods}
\label{sec:methods}

\subsection{Stochastic interpolation model}
\label{sec:model}

\begin{figure*}
    \centering
    \includegraphics[width=0.95\linewidth]{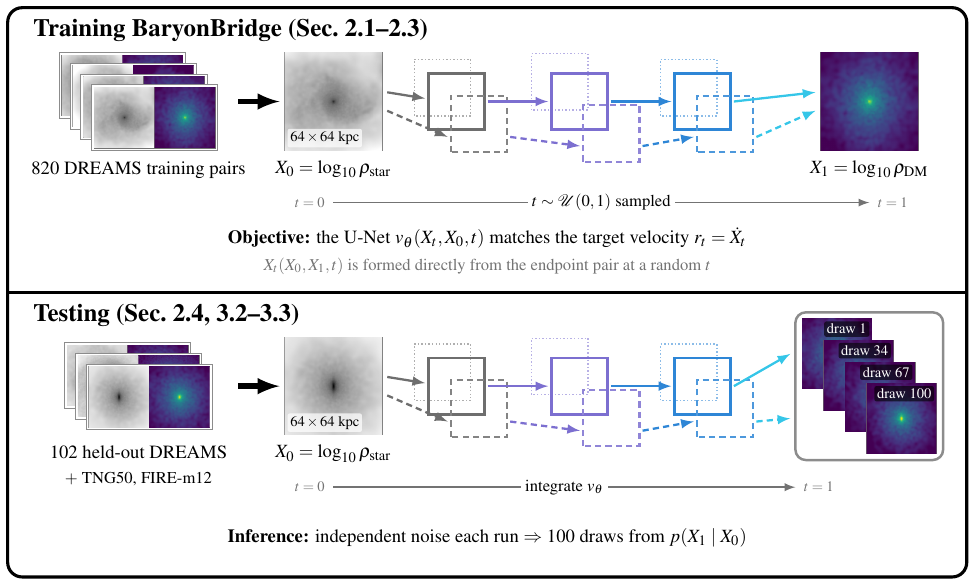}
    \caption{
    Schematic of the training (top) and inference (bottom) procedures. 
    Relevant sections are noted accordingly.
    }
    \label{fig:schematic}
\end{figure*}

We use a stochastic-interpolant generative model based on the BaryonBridge framework of \citet{horowitz25}.
BaryonBridge was developed to learn conditional mappings between dark-matter-only simulations and hydrodynamical baryonic fields by constructing a stochastic bridge between two physically related density fields.
Our implementation adopts the same broad framework, but applies it to a different physical regime and reverses the direction of the astrophysical inference problem.
Rather than mapping from dark matter fields to baryonic fields on cosmological volumes, we map from the stellar density field of an individual Milky Way-mass galaxy to its dark matter density field.
Compared to the original application, the task in this study focuses to galactic scales and thus is expected to be more challenging as the mapping is more sensitive to non-linear effects from baryonic physics.
A schematic is shown in Figure~\ref{fig:schematic}.

The goal of the model is to sample from the conditional distribution
\begin{equation}
p(X_1 \mid X_0),
\end{equation}
where $X_0 = \log_{10}(\rho_{\rm star})$ is the stellar density map and $X_1 = \log_{10}(\rho_{\rm DM})$ is the corresponding dark matter density map.
Both fields are represented on the same two-dimensional grid, with identical centering and face-on orientation.
In the fiducial setup used here, the maps are discretized onto $128\times128$ pixels.
The stellar map is used as the conditioning field, while the dark matter map is the target field.
Thus, each galaxy provides one pair for model training, validation, and testing. 

The stochastic-interpolant formulation defines a continuous stochastic path between an input field and a target field.
In our application, this path connects the stellar density map to the dark matter density map, rather than connecting a generic noise distribution to the target field.
A neural network is trained to learn the drift or velocity field that transports samples along this bridge.
At inference time, the learned stochastic differential equation is integrated from the stellar-density endpoint to the dark-matter-density endpoint using a stochastic Heun scheme \citep{kloeden92} with 100 discretization steps to reduce discretization error relative to a single-stage Euler-Maruyama integrator at fixed step count. 
Each step takes an Euler-Maruyama predictor and then a trapezoidal corrector that reuses the same Wiener increment.
The process produces a sample from the predicted dark matter distribution.
Repeating this sampling procedure for the same stellar density map for a given test galaxy yields an ensemble of plausible dark matter maps, which captures the non-uniqueness of the stellar-to-dark-matter mapping arising from halo-to-halo variance, assembly history, and baryonic feedback the model has learned from the training galaxies.

The drift model is implemented with a conditional convolutional U-Net architecture \citep{ronneberger15,ho20}, following the fully convolutional design used in BaryonBridge.
The network operates on the two-dimensional pixel grid at every layer, and the output is returned at the input resolution with each output pixel spatially aligned with the corresponding input pixel, so the geometry of the image is carried through the network by construction.
The U-Net encoder-decoder structure and skip connections allow the network to combine local density information with larger-scale morphology.
The interpolation time is embedded and supplied to the network, and the stellar density map is provided as the conditioning field throughout the generative process.
Additional scalar information, such as cosmological or feedback parameters, can be embedded following \citet{horowitz25}.
In this study, we focus on models trained with five simulation parameters from the DREAMS CDM MW suite (see Section~\ref{sec:training_data}), and perform inference with both the true parameters and arbitrary parameter choices to test how the presence or absence of accurate scalar conditioning affects model performance. 
Alternatively, the model can be trained without additional scalar features, focusing on the information contained in the stellar density field itself.
We show in Appendix~\ref{sec:appendix_noh} that such a model, trained on the stellar density map alone, still recovers most of the galaxy-to-galaxy variance in the dark matter profile, so the scalar parameters refine rather than drive the mapping.

As our science case is especially sensitive to the inner dark matter density profile, we introduce one modification to the original BaryonBridge training objective.
In the original implementation, a pixel-wise mean-squared-error objective weights each map pixel equally.
For approximately Cartesian maps of fixed pixel size, this implicitly gives larger total weight to the outer regions of the galaxy because annuli at larger radii contain more pixels, while the central region occupies only a small fraction of the image.
This geometric weighting is not aligned with the main goal of this work, which is to recover the radial structure of the dark matter profile, particularly in the inner galaxy where baryonic contraction and feedback-driven expansion are most relevant.
We therefore replace the uniform pixel weighting in the velocity-matching loss with a radially dependent weight function, $w(R)$, so that
\begin{equation}
\mathcal{L}_{w} = E\left[w(R)\,|v_\theta - r_t|^2\right],
\end{equation}
where $r_t$ is the target velocity connecting the stellar and dark matter maps, and $v_\theta$ is the drift model from the preceding paragraphs, with trainable weights $\theta$.
The expectation $E$ runs over training pairs.
For a pixel at projected galactocentric radius $R$ (measured from the center of the image), we adopt
\begin{equation}
w(R) = \frac{1}{\max(R, R_{\rm min})},
\end{equation}
with $R_{\rm min}=2$\,kpc, normalized to unit mean over the pixel grid before use in the loss.
This inverse-radius weighting approximately equalizes the total loss contribution of each annulus, which otherwise scales as $\sim 2\pi R\,dR$ under uniform pixel weighting. 
The floor $R_{\rm min}$ regularizes the central $1/R$ divergence so that the innermost pixels do not dominate the loss numerically.
Because all maps share a common centering and field of view, the projected radius $R$ of every pixel is a fixed property of the grid. 
$w(R)$ is precomputed once as a $128\times128$ array from the pixel-center coordinates and multiplied element-wise into the map of squared residuals before averaging. 
The pixel positions therefore enter through the loss rather than through the network input so the weighting changes only how much each location contributes to the gradient, not what the network is shown.
Thus, this weighting does not change the stochastic-interpolant formulation or the inference procedure.

Except for the modified spatial weighting of the training objective described above, we follow the BaryonBridge stochastic-interpolant implementation and refer the reader to \citet{horowitz25} for mathematical details, and to \citet{albergo23,chenyf24} for the general stochastic-interpolant framework.
The purpose of the present work is not to introduce a new generative architecture, but to test whether this class of conditional stochastic-interpolant models can be transferred from cosmological-field reconstruction to the internal structure of individual galaxies.
By applying the framework to stellar and dark matter density maps of Milky Way-mass galaxies, we isolate the astrophysical question of whether the observable stellar mass distribution contains sufficient information to probabilistically constrain the underlying dark matter density field.

\subsection{DREAMS simulation as training data}
\label{sec:training_data}

The model training is performed using the Milky Way-mass cold dark matter simulation suite from the DaRk mattEr and Astrophysics with Machine learning and Simulations (DREAMS) Project~\citep{rose25a}. 
The simulations are performed within a $\Lambda$CDM framework using the moving-mesh code \texttt{AREPO} \citep{springel10,springel19,weinberger20}.
The suite~\citep{rose25b} consists of 1024 cosmological zoom-in simulations with varying initial conditions to study the effect of halo-to-halo variance. 
We refer the reader to previous works for comprehensive discussions of the simulation methodology and validation \citep{rose25b,rose25c,garcia26,lilie26}. 

We describe features of the simulations that are important for interpreting the results of the trained diffusion models.
Each simulated Milky Way-mass galaxy has a dark matter particle mass resolution of $1.8 \times (\Omega_m / 0.314) \times 10^{6} \, M_\odot$ and a baryonic mass resolution of $2.8 \times 10^{5} \, M_\odot$, with a physical gravitational softening length of 0.441 kpc at $z = 0$. 
Baryonic processes are modeled using the IllustrisTNG galaxy formation model \citep{weinberger17,pillepich18a}, which is an extension of the Illustris galaxy formation model \citep{vogelsberger13, torrey14}, and includes prescriptions for gas cooling \citep{katz96}, star formation \citep{springel03}, 
stellar feedback, black hole growth \citep{sijacki15}, and active galactic nucleus feedback~\citep{weinberger18}. 
Each simulation is run with a unique combination of IllustrisTNG model parameters drawn from a predefined parameter space. 
The parameter combinations are designed to capture the impact of cosmological and baryonic feedback model variations similar to the CAMELS project~\citep{villaescusanavarro21}, with the CAMELS having a wider variation in cosmology and an additional active galactic nuclei parameter for probing a broader range of halo masses in the high-redshift universe.
Unlike the original CAMELS suites, which sample the parameter space using a Latin-hypercube design, DREAMS adopts a Sobol low-discrepancy sequence, providing a more uniformly space-filling and readily extensible sampling of the multidimensional parameter space \citep{sobol67,rose25a}.
The varied parameters in DREAMS include two cosmological parameters, the total matter density $\Omega_m$ and the amplitude of matter fluctuations $\sigma_8$, as well as three baryonic parameters associated with the TNG feedback model. 
These baryonic parameters control the overall strength of baryonic feedback: specific energy injected by supernova feedback (hereafter $\bar{e}_w$), the normalization of supernova wind speed ($\kappa_w$), and the coupling efficiency of high-accretion-rate AGN feedback ($\epsilon_{f,\,{\rm high}}$). 
A comprehensive review of the impact of these parameters on the dark matter density profile is presented by \citet{garcia26}.

For the diffusion model introduced in the previous section, the DREAMS suite is an ideal training dataset.
The varying initial conditions and simulation parameters (see e.g., Table~1 of \citealt{garcia26}) allow the model to learn the stochastic mapping in the context of possible halo-to-halo variances and systematic uncertainties in the simulation model.
These variations in the training galaxies are crucial for preventing the model from overfitting to Milky Way-mass galaxies that are defined by the default simulation parameters.
We split the 1024 DREAMS galaxies into 820 training, 102 validation, and 102 test samples. 

\subsection{Density map construction}
\label{sec:map_gen}

For each simulated galaxy in the DREAMS suite, we construct paired stellar and dark matter density maps with each galaxy oriented face-on, by aligning the stellar angular momentum vector within $10$\,kpc to be along the line of sight. 
We use density maps evaluated at the mid-plane of the galaxy, rather than line-of-sight-integrated surface-density maps.
This choice isolates the idealized density-to-density mapping and avoids additional projection choices.
Observational projection effects are deferred to future studies.

The maps are evaluated on a regular two-dimensional grid in the rotated $x$-$y$ plane.
In the fiducial setup, we use a field of view of $64\times 64\,{\rm kpc}$ and a resolution of $128\times128$ pixels, corresponding to a pixel scale of $0.5$\,kpc.
The field of view is chosen based on the typical extent of stellar components, e.g. the disk and inner stellar halo, in Milky Way-mass galaxies, while the resolution is driven partially by the spatial resolution of DREAMS (e.g., softening lengths at 0.441\,kpc).
To first order, the resolution can also be interpreted as the pixel resolution scale for a given instrument used to obtain the galaxy images, which are directly tied to the stellar density profiles.
Assuming an instrument pixel scale of $0.1$\,arcsec, typical for current and upcoming space telescopes such as Euclid/VIS and Roman/WFI, these images correspond to galaxies at a redshift of $\sim0.33$, well within the expected source redshift distributions of upcoming surveys \citep{eifler21,mellier25}.

We estimate both the stellar density field and the dark matter density field with the same adaptive kernel-smoothing procedure as described in \citet{garcia26}.
For each grid point, we build a KD-tree \citep{bentley75} from the corresponding particle distribution and identify the nearest neighboring particles.
The local smoothing length is set by the distance to the outermost of 128 neighbors, and the density at the grid point is computed by summing the particle masses weighted by a cubic-spline kernel.
The smoothing prescription for the input and target fields provides a continuous, non-zero density estimate even in low particle number regions.

The resulting stellar and dark matter maps are stored as logarithmic density fields, $\log_{10}(\rho_{\rm star})$ and $\log_{10}(\rho_{\rm DM})$, respectively.
These logarithmic maps are then normalized by the data loader before being passed to the diffusion model.
Thus, each training example consists of a paired stellar density condition and dark matter density target constructed from the same galaxy, in the same coordinate frame, and with identical spatial sampling.

\subsection{Validation}
\label{sec:validation}

We validate the model in two complementary ways. 
First, we test the model on held-out DREAMS galaxies that are not used in training. 
This evaluates interpolation within the same simulation framework and quantifies how well the model captures halo-to-halo variance within the DREAMS suite. 
Second, we apply the trained model to galaxies drawn from independent hydrodynamical simulations. 
These out-of-domain tests are designed to determine whether the learned stellar-to-dark-matter mapping is robust to changes in numerical resolution, simulation volume, subgrid feedback implementation, and galaxy assembly history. 
We use the IllustrisTNG and \Fire simulation suites for this purpose.

\subsubsection{IllustrisTNG simulations}

The IllustrisTNG simulations provide a controlled out-of-domain test because they share the same moving-mesh code, \texttt{AREPO}, and the same broad galaxy-formation model family as DREAMS, but differ in simulation volume, mass resolution, and sample construction. 
The full IllustrisTNG suite consists of three cosmological volumes, TNG50, TNG100, and TNG300 \citep{nelson18,pillepich18b,naiman18,marinacci18,springel18,pillepich19,nelson19}. 
The IllustrisTNG cosmological parameters are consistent with $\Lambda$CDM model determined by Planck XIII \citep{planck16} to be $\Omega_\mathrm{m} = 0.3089,\, \Omega_\mathrm{b} = 0.0486,\, \Omega_{\Lambda} = 0.6911,\, H_{0} = 100 \, h \; \rm km \; \rm s^{-1} \; \rm Mpc^{-1}$ with $h = 0.6774,\, \sigma_{8} = 0.8159,$ and $n_\mathrm{s} = 0.9667.$
In this work, we use TNG50, with its higher mass and spatial resolution, to test whether the DREAMS-trained model depends on the effective resolution of the training simulations. 
TNG50 has a dark matter mass resolution of $4.5\times10^{5}\,M_\odot$, a baryonic mass resolution of $8.5\times10^{4}\,M_\odot$, and a physical gravitational softening length of $0.288$\,kpc at $z=0$. 

We select Milky Way-mass TNG50 galaxies using criteria similar to those used for the DREAMS sample. 
Specifically, we require halos to satisfy $M_{\rm halo}=5\times10^{11}$-$2\times10^{12}$\,\msun and to be isolated such that they are at least $1$\,Mpc away from the virial radius of the closest halo with $M_{\rm halo}>5\times10^{11}$\,\msun. 
A total of $264$ galaxies are selected following the criteria.
For each selected galaxy, we construct stellar density and dark matter density maps following the same procedure used for the DREAMS maps (see Section~\ref{sec:map_gen}). 
The resulting maps are then passed through the DREAMS-trained model without retraining or fine-tuning.
For models trained with additional simulation parameters, we adopt the true fiducial values from the original IllustrisTNG simulation, as the underlying \texttt{AREPO} code and galaxy-formation model are identical to DREAMS. 

This comparison is designed to isolate the effect of simulation resolution and sample selection. 
Because the TNG galaxy-formation model lies within the same broad model family used by DREAMS, and because the fiducial TNG cosmological and feedback parameters are contained within the DREAMS parameter ranges, large systematic failures on TNG50 would indicate that the learned map is sensitive to resolution or the TNG simulation parameters.

\subsubsection{\Fire simulations}

We also test the model on Milky Way-mass galaxies from the \Latte \citep{wetzel16} suite of the Feedback In Realistic Environments \Fire-2 simulations \citep{hopkins18}. 
We use eight publicly available zoom-in simulations, m12b, m12c, m12f, m12i, m12m, m12r, m12w, and m12z \citep{wetzel23}. 
These simulations are run with the \texttt{GIZMO} code \citep{hopkins15} and use the \Fire-2 galaxy-formation model, which explicitly models stellar feedback from supernovae, stellar winds, radiation pressure, photoionization, and photoelectric heating  \citep[for a detailed discussion of the simulations, see][]{wetzel16,hopkins18,sanderson18,wetzel23}. 
The eight \Fire galaxies are selected because they are designed to form Milky Way-mass systems and be cosmologically isolated from halos of similar or greater mass, with halo masses ranging between $0.925$-$1.71\times10^{12}\,M_\odot$. 
The dark matter mass resolution, baryonic mass resolution, dark matter softening length, and baryonic gravitational softening length range from $2.1\times10^{4}\,M_\odot$ to $3.9\times10^{4}\,M_\odot$, from $4.2\times10^{3}\,M_\odot$ to $7.1\times10^{3}\,M_\odot$, from $33$\,pc to $40$\,pc, and from $3.2$\,pc to $4.0$\,pc at $z=0$, with three sets of cosmological parameters used (see Table~1 of \citealt{wetzel23}).
We process the \Fire snapshots into stellar density and dark matter density maps using the same procedure adopted for DREAMS and TNG50 (see Section~\ref{sec:map_gen}). 
When applying the model trained with DREAMS simulation parameter conditioning to the FIRE suite, we marginalize over the DREAMS-specific baryonic feedback parameters as there are no direct mapping of these parameters to the FIRE galaxy formation models.
Specifically, we sample values for the three baryonic parameters following a flat prior distribution within the sampled range in DREAMS. 
In practice, we draw $100$ parameter configurations from this flat prior for each \Fire galaxy, one stochastic realization each, when generating the prediction posterior for the \Fire tests.

The \Fire simulations therefore provide a substantially stronger domain-shift test than TNG50. 
They differ from DREAMS not only in mass resolution and hydrodynamic solver, but also in the implementation and time variability of baryonic physics.
The resulting test therefore probes whether a model trained only on DREAMS can recover dark matter profiles in galaxies whose baryonic structures arise from a different feedback model.

Specifically, the inner dark matter profiles of \Fire galaxies are known to differ systematically from those in IllustrisTNG-like models \citep{lazar20}. 
Bursty stellar feedback redistributes dark matter in the central regions and produces shallower inner profiles than those expected from the DREAMS training set. 
Specifically, the bursty feedback interrupts adiabatic contraction of the halos more than smooth feedback models, which are not captured even with the DREAMS physics variation factoring in high feedback (see e.g., \citealt{hussein25,garcia26}).
Thus, disagreement in the inner regions of \Fire galaxies should not necessarily be interpreted as a numerical failure of the diffusion model. 
Instead, it provides a physically meaningful diagnostic of where the stellar-density-only mapping becomes non-unique across galaxy-formation models.

\section{Results}
\label{sec:results}

\subsection{Metrics}
\label{sec:metrics}

\subsubsection{Uncertainty-weighted residuals}

To evaluate BaryonBridge, we generate predictions by applying the trained stochastic interpolant to held-out simulated galaxies without any data randomization or augmentation. 
Since the sampler is stochastic, we generate independent predictions for each galaxy by repeating the same inference call with identical conditioning inputs. 
These repeated stochastic draws sample the model's predictive distribution for a fixed galaxy. 
Unless stated otherwise, we use $100$ inference draws per galaxy throughout.
Prediction performance is first quantified at the pixel level using the ensemble of stochastic predictions generated for each galaxy. 
For each pixel $i$, we take the mean of the stochastic predictions of the logarithmic dark matter density, $\overline{\log ({\rho}_{{\rm pred},i})}$, as the fiducial predicted dark matter density, and estimate the corresponding predictive uncertainty from the dispersion of the same ensemble. 
Specifically, we define a compact uncertainty amplitude,
\begin{equation}
    \sigma_{{\rm pred},i} = \left[\frac{1}{N_{\rm draw}-1}\sum_{j=1}^{N_{\rm draw}}\left(\log(\rho_{{\rm pred},i}^{(j)}) - \overline{\log(\rho_{{\rm pred},i})}\right)^{2}\right]^{1/2},
\end{equation}
where $\log(\rho_{{\rm pred},i}^{(j)})$ is the $j$-th of the $N_{\rm draw}$ stochastic predictions of the dark matter density at that pixel and $\overline{\log(\rho_{{\rm pred},i})}$ denotes their mean over the ensemble.
We then define the uncertainty-weighted residual,
\begin{equation}
    \Delta_i = \frac{\overline{\log({\rho}_{{\rm pred},i})} - \log(\rho_{{\rm true},i})}{\sigma_{{\rm pred},i}},
\end{equation}
where $\log(\rho_{{\rm true},i})$ is the true dark matter density in the target map. 
This quantity measures the prediction error in units of the model-estimated uncertainty. 
Pixels with narrow predictive distributions are therefore required to match the target more closely, while pixels with broader predictive distributions are allowed larger absolute deviations without being assigned the same statistical significance.

For a given galaxy, the full set of $\Delta_i$ values across all pixels provides a compact diagnostic of both the accuracy and calibration of the model prediction. 
If the model is unbiased and its stochastic samples provide a well-calibrated estimate of the predictive uncertainty for that galaxy, the combined distribution of pixel-level uncertainty-weighted residuals should be approximately consistent with a standard normal distribution. 
A nonzero mean indicates an overall bias in the predicted dark matter density map, positive for systematic over-prediction and negative for under-prediction.
The standard deviation measures uncertainty calibration: a value greater than unity means the true residuals are broader than the model uncertainties imply, so the model is underestimating its uncertainty, while a value less than unity means the opposite.

\subsubsection{Radial profiles}

The same uncertainty-weighted residual map can also be binned spatially to identify where the model succeeds or fails within an individual galaxy.
We consider radial and/or azimuthal bins on the two-dimensional image and compute the mean and standard deviation of $\Delta_i$ within each bin, interpreted exactly as above but now locally.
The spatially resolved diagnostic allows us to determine whether the model performance depends systematically on position within the galaxy, for example whether the prediction is biased in the central region, in the outskirts, or along particular azimuthal structures.

As dark matter studies are often most directly concerned with the radial density structure, we further use the radial bins to construct a population-level diagnostic.
Here the residual is formed at the level of the azimuthally averaged profile rather than pixel by pixel.
For galaxy $g$ and radial bin $b$ we take the annulus-averaged predicted and true densities, and normalize their difference by the spread of the annulus-averaged profile across the stochastic draws,
\begin{equation}
    \Delta_{g,b} = \frac{\langle\overline{\log({\rho}_{{\rm pred}})}\rangle_b - \langle\log(\rho_{{\rm true}})\rangle_b}{{\rm std}_{j}\left[\langle\log(\rho_{{\rm pred}}^{(j)})\rangle_b\right]},
\end{equation}
where $\langle\cdot\rangle_b$ denotes the mean over the pixels in bin $b$ and $j$ indexes the stochastic draws.
We then combine $\Delta_{g,b}$ across a sample of galaxies, such as the DREAMS test set or the out-of-domain TNG50 and \Fire samples.
The mean and width of the resulting distribution at fixed radius carry the same meaning as before, now at the population level: a radial trend in the mean indicates a population-level bias in the recovered dark matter density profile, and a radial trend in the width indicates radius-dependent miscalibration.

\subsubsection{Anisotropy in $m=2$ azimuthal mode}

The radial density profile is azimuthally symmetric by construction, so we add a third diagnostic that measures how well the model recovers the anisotropy of the target dark matter maps.
It decomposes the dark matter density into azimuthal Fourier modes, and we report the $m=2$ mode for simplicity.
For galaxy $g$ and radial bin $b$,
\begin{equation}
    c_{m}(g,b) = \frac{\sum_{i \in b} \rho_{i}\, e^{\,\mathrm{i} m \theta_{i}}}{\sum_{i \in b} \rho_{i}},
    \label{eq:fourier}
\end{equation}
where $i$ indexes the pixels of annulus $b$, $\rho_{i}$ is the linear dark matter density of pixel $i$, and $\theta_{i}$ is its position angle in the plane of the map, measured as $\arctan(x/y)$ so that $\theta=0$ lies along the $+y$ axis and increases towards $+x$.
The sum runs over pixels at their exact $\theta_{i}$, without azimuthal binning.
The $m=2$ amplitude $A_{2}=\left|c_{2}\right|$ measures the Fourier mode strength and $\psi_{2}=2^{-1}\arg c_{2}$ is the position angle of the major axis of that mode.

The amplitude is scored with the same uncertainty-weighted residual used above,
\begin{equation}
    \Delta^{A_2}_{g,b} = \frac{\overline{A_2^{\rm pred}}(b) - A_2^{\rm true}(b)}{{\rm std}_{j}\left[A_2^{{\rm pred},(j)}(b)\right]},
\end{equation}
with the overbar again denoting the mean over the $N_{\rm draw}$ stochastic draws indexed by $j$, and the orientation with an alignment statistic,
\begin{equation}
    \mathcal{A}_{g} = \frac{1}{N_{b}}\sum_{b} \cos\left[2\left(\overline{\psi_2^{\rm pred}}(b) - \psi_2^{\rm true}(b)\right)\right],
    \label{eq:alignment}
\end{equation}
in which $\overline{\psi_2^{\rm pred}}$ is the position angle of the ensemble-averaged moment $\overline{c_2^{\rm pred}}$ and $N_{b}$ is the number of radial bins.
$\mathcal{A}_{g}=1$ if the predicted major axis coincides with the true one at every radius, $-1$ if the two are everywhere perpendicular, and $0$ if the predicted orientation is unrelated to the truth.
In other words, if the model only recovers the azimuthally averaged radial profile, we would expect $\mathcal{A}_{g}\simeq0$.

We evaluate this diagnostic over $2 < r < 32$\,kpc.
The outer limit is the largest circle inscribed in the $64\times64$\,kpc field of view, beyond which the annulus samples only the corners of the image, at $\theta=\pm45^{\circ}$ and $\pm135^{\circ}$, which is a purely geometric $m=4$ signal.
The inner limit removes annuli containing too few pixels for a position angle to be meaningful.

\subsection{Performance on held-out DREAMS galaxies}
\label{sec:dreams_validation}

\subsubsection{Individual galaxies}
\label{sec:ind_gal}

We first evaluate the model on $102$ held-out Milky Way-mass galaxies from the DREAMS suite. 
These galaxies are drawn from the same simulation suite as the training data but are not used during training. 
This test therefore measures the model's interpolation performance within the DREAMS domain and provides the baseline against which the out-of-domain TNG50 and \Fire tests are compared.
For the majority of the galaxies in the test set, the median per-pixel model uncertainty for the predicted dark matter density is $\sim0.096$\,dex, across the 102 held-out galaxies and only a weak dependence on projected radius.
Over the same 102 galaxies, the pixel-to-pixel scatter of $\overline{\log({\rho}_{{\rm pred}})}-\log(\rho_{{\rm true}})$ has a median of $0.085$\,dex, $\sim11\%$ smaller than the median predicted per-pixel uncertainty.
The stochastic ensemble therefore returns an uncertainty comparable to the size of the error it is meant to describe, indicating that the model is well calibrated and even mildly conservative at the pixel level.

\begin{figure*}
    \centering
    \includegraphics[width=0.95\linewidth]{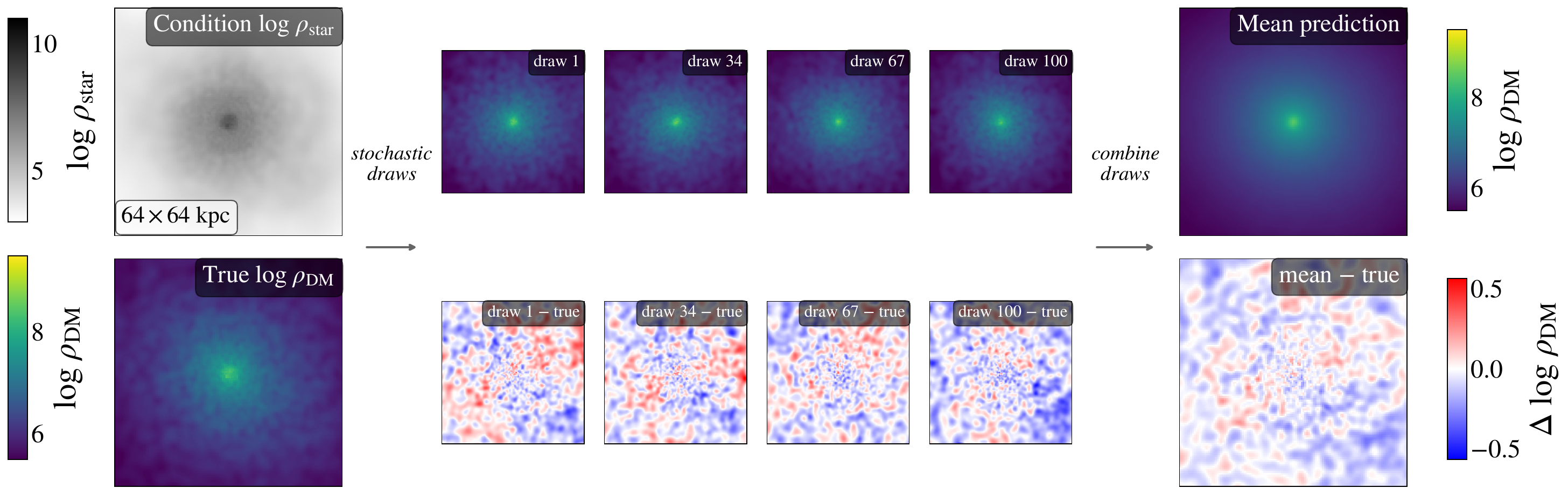}
    \caption{
    The left two panels show the condition stellar density input and true dark matter density map in $64\times64$\,kpc field of view for an example DREAMS galaxy in the held-out sample.
    The middle panels show four of the $100$ individual stochastic predicted dark matter maps for this galaxy.
    The right panels show the combined predicted dark matter density map and $\Delta$-residuals.
    Read from left to right, the panel arrangement therefore also serves as a schematic of the inference procedure described in Section~\ref{sec:model}: the conditioning stellar density map is passed to the trained model, which is sampled repeatedly to generate an ensemble of stochastic draws, and these draws are then combined pixel-wise into the single mean prediction shown at right.
    }    
    \label{fig:dreams_map}
\end{figure*}

We next examine the prediction on individual galaxies.
Figure~\ref{fig:dreams_map} shows a randomly selected example of the stellar density input, true dark matter density map, predicted dark matter density maps (including individual draws and final mean), and $\Delta$-residuals for held-out DREAMS galaxies. 
For this example galaxy, the predicted map reproduces the overall structure of the true dark matter distribution. 
The model achieves a mean $\Delta$-residual of $-0.37$, with a pixel-to-pixel scatter of $0.87$ for this galaxy. 
In absolute terms, the predicted and true dark matter densities differ by $-0.04\pm0.08$\,dex across pixels, against a median predicted uncertainty of $0.096$\,dex for this galaxy.

\begin{figure}
    \centering
    \includegraphics[width=0.95\linewidth]{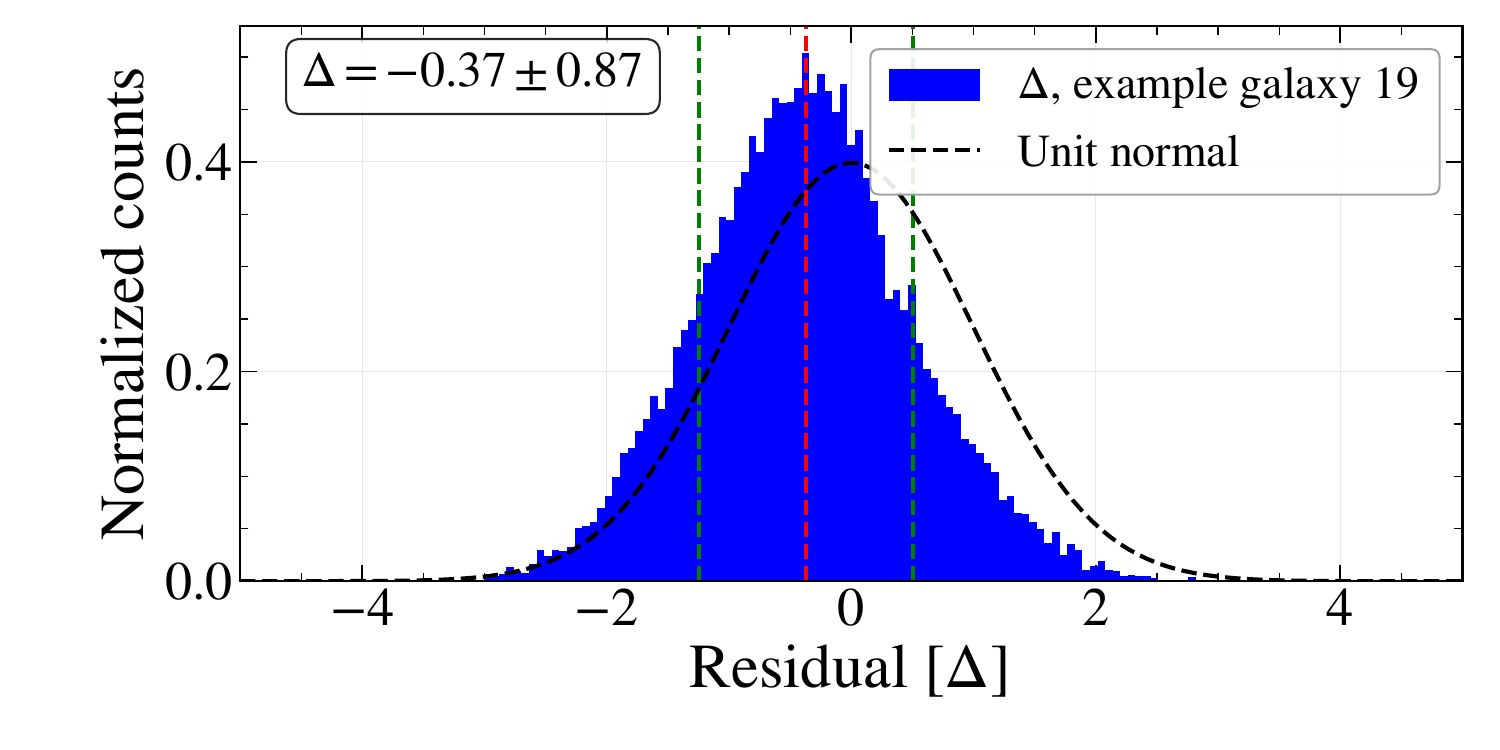} \\
    \includegraphics[width=0.95\linewidth]{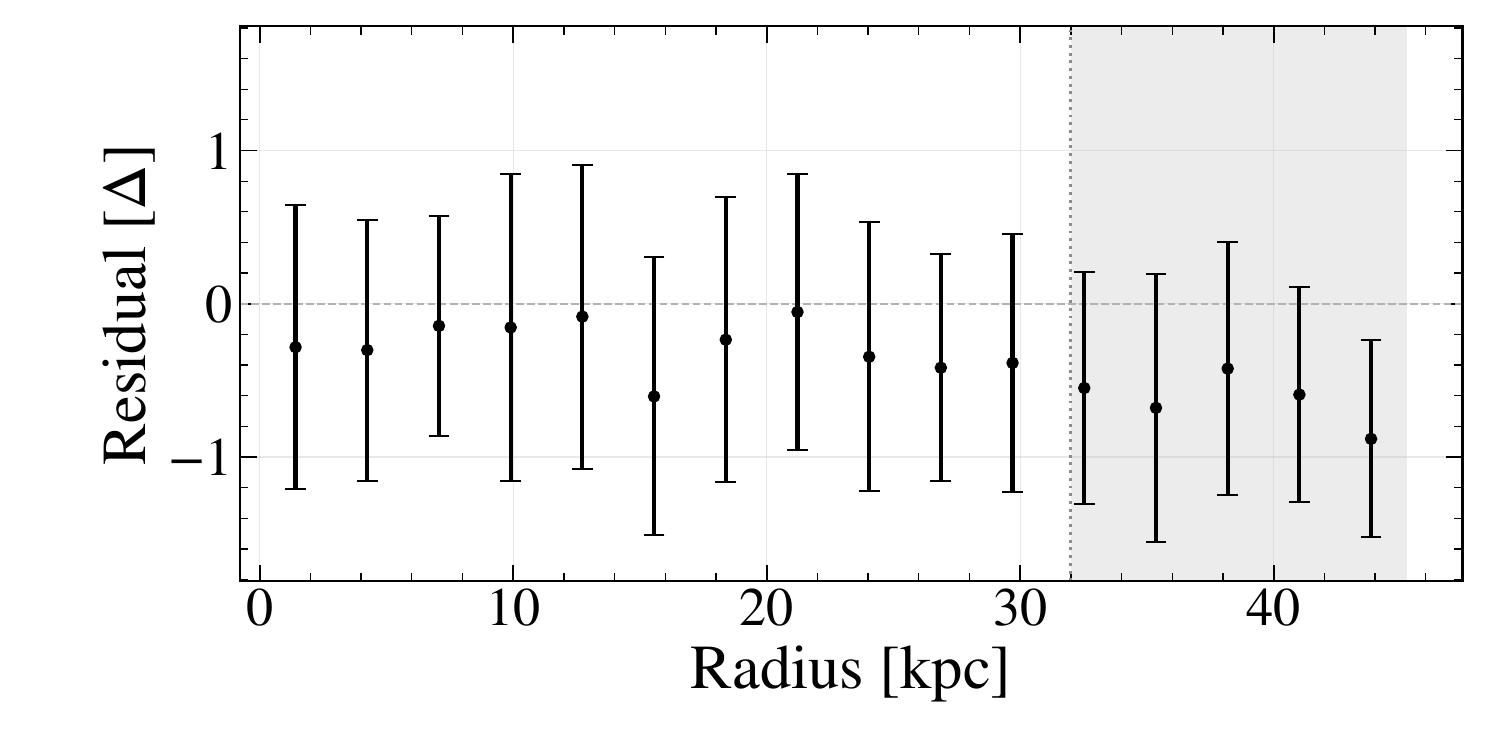} \\
    \includegraphics[width=0.95\linewidth]{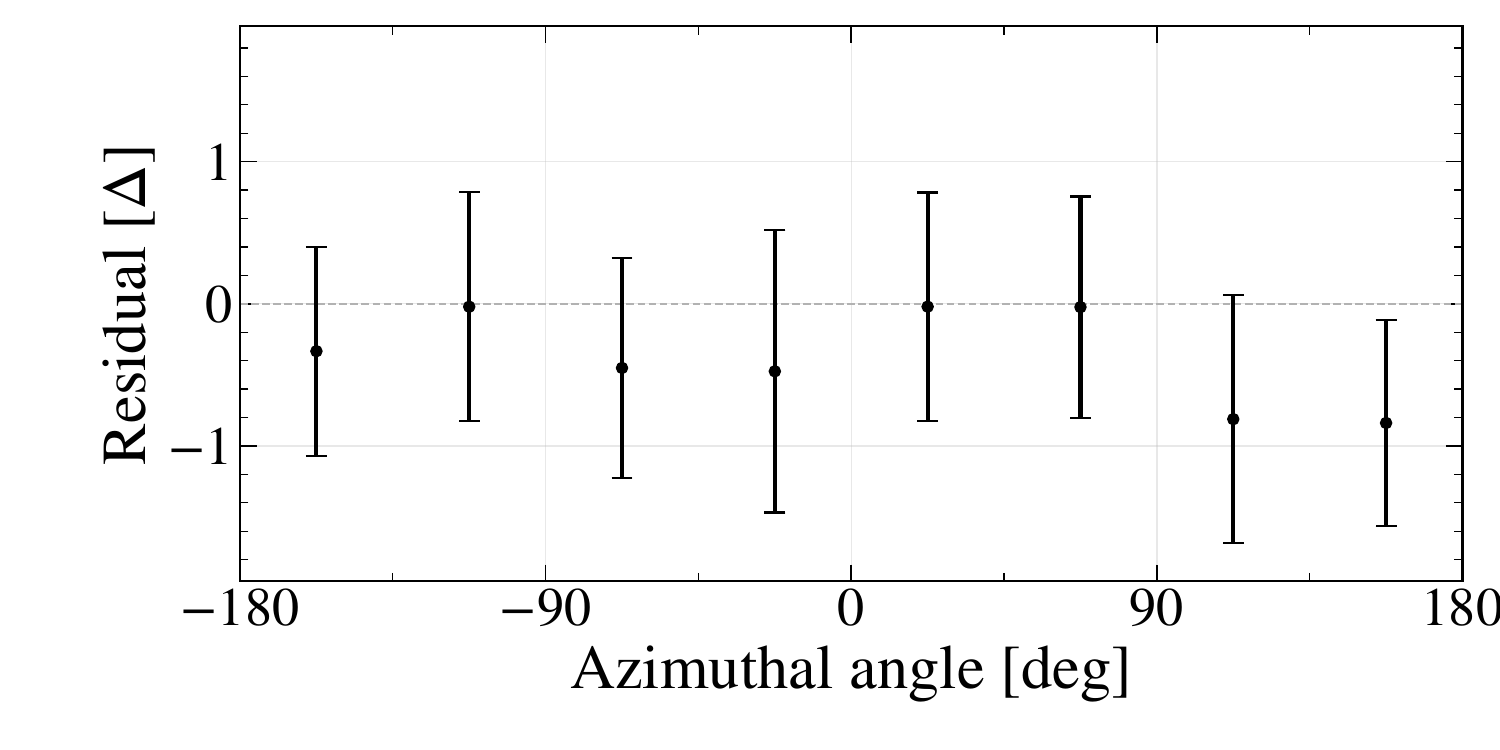}
    \caption{
    The top panel shows the distribution (blue) of the $\Delta$-residuals for the same galaxy in Figure~\ref{fig:dreams_map}, with the mean (red dashed line) and $\pm1$ standard deviation (green dashed lines).
    A unit normal distribution is shown in black dashed curve for reference.
    The middle and bottom panels show the means and standard deviations of the $\Delta$-residuals in sixteen radial bins and eight azimuthal bins, respectively. 
    The dashed horizontal lines mark $\Delta = 0$.
    In the middle panel the shaded region beyond $r=32$\,kpc marks radii at which an annulus is no longer fully contained in the $64\times64$\,kpc field of view, so those bins average over the corners of the image alone.
    Azimuthal angle in the bottom panel is measured in the plane of the map as $\arctan(x/y)$, so that $\theta=0$ lies along the $+y$ axis and $\theta$ increases towards $+x$.
    In both coordinates the binned residuals scatter about zero with amplitudes comparable to their own uncertainties, indicating no strong radial or azimuthal bias for this galaxy.
    }
    \label{fig:dreams_residual}
\end{figure}

We then examine the distribution of the $\Delta$-residual, and the radial and azimuthal dependence for this particular galaxy as described in Section~\ref{sec:metrics}. 
In Figure~\ref{fig:dreams_residual}, we split the images of same galaxy from Figure~\ref{fig:dreams_map} into sixteen radial bins and eight azimuthal bins and repeat the mean and standard deviation calculation for the $\Delta$-residuals within each bin.

Across all test DREAMS galaxies, the binned $\Delta$-residual distributions allow us to identify failure modes within the DREAMS domain.
We refer to the combination of the residual map, the full pixel-level $\Delta$ distribution, and the radial and azimuthal dependence of the binned $\Delta$ statistics as the ``individual-galaxy residual morphology''.
Inspecting these together is informative because they are not equivalent: a galaxy can have an acceptable global $\Delta$ distribution while still showing localized radial or azimuthal structure, and conversely a nonzero global mean $\Delta$ with little spatial dependence indicates a nearly uniform normalization offset rather than a shape error in the recovered profile.
By visual inspection, we find four broad, non-exclusive modes across the held-out sample: well-calibrated cases with little coherent structure, globally over- or under-predicted cases, cases with coherent radial trends, and cases with azimuthal structure. 
The last case is a useful diagnostic of substructure, asymmetry, or environmental features that a purely radial comparison would miss.
We treat these as descriptive rather than as a formal classifier.
Appendix~\ref{sec:appendix_stamps} shows examples of all four across 16 randomly selected held-out galaxies, which highlight that most of them are well-calibrated.

\begin{figure}
    \centering
    \includegraphics[width=0.95\linewidth]{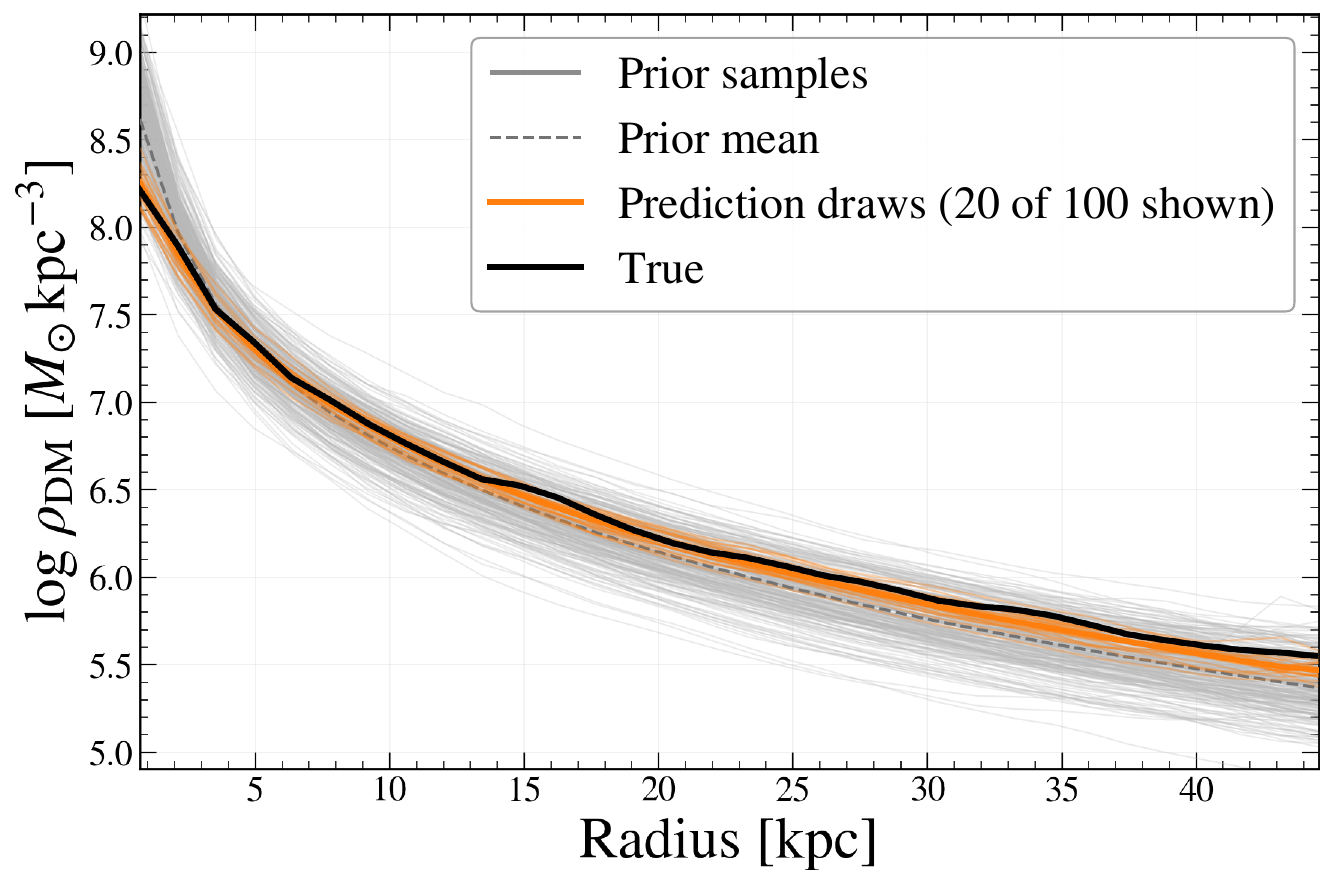} 
    \caption{ 
    Azimuthally averaged dark matter density profile for the same held-out DREAMS galaxy shown in Figure~\ref{fig:dreams_map}.
    Gray curves show a random subset of radial profiles from the DREAMS training set, representing the training-set prior at fixed Milky Way-mass selection.
    Orange curves show the stochastic BaryonBridge prediction ensemble for the held-out galaxy, and the black curve shows the true dark matter density profile.
    The narrowing of the posterior relative to the training-set prior illustrates that the conditioning stellar density map provides galaxy-specific information about the dark matter profile.
    }
    \label{fig:dreams_profile_comp}
\end{figure}

Despite the various residual morphology, the model is recovering the radial dark matter density profiles of the DREAMS galaxies well in most cases.
Figure~\ref{fig:dreams_profile_comp} places the same example galaxy in the context of the training-set prior.
The gray curves show the radial dark matter density profiles of the DREAMS galaxies, while the orange curves show the posterior ensemble of stochastic BaryonBridge predictions for the held-out galaxy.
The posterior is substantially narrower than the training-set prior and is centered close to the true radial profile, demonstrating that the stellar density map provides information beyond the average dark matter profile of Milky Way-mass DREAMS galaxies.
For this example, the posterior ensemble spans $\sim0.05$\,dex ($1\sigma$) at fixed radius and tracks the true azimuthally averaged profile to within $0.03$\,dex over $r\lesssim30$\,kpc, whereas the prior has a $1\sigma$ width of $0.14$-$0.15$\,dex at the same radii, broader by a factor of $\sim3$.
Thus, the model is using the input stellar structure to predict the dark matter profile within the broader halo-to-halo diversity of the training suite.

\begin{figure}
    \centering
    \includegraphics[width=0.95\linewidth]{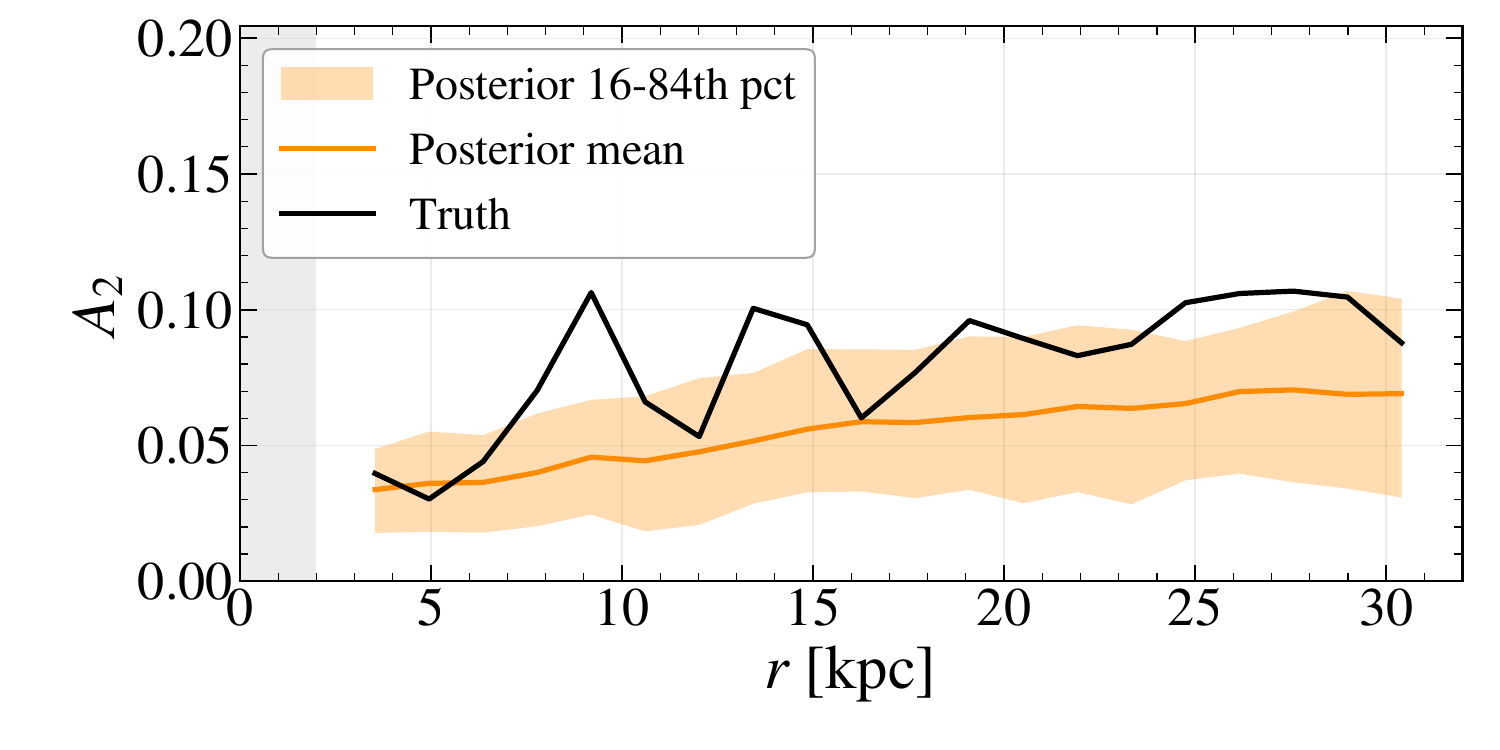} \\
    \includegraphics[width=0.95\linewidth]{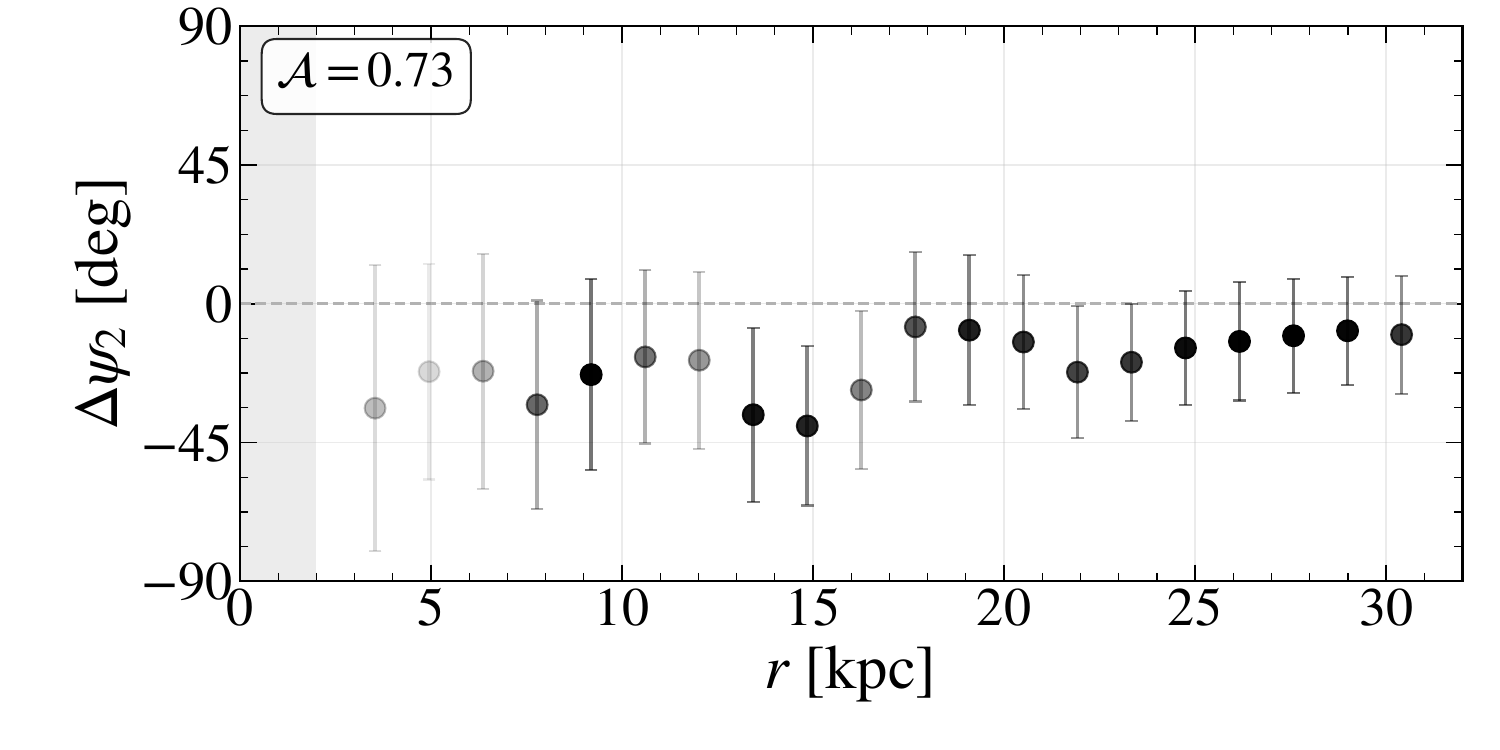}
    \caption{
    Azimuthal $m=2$ structure of the projected dark matter density for the same held-out DREAMS galaxy shown in Figure~\ref{fig:dreams_map}.
    The top panel shows the anisotropy amplitude $A_{2}$ against projected radius: black is the truth, orange the mean over the $100$ stochastic draws, and the shaded band the $16$th-$84$th percentile range of the ensemble.
    The bottom panel shows the offset $\Delta\psi_{2}$ between the predicted and true major-axis position angles, with error bars giving the circular standard deviation of $\psi_{2}$ across draws.
    Marker opacity is scaled by the true $A_{2}$, so annuli that are nearly axisymmetric appear faded.
    Both panels are restricted to $2 < r < 32$\,kpc as described in Section~\ref{sec:metrics}.
    }
    \label{fig:dreams_anisotropy}
\end{figure}

The radial profile is, however, only the $m=0$ part of the comparison.
Figure~\ref{fig:dreams_anisotropy} applies the azimuthal decomposition of Section~\ref{sec:metrics} to the same galaxy.
Its projected dark matter distribution has mild $m=2$ bar-like anisotropy, with a true $A_{2}$ ranging between $0.03$ and $0.11$ across the twenty annuli and averaging $0.08$, and the model reproduces that amplitude to within the ensemble spread, under-predicting the amplitude by $\Delta^{A_2}=-0.98$ on average.
The orientation is recovered considerably more sharply than the amplitude: the predicted major axis lies within $30^{\circ}$ of the true one in sixteen of the twenty annuli, with a median offset of $19^{\circ}$, giving $\mathcal{A}=0.73$ for this galaxy.
Appendix~\ref{sec:appendix_stamps} shows both panels of Figure~\ref{fig:dreams_anisotropy} repeated for 16 randomly selected held-out galaxies.

\subsubsection{Population-level DREAMS calibration}
\label{sec:all_gal}

\begin{figure*}
    \centering
    \includegraphics[width=0.9\linewidth]{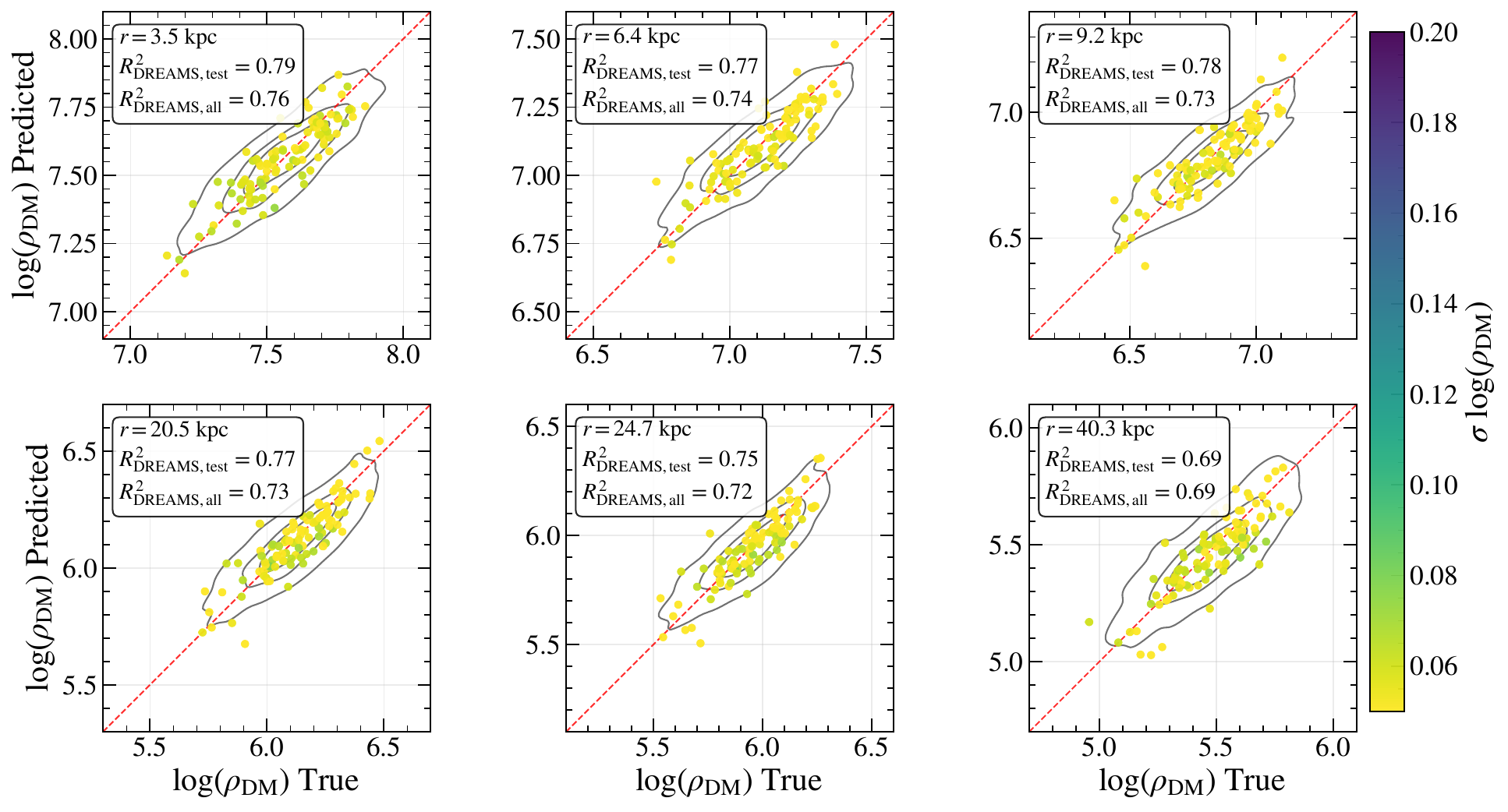} 
    \caption{
    Predicted versus true azimuthally averaged dark matter density at six radial locations, for the in-domain DREAMS sample.
    The 102 held-out test galaxies are drawn as filled circles, and the full 1024-galaxy DREAMS population is shown behind them as grey density contours enclosing $39$, $68$ and $95$ percent of the sample. 
    $R^2$ values are quoted separately for the two.
    Each galaxy is color-coded by the model uncertainty on the azimuthally averaged profile in the corresponding radial bin.
    The held-out DREAMS galaxies track the one-to-one line as tightly as those seen during training at every radius ($R^2_{\mathrm{test}} = 0.69$-$0.79$), whereas the out-of-domain samples scatter more widely, with TNG50 degrading toward large radii and all eight \Fire galaxies lying above the line in the innermost bin.
    }
    \label{fig:dreams_profiles}
\end{figure*}

\begin{figure*}
    \centering
    \includegraphics[width=1.0\linewidth]{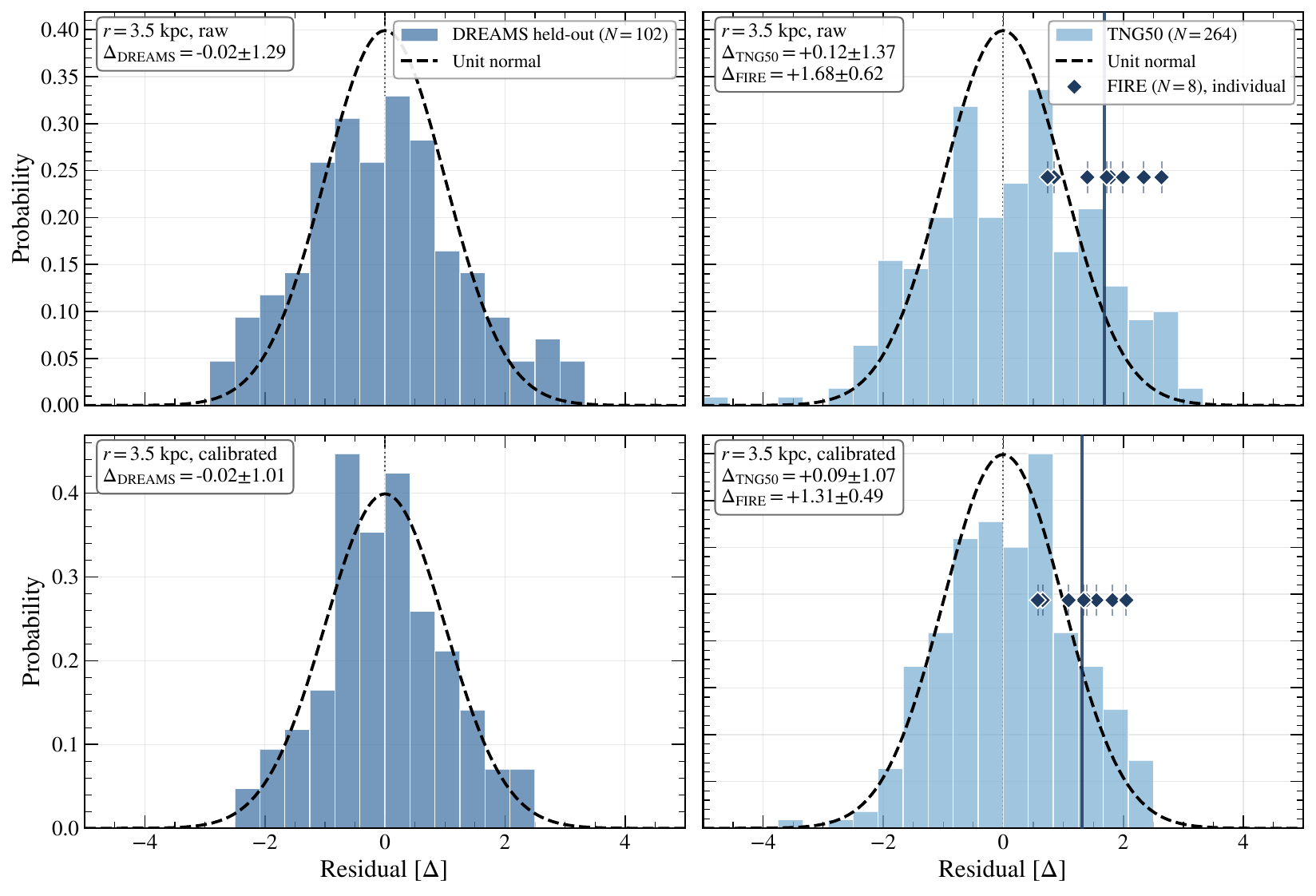}
    \caption{
    Distribution of the profile-level $\Delta$-residual at $r=3.5$\,kpc, before (top) and after (bottom) calibration, for the held-out DREAMS sample (left) and the out-of-domain TNG50 and \Fire samples (right).
    The dashed curve is a unit normal in every panel.
    In the raw distributions the residuals are broader than unity, i.e. the stochastic ensemble is over-confident about the azimuthally averaged profile.
    The DREAMS calibration is performed via \textit{leave-one-out}: the factor applied to each held-out galaxy is derived from the other 101 held-out galaxies only, so that no galaxy contributes to its own correction.
    The recovered width of $1.01$, rather than exactly unity, reflects this and shows that the correction generalizes rather than being fitted.
    The same factor, measured entirely in-domain, is applied to the out-of-domain samples.
    \Fire is shown as individual galaxies rather than as a histogram because the sample size is small, with the vertical line marking the mean across the eight galaxies.
    }
    \label{fig:calibration}
\end{figure*}

We examine the radial residual behavior across the full DREAMS population. 
This population-level diagnostic tests whether the model is systematically biased at a given radius after marginalizing over halo-to-halo variation and simulation-parameter variation. 
Figure~\ref{fig:dreams_profiles} shows the comparison between the mean predicted and true dark matter density at six selected radial bins, whereas Figure~\ref{fig:calibration} focuses on the $\Delta$-residual in the $r=3.5$\,kpc bin.
The $\Delta$-residual shown in Figure~\ref{fig:calibration} should be distinguished from the individual-galaxy radial residual trends discussed above: here, each point in the $\Delta$ distribution represents the mean residual of one galaxy at a fixed radial bin, so the result measures population-level calibration at that radius rather than the radial shape of the residual pattern within any single galaxy.
Specifically, for each galaxy, we compute the mean $\Delta$-residual in each radial bin and then combine these measurements across the $1024$ DREAMS galaxies, of which $102$ are the held-out test split and the remaining $922$ were seen during training or validation.
We quote statistics for the held-out split throughout, and report the full-sample values alongside them to show that the two agree.
We find that the population residual distribution at fixed radius is generally consistent with a Gaussian distribution centered close to zero, with a small systematic offset that grows with radius, from $\Delta\simeq-0.02$ at $r=3.5$\,kpc to $\simeq-0.35$ in the outermost bin sampled.
The widths of the distributions, however, are consistently greater than unity: for the held-out galaxies the standard deviation rises from $1.29$ at $r=3.5$\,kpc to $1.99$ at $r=24.7$\,kpc, and is $1.73$ in the outermost bin sampled ($r=40.3$\,kpc, i.e., the corners of the field of view).
Six further residual distributions at different radial bins in Appendix~\ref{sec:appendix_radial}.
The stochastic ensemble therefore underestimates the uncertainty on the azimuthally averaged profile by a factor up to two, with the underestimate growing mildly with radius.

The underestimated uncertainty is a specifically \textit{profile-level} miscalibration, and it coexists with good calibration at the pixel level (Section~\ref{sec:metrics}).
The reason is that the model's residuals are spatially coherent while a substantial part of its stochastic variability is not.
Averaging over an annulus therefore suppresses the draw-to-draw scatter much faster than it suppresses the actual error. 
The raw ensemble spread of the azimuthally averaged profile is only $\sim0.05$\,dex, whereas the observed profile errors run from $0.072$\,dex in the center to $0.092$\,dex in the outermost bin.
Rescaling the predictive uncertainty by the measured widths above gives a \textit{calibrated} profile-level uncertainty of $\sim0.09$\,dex, rising from $0.07$\,dex at $r=3.5$\,kpc to $0.10$\,dex at $r=40.3$\,kpc.
Notably this is comparable to the typical per-pixel uncertainty of $0.096$\,dex, quoted in Section~\ref{sec:ind_gal}. 
Because the residuals are coherent across an annulus, azimuthal averaging does not in practice reduce the uncertainty on the recovered profile at all.
We adopt $\sim0.09$\,dex as the realistic profile-level uncertainty of the model and use it when assessing the out-of-domain tests below.

Figure~\ref{fig:calibration} (left panels) shows this correction directly.
To verify that it generalizes rather than simply being fitted to the sample it is measured on, we apply it in a leave-one-out fashion: the factor used for each held-out galaxy is derived from the remaining $101$ held-out galaxies, so no galaxy contributes to its own correction and no training galaxy enters at any stage.
The corrected residual distribution has a standard deviation of $1.01$, i.e. consistent with unity but not identically unity as it would be if the factor had been fitted in place.
The same in-domain factor is what we apply to the out-of-domain samples in Section~\ref{sec:domain_adaptation}.

Crucially, the held-out galaxies are recovered at least as well as the sample as a whole. 
The coefficient of determination ($R^2=1$ for perfect correlation) between the predicted and true densities ranges from $R^2=0.69$ to $0.79$ across the six radial bins for the test split, compared with $R^2=0.69$-$0.76$ for all $1024$ galaxies, and the $\Delta$ distributions of the two sets are statistically indistinguishable.
The population-level performance reported here is therefore not inflated by the inclusion of training galaxies.
Taken together, these results indicate that the model is essentially unbiased as a population-level predictor within the DREAMS domain, recovering the radial profile to $\sim0.09$\,dex with a mean offset well below that, but that its raw stochastic ensemble should not be read directly as the uncertainty on an azimuthally averaged profile without the factor $1.3$-$2$ correction derived above.

The same population view can be taken of the azimuthal structure introduced in Section~\ref{sec:metrics}, and it behaves differently from the profile.
Across the $102$ held-out galaxies the same statistic gives $\mathcal{A}=0.41\pm0.04$, with $84\%$ of galaxies scoring $\mathcal{A}>0$ and a median of $0.43$.
Additionally, weighting each annulus by $A_{2}$ raises the median to $0.68$.
Scoring each galaxy's prediction against a \textit{different} galaxy's truth instead returns $\mathcal{A}=0.03\pm0.05$, consistent with zero, confirming on the maps themselves that the signal is specific to the galaxy being predicted rather than an artifact of the field of view shared by every image.
As expected, the performance depends on how anisotropic the galaxy actually is.
Specifically, in the quartile with the roundest true dark matter distributions, the median $\mathcal{A}$ is $0.03$, whereas in the most elongated quartile it reaches $0.91$, with $96\%$ of those galaxies scoring positively.
Because $\mathcal{A}$ has an expectation of exactly zero for any axisymmetric predictor, this is direct evidence that the model uses the stellar image to place dark matter at particular position angles, and not merely to set the radial profile.

\begin{figure}
    \centering
    \includegraphics[width=0.95\linewidth]{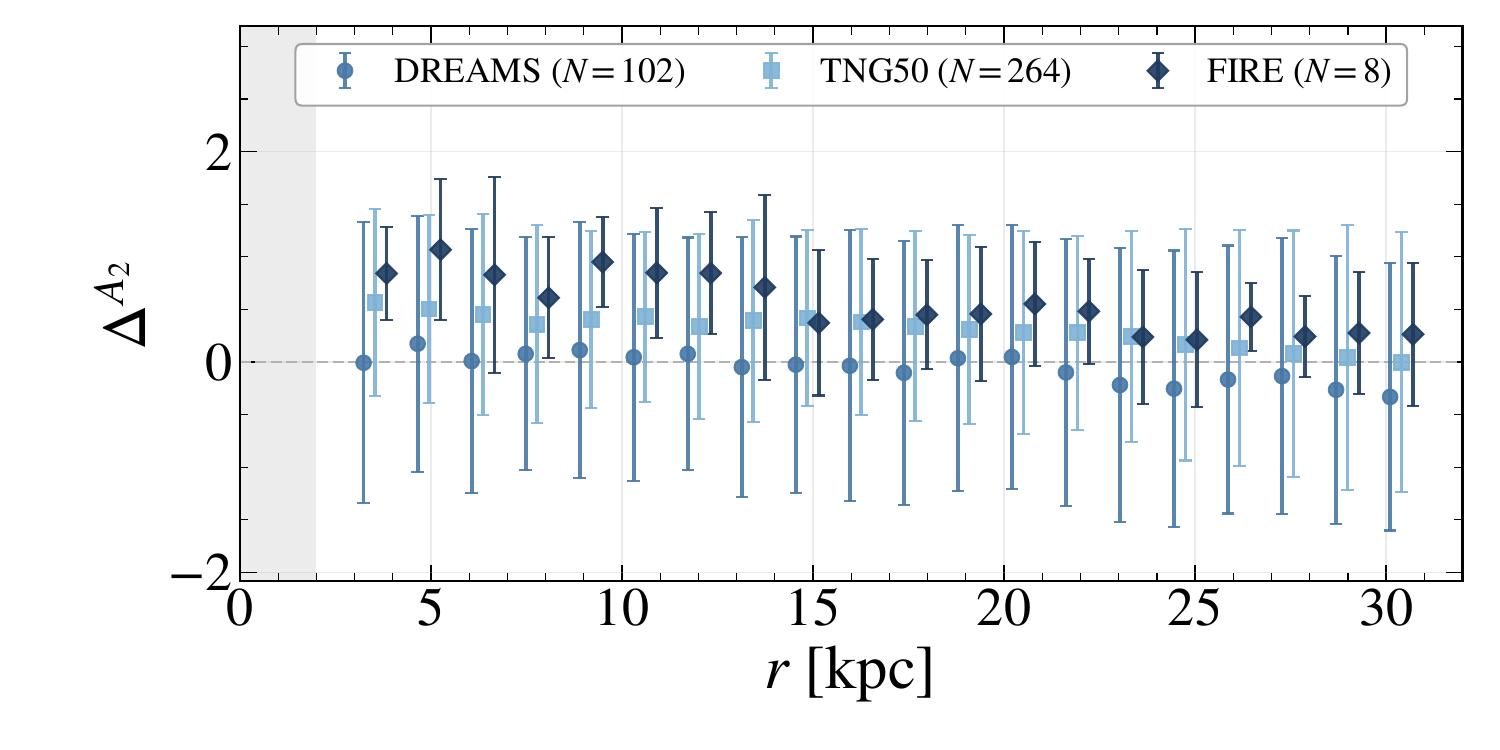} \\
    \includegraphics[width=0.95\linewidth]{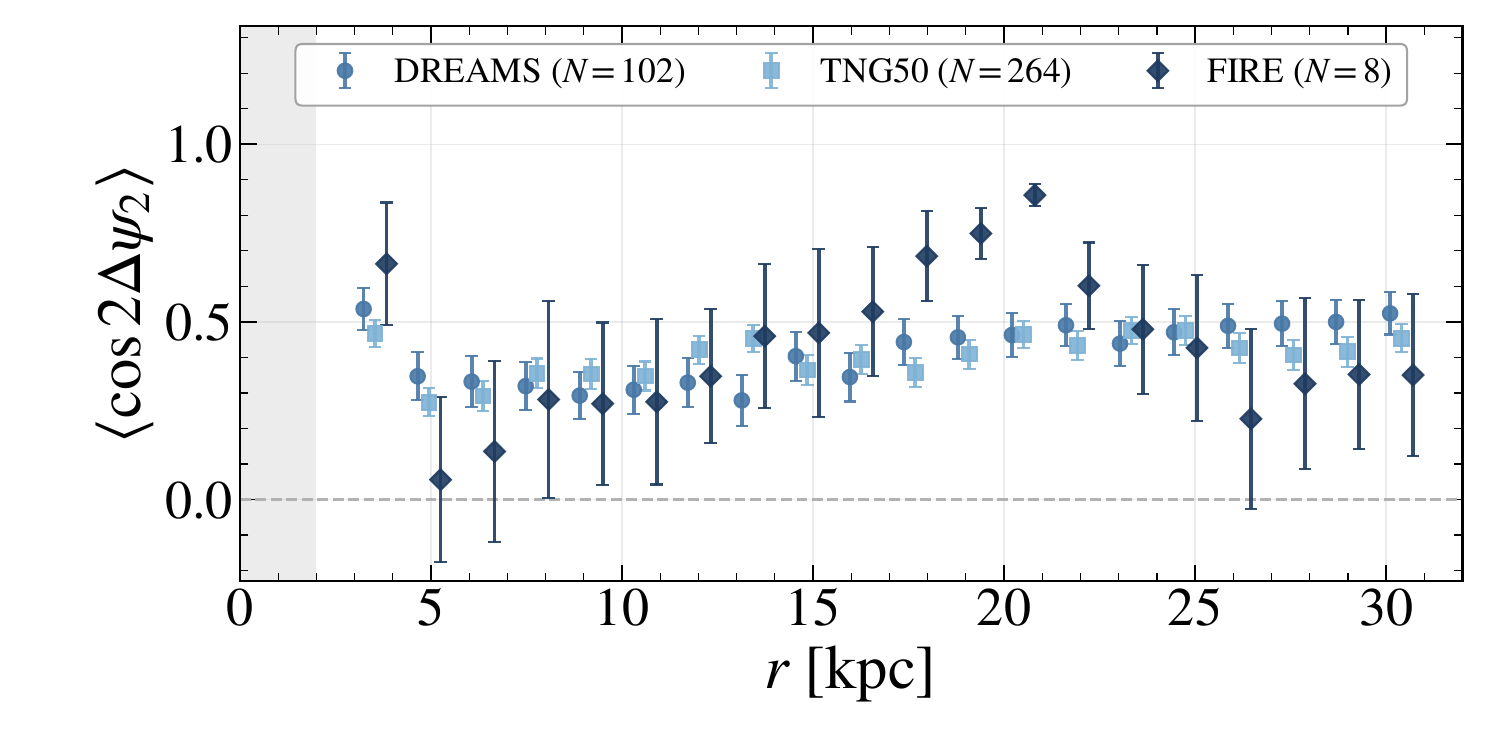}
    \caption{
    Azimuthal $m=2$ diagnostics across the $102$ held-out DREAMS galaxies and for the two out-of-domain samples, TNG50 and \Fire.
    The top panel shows the mean and standard deviation over galaxies of the amplitude residual $\Delta^{A_2}$.
    The dashed line marks perfect agreement between the predicted and true anisotropy amplitude.
    The bottom panel shows the mean and standard deviation of the alignment between the predicted and true major axes.
    The dashed line marks the value expected of a predictor carrying no azimuthal information.
    }
    \label{fig:dreams_anisotropy_pop}
\end{figure}

Figure~\ref{fig:dreams_anisotropy_pop} shows the two anisotropy diagnostics against radius for the held-out sample.
The mean alignment $\langle\cos 2\Delta\psi_{2}\rangle$ is positive and highly significant at every radius, running between $0.28$ and $0.54$ with a shallow minimum near $r\simeq13$\,kpc and rising outwards to $\simeq0.52$ in the outermost annulus, against a mismatched-galaxy null of $0.03$.
The model therefore recovers the orientation of the dark matter distribution across the whole radial range probed, not only where the stellar light is brightest.
The amplitude residual $\Delta^{A_2}$ is consistent with zero inside $r\simeq20$\,kpc and drifts mildly negative outside it, reaching $-0.33$ in the outermost annulus, so the model slightly under-predicts how elongated the outer halo is.
Its population width at fixed radius is $1.11$-$1.34$, mildly greater than unity in the same sense, and for the same reason, as the profile-level widths quoted above: the ensemble is somewhat overconfident about $A_{2}$ at fixed radius, though by a considerably smaller factor than it is about the azimuthally averaged density.
Figure~\ref{fig:dreams_anisotropy_pop} also carries the two out-of-domain samples, discussed in Section~\ref{sec:domain_adaptation}.

\subsubsection{Simulation-parameter conditioning}
\label{sec:sim_par_condition}

To isolate the effect of simulation-parameter conditioning, we perform an additional inference test in which the trained model is given deliberately incorrect feedback parameters while the stellar density input is held fixed.
This experiment tests whether the model has learned to use the scalar conditioning variables to adjust the stellar-to-dark-matter mapping, rather than only relying on the image-level stellar density information.

\begin{figure*}
    \centering
    \includegraphics[width=1.0\linewidth]{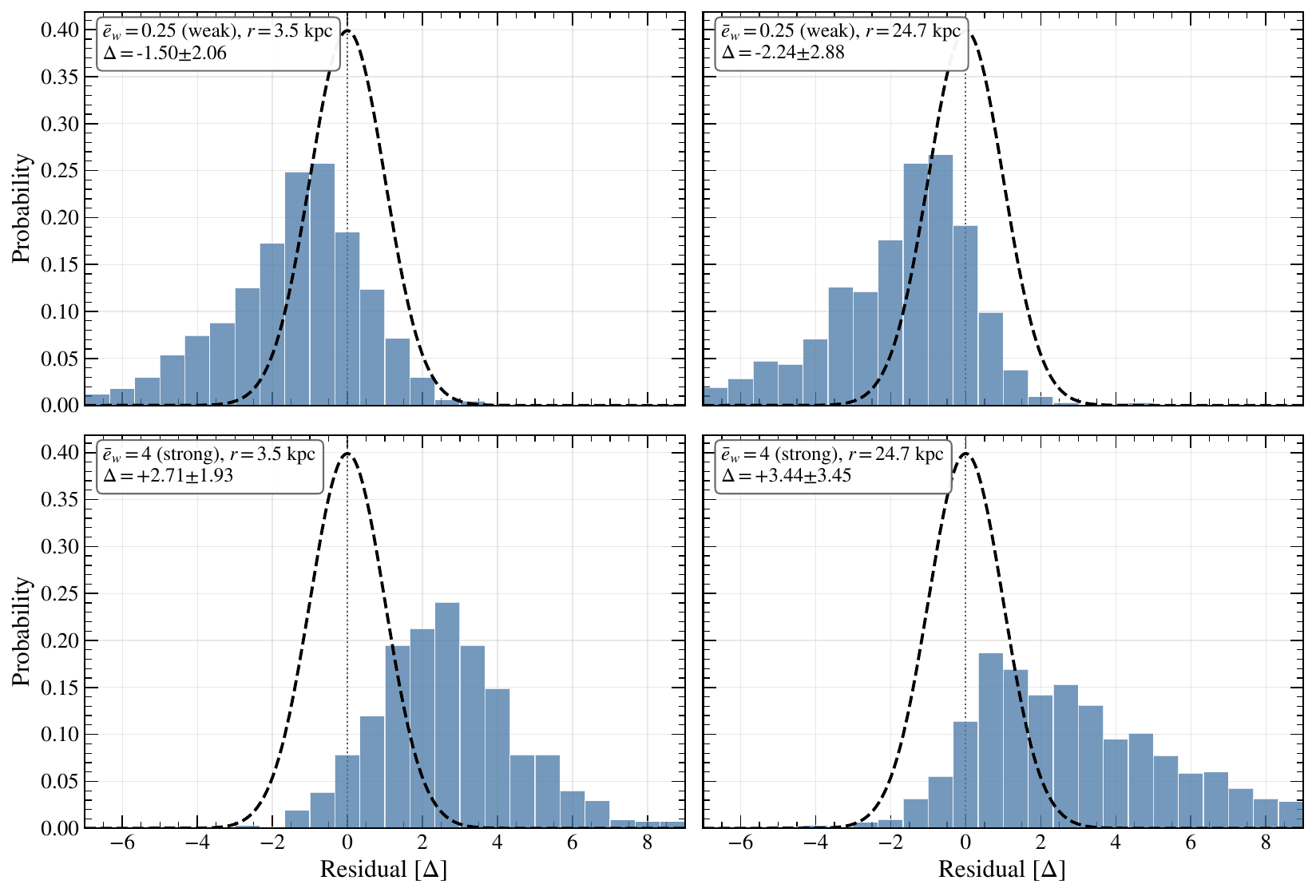}
    \caption{
    Population-level $\Delta$-residual distributions when the feedback parameter $\bar{e}_w$ is intentionally set to the minimum (top) and maximum (bottom) values of the DREAMS range during inference, at an inner (left) and an outer (right) radius, while the stellar density inputs are held fixed.
    All panels use the calibrated uncertainty of Figure~\ref{fig:calibration}, and the dashed curve is a unit normal.
    The coherent shifts demonstrate that the scalar feedback conditioning changes the inferred dark matter density normalization, and comparing the two columns shows that the response grows with radius.
    }
    \label{fig:sim_par_test}
\end{figure*}

As shown in Figure~\ref{fig:sim_par_test}, setting the feedback parameters to values corresponding to stronger feedback shifts the residual distributions toward positive $\Delta$, while setting them to values corresponding to weaker feedback shifts the residual distributions toward negative $\Delta$.
The response is monotonic across the five sweep values, of which Figure~\ref{fig:sim_par_test} shows the two extremes.
Quoted against the \textit{calibrated} profile-level uncertainty derived above, the mean residual at $r=3.5$\,kpc runs from $\Delta=-1.50$ at $\bar{e}_w=0.25$ to $+2.71$ at $\bar{e}_w=4$, a total change of $\Delta\simeq4.2$, where as at $r=24.7$\,kpc, the change grows to $\sim5.7$, from $-2.24$ to $+3.44$.
The response therefore strengthens with radius, and at both radii the shift between the extreme parameter values is several times the residual scatter of the fiducial run, which after calibration is of order unity.
This behavior is qualitatively consistent with the expected degeneracy between stellar mass assembly and the dark matter response: weaker feedback generally permits more efficient central star formation and a different baryonic modification of the halo, while stronger feedback can produce a different stellar-to-dark-matter relation at fixed present-day stellar density.
To first order, the model infers that the dark matter density across the full DREAMS population must be higher to form the same baryon content if the galaxy formation model has high stellar feedback, thus producing positive residuals when $\bar{e}_w$ is increased from their true values.
The parameter-sweep test therefore supports the interpretation that at least part of the residual structure seen in individual DREAMS galaxies reflects feedback-dependent variation in the baryon-dark matter mapping.
Appendix~\ref{sec:appendix_sweeps} shows the same tests for the other four parameters.
We find that the model is most sensitive to the stellar feedback parameters, $\bar{e}_w$ and $\kappa_w$, followed by the AGN parameter, and the two cosmological parameters, $\Omega_m$ and $\sigma_8$, being the least sensitive. 
Appendix~\ref{sec:appendix_noh} provides the complementary measurement, quantifying how much performance is lost when the conditioning is removed altogether.

Overall, the held-out DREAMS test establishes that the stellar-density-to-dark-matter-density mapping is learnable, with sensible sensitivity to the simulation parameter conditioning, within the simulation domain. 
With the full sample spanning $\sim0.9$-$1.1$\,dex in dark matter density at different radial distances from the center ($0.45$-$0.55$\,dex for the central $90$ per cent of the sample), the model recovers both two-dimensional maps and azimuthally averaged radial profiles with accuracy of order $\sim0.1$\,dex, while producing a stochastic prediction ensemble that can be used as an uncertainty estimate. 
This in-domain performance provides the reference point for the out-of-domain tests below, where we ask whether the same learned mapping remains valid when the input galaxies are drawn from independent simulations.

\subsection{Domain adaptation with varying simulations}
\label{sec:domain_adaptation}

\begin{figure*}
    \centering
    \includegraphics[width=0.9\linewidth]{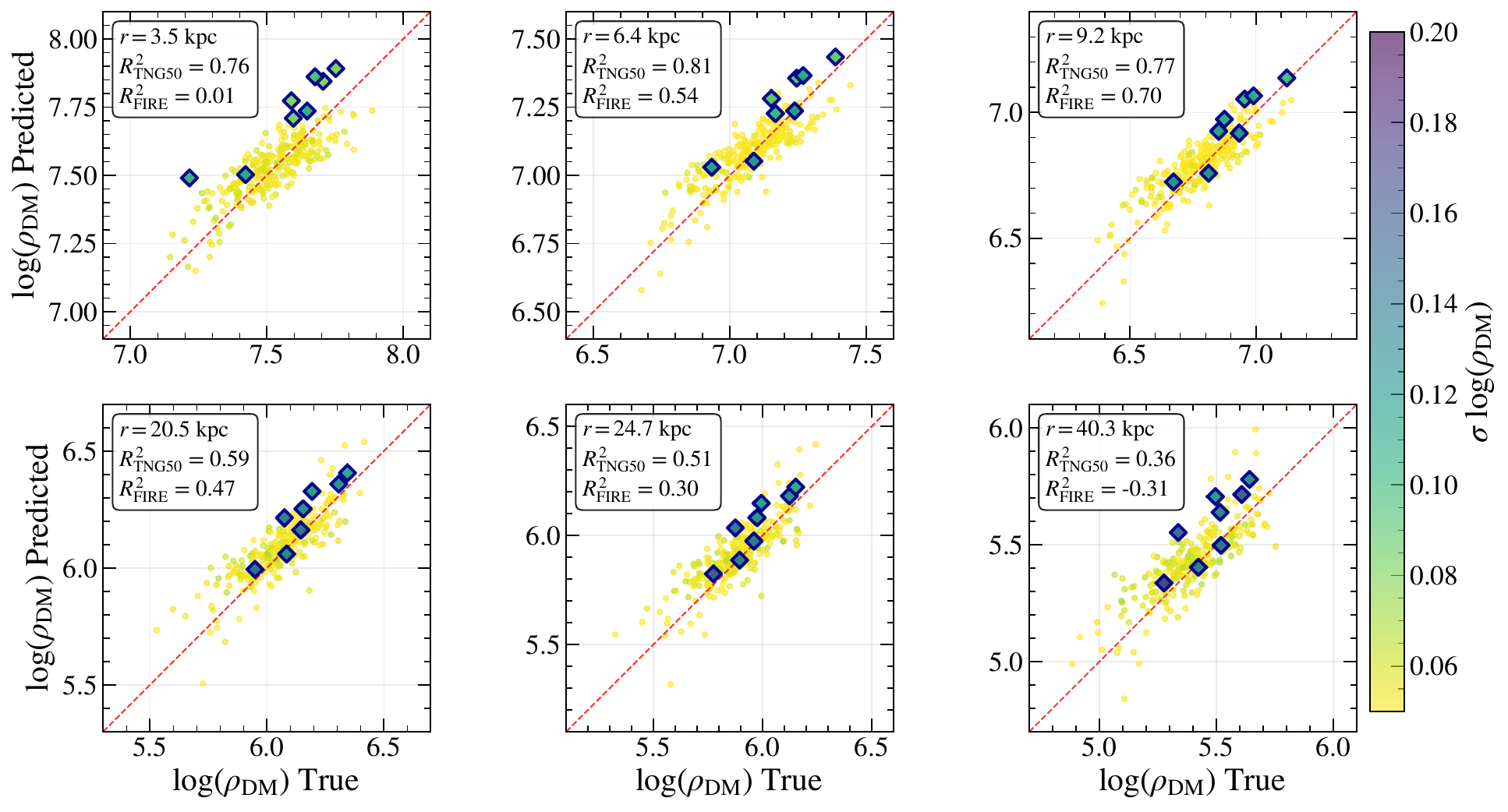}
    \caption{
    Same as Figure~\ref{fig:dreams_profiles}, but for the out-of-domain TNG50 and \Fire samples (circles and diamonds respectively).
    The out-of-domain samples scatter more widely, with TNG50 showing slight over-prediction toward large radii and all eight \Fire galaxies lying above the line in the innermost bin.
    }
    \label{fig:tng50_fire_profiles}
\end{figure*}

Having established the in-domain performance of the model on held-out DREAMS galaxies, we now test whether the same learned mapping generalizes to independent hydrodynamical simulations.
We first evaluate the DREAMS-trained model on the test galaxies from TNG50. 
Because TNG50 shares the same broad galaxy-formation model family as DREAMS but has higher mass resolution, this comparison primarily tests whether the learned mapping is robust to resolution and simulation volume differences. 
Figure~\ref{fig:tng50_fire_profiles} compares the corresponding azimuthally averaged radial profiles.

The model performs comparably on TNG50 and held-out DREAMS galaxies, except for a mild positive bias that increases with radius, from $\Delta\simeq0.12$ at $r=3.5$\,kpc to $\Delta\simeq0.83$ at $r=40.3$\,kpc, peaking at $\Delta\simeq1.04$ at $r=24.7$\,kpc. 
Expressed against the calibrated profile-level uncertainty of Section~\ref{sec:all_gal} this corresponds to $\Delta\simeq0.09$ to $\sim0.52$, i.e., an absolute density difference reaching at most $\sim0.05$\,dex and remaining well within the $\sim0.09$\,dex profile-level uncertainty at every radius.
The $\Delta$-residual distributions in the outer bins beyond $\sim20$\,kpc are nonetheless the most discrepant, with both the largest positive bias and the broadest spread ($\Delta=1.04\pm2.99$ at $r=24.7$\,kpc).
The discrepancies indicate that the model overestimates the dark matter density in the TNG50 halo outskirts.
This result suggests that the diffusion model is sensitive to simulation resolution in regions of low stellar and dark matter densities. 
In particular, for regions within $\sim20$\,kpc, the model captures a stellar-to-dark-matter mapping that is robust across the DREAMS and TNG50 realizations of Milky Way-mass galaxies.
For regions beyond $\sim20$\,kpc, where stellar density becomes exponentially low for typical Milky Way-mass galaxies, the predicted dark matter density profiles show positive bias.

It is thus important to be cautious with model prediction in these outer regions where the input stellar density map may be sensitive to the simulation resolution.
Nonetheless, we note that, for practical application of the model to observation data, such low stellar density regions are most likely too faint to contribute any meaningful signal to the model prediction, and the model performance will be driven by the expected level of flux noise floor for a given instrument.

We next apply the same model to the \Fire galaxies. 
This is a more stringent test because \Fire differs from DREAMS in hydrodynamic method, mass resolution,  and feedback implementation.
The dark blue outlined diamonds and histograms in Figure~\ref{fig:tng50_fire_profiles} shows their radial dark matter density profiles. 
Despite the stronger domain shift, the model recovers the broad dark matter distribution for most \Fire galaxies, except in the central few kpc.
All eight \Fire galaxies are over-predicted in the central dark matter density: the median residual is $+0.41$\,dex in the innermost bin ($r\simeq1$\,kpc) and $+0.19$\,dex inside $r=5$\,kpc, ranging from $+0.13$\,dex for m12r to $+0.35$\,dex for m12z across the sample.
This is two to four times the $\sim0.09$\,dex profile-level uncertainty the model achieves in domain (Section~\ref{sec:all_gal}), and so represents a genuine failure of the mapping rather than a reflection of the broader predictive spread ($\sigma_{\rm pred}\simeq0.12$-$0.16$\,dex) produced by marginalizing over the DREAMS baryonic parameters. 
See Appendix~\ref{sec:appendix_noh} for results when the model is not conditioned on the simulation parameters.
Between $r\simeq5$ and $20$\,kpc the predicted profiles agree with the true profiles to within $0.1$\,dex, with a median offset of only $+0.04$\,dex.
Beyond $20$\,kpc the agreement degrades mildly, to a median offset of $+0.07$\,dex, driven by m12b and m12w, which reach $\sim0.17$\,dex.

The residual outer-region discrepancies, which affect only a minority of the \Fire galaxies, plausibly originate from differences in the simulation resolution, similar to the case of TNG50.
The much larger discrepancy in the inner regions, on the other hand, is consistent with the physical difference between the DREAMS/IllustrisTNG and \Fire feedback models. 
The DREAMS training set samples variations in the IllustrisTNG feedback parameters, but those variations do not necessarily span the same baryonic response produced by explicitly bursty stellar feedback in \Fire. 
As a result, a model trained only on DREAMS learns the range of inner dark matter responses available within an IllustrisTNG-like framework and extrapolates poorly when the true galaxy has a more strongly cored central profile.
Addressing this inner region bias would thus require expanding the training sample to include the simulation models with bursty stellar feedback, such as \Fire. 
Upcoming \Fire suite from the DREAMS collaboration (Kho et al. in prep.) will address this issue by providing a current DREAMS-equivalent training sample for the diffusion model.

Additionally, this failure mode is scientifically informative for the model input condition maps. 
It indicates that the stellar density field alone can predict the large-scale dark matter structure across multiple simulation suites, but it does not uniquely determine the central dark matter profile when the underlying feedback physics changes significantly. 
This interpretation is reinforced by the test shown in Appendix~\ref{sec:appendix_noh}, in which a model given no simulation parameters at all still recovers most of the profile variance from the stellar map, confirming that the image is the dominant source of information.
The \Fire comparison therefore identifies the central region as the part of the mapping most sensitive to galaxy-formation physics. 
It also motivates future models to include additional conditioning information, such as stellar velocity moments, gas structure, star-formation history, or other observables that encode the impact of feedback on the dark matter distribution.

We then inspect the model performance based on the anisotropic metric, as shown in Figure~\ref{fig:dreams_anisotropy_pop}.
The predictive power on the orientation transfers essentially intact. 
The mean alignment is $\mathcal{A}=0.40\pm0.02$ for the $264$ TNG50 galaxies and $0.43\pm0.12$ for the eight \Fire galaxies, against $0.41\pm0.04$ in domain.
In both samples the alignment is positive at every radius, and $82\%$ and $88\%$ of galaxies individually score $\mathcal{A}>0$, compared with $84\%$ per cent in domain.
The model is therefore recovering anisotropic structure in out-of-domain simulations.
Additionally, the \Fire galaxies show particularly well alignment between predicted and truth at $r\simeq20$\,kpc, but it should be interpreted with caution as we only have eight \Fire galaxies and the prediction draws are sampled while marginalizing over the baryon physics parameters.

The amplitude, by contrast, shows worse performance in similar radius bins as the radial profile case.
The $\Delta^{A_2}$ rises from $\sim1.09$ in domain to $\sim1.45$ for \Fire, and in both out-of-domain samples the excess is larger inside $r=20$\,kpc than outside it.
The model therefore places the anisotropy correctly but overstates how strong it is, most severely in the central regions where the \Fire feedback physics departs furthest from the DREAMS training range.
This behavior reinforces the interpretation from the radial profile that the geometry of the dark matter distribution in inner region is sensitive to the baryonic model.

Overall, the out-of-domain tests show that the learned mapping generalizes well across simulations at galactocentric distances $5$-$20$\,kpc, where the stellar and dark matter density fields are not sensitive to simulation resolution and baryonic models.
Even in regions where the model predictions show significant bias, measured by $\Delta$, the absolute deviation between the predicted and true dark matter density is within $\sim0.1$\,dex.
For dark matter density profile studies, the expected signals from alternative dark matter models can easily exceed this scale, meaning the model is still useful at the current state. 

\section{Discussion}
\label{sec:discussion}

Our results demonstrate that a conditional diffusion model can recover the dark matter density of Milky Way-mass galaxies from their stellar mass distribution. 
The success of this mapping is non-trivial. 
The stellar density field is not a direct tracer of the dark matter field, and the relation between the two depends on assembly history, baryonic feedback, halo concentration, angular momentum, and the stochasticity of galaxy formation. 
Nevertheless, the model recovers the dark matter density maps of held-out DREAMS galaxies with typical accuracy of $0.09$\,dex, both per pixel and on the azimuthally averaged profile.
This suggests that the stellar density field contains substantial information about the underlying dark matter distribution, at least for Milky Way-mass systems drawn from the range of halo-to-halo variance and feedback variation represented in DREAMS.

The out-of-domain validation clarifies what the model has and has not learned, in different regimes. 
The performance on TNG50 indicates that the model is not strongly tied to the mass resolution of the DREAMS training simulations within regions with sufficient particle counts to be resolution independent, where are also the regions where we expect measurable flux in extragalactic photometry surveys. 
Although TNG50 has higher resolution, the recovered dark matter maps and radial profiles remain comparable to the DREAMS test performance within $\sim20$\,kpc. 
This result supports the interpretation that the model is learning a physically meaningful baryon-dark matter relation rather than a simulation-specific numerical artifact.

For the \Fire comparison, the model recovers the broad dark matter distribution in most \Fire galaxies, but over-predicts the central dark matter density in all eight of them. 
This is expected as the central dark matter response in \Fire is driven by bursty feedback histories that are not fully represented by the IllustrisTNG-like feedback variations in DREAMS \citep{hussein25,garcia26}.
The inner discrepancy therefore should not be viewed only as a limitation of the model architecture. 
It is also a diagnostic of the physical non-uniqueness of the stellar-density-to-dark-matter-density mapping. 
Two galaxies with similar present-day stellar density profiles may have different central dark matter profiles if their feedback histories differ. 
In this sense, the model identifies the central $r\lesssim 5$\,kpc region as the regime where additional observables are most needed.

This narrowed interpretation is important for the scope of the present paper. 
We do not claim that stellar density alone provides a complete observationally ready estimator of dark matter density. 
Rather, we show that the idealized stellar-density-to-dark-matter-density problem is learnable, probabilistic, and testable across simulations. 
This is the first stage of a broader inference pipeline. 
By isolating the physical mapping from the observational forward model, we can identify which failures originate from galaxy-formation physics and which failures arise later from observational systematics.

\subsection{Future directions}

The next step is to move from idealized stellar density maps to mock observable galaxy images. 
Real imaging data include surface-brightness limits, sky noise, point-spread functions, unknown inclination, crowding, dust attenuation, stellar-population effects, bandpass dependence, and survey-specific selection functions. 
These effects can erase or distort the low-density stellar outskirts that are particularly informative about halo structure. 
A realistic observational model is therefore necessary before applying this framework to Euclid, Roman, Rubin, or other imaging surveys. 
One way to achieve this goal is by utilizing forward modeling tools such as Synthesizer \citep{lovell25} to generate mock images from simulation suites, which will be explored in an upcoming follow-up study.

A second direction is to expand the set of conditioning fields. 
The present model uses only the stellar density along the galaxy mid-plane, which is intentionally restrictive. 
The \Fire comparison shows why this restriction matters: present-day stellar density alone may not fully encode the feedback history that determines the central dark matter response. 
Additional observables could help break this degeneracy. 
Velocity moments, stellar age or metallicity maps, gas morphology, star-formation-rate maps, or multi-band photometry may provide complementary information about the assembly and feedback history of each galaxy. 
In future work, these quantities can be included as additional conditioning channels in the same diffusion framework. 
The resulting model would test which observables most improve dark matter recovery, and on which physical scales.

A third direction is to move beyond isolated Milky Way-mass halos. 
The current model is trained on Milky Way-mass systems, so its learned mapping is tied to a relatively narrow halo-mass range. 
However, many of the most interesting applications of spatially resolved dark matter inference involve lower-mass galaxies, massive disk galaxies, and statistical samples spanning a broad range of stellar and halo masses. 
The DREAMS mass-varied suite (Garcia et al. in prep.) provides a natural next training set for this extension. 
Since halo mass changes both the stellar-to-halo-mass relation and the characteristic scale of the dark matter profile, applying the current Milky Way-mass model directly to a broad halo-mass sample would be an extrapolation. 
A mass-varied training set will allow us to test whether the diffusion model can learn a unified mapping across halo mass, and whether explicit halo-mass or stellar-mass conditioning is required.
Additionally, the isolation selection criterion limits the occurrence of major merger events in the training set and thus the model does not capture the dark matter subhalos in the images in cases of recent merger events.
Future studies should either lift the isolation criterion or perform class rebalancing with existing dataset by injecting subhalos into existing training simulations. 

A fourth direction concerns the structure of the predictive uncertainty itself.
As shown in Section~\ref{sec:all_gal}, the stochastic ensemble is well calibrated pixel by pixel but underestimates the uncertainty on the azimuthally averaged profile by a factor of $\sim2$, because the model's residuals are more spatially coherent than its stochastic draws.
This is a natural consequence of a training objective defined pixel by pixel: the loss constrains the marginal predictive distribution at each pixel, but places no explicit requirement on the joint distribution across pixels, and it is the joint structure that determines the uncertainty on any aggregated quantity.
The correction we derive is straightforward to apply and is measured directly from held-out data, so it does not limit the present results.
Nonetheless, a model whose ensemble was calibrated at both the pixel and profile level would be preferable, since the quantities of greatest astrophysical interest, such as enclosed mass, inner slope, or core radius, are all aggregated.

Ultimately, this framework can be developed into an experimental-design tool for dark matter inference. 
Given a proposed survey configuration, one could forward model galaxy images beyond existing/planned surveys, with arbitrary depth, angular resolution, filter coverage, redshift range, and noise properties, then ask what aspects of the dark matter distribution can be recovered and with what uncertainty. 
This would connect simulation-based inference directly to survey design. 
Instead of asking only whether a trained model performs well on a fixed dataset, we could ask how the expected dark matter constraints change as a function of observing strategy, target selection, instrument capability, and available ancillary data. 
Such a tool would be especially useful for planning how deep imaging, spectroscopic follow-up, and multi-band information should be combined to maximize sensitivity to dark matter structure in galaxies.

\section{Conclusions}
\label{sec:conclusion}

We have presented an application of BaryonBridge for inferring dark matter density maps from stellar density maps along the galaxy mid-planes in Milky Way-mass galaxies. 
\footnote{The code for training the model and reproducing the analysis is available as a repository on GitHub (\url{https://github.com/xou-mit/star2dm}).}
The model is trained on the DREAMS suite of hydrodynamical zoom-in simulations, which samples halo-to-halo variance as well as variations in cosmological and baryonic feedback parameters. 
By focusing on the idealized stellar-density-to-dark-matter-density mapping, we train and test the stochastic interpolation model on learning the physical relation between baryons and dark matter.

Our main conclusions are as follows:

\begin{itemize}
\item On held-out DREAMS galaxies, the model recovers the two-dimensional dark matter density field with a typical accuracy of $0.09$\,dex, with the strongest residuals appearing in the outer halo regions where the stellar-density field provides limited information (Section~\ref{sec:ind_gal}). 
This ensemble is well calibrated pixel by pixel, and the per-pixel uncertainties serve as a conservative floor for the azimuthally averaged profile uncertainties  (Section~\ref{sec:all_gal}).

\item The model recovers structure beyond the azimuthally averaged profile.
Decomposing the predicted and true maps into azimuthal Fourier modes shows that the orientation of the $m=2$ anisotropy is recovered with a mean alignment of $\mathcal{A}=0.41\pm0.04$, and $84\%$ of held-out galaxies score $\mathcal{A}>0$ (Section~\ref{sec:all_gal}).
The median $\mathcal{A}$ reaches $0.91$ in the most elongated quartile of the test DREAMS sample.

\item The model prediction is most robust when conditioned on both the stellar density map and simulation parameters but still performs well when only conditioned on the map (Appendix~\ref{sec:appendix_noh}). 
When the simulation parameters are available, the model is most sensitive to the baryonic feedback parameters, as expected (Section~\ref{sec:sim_par_condition} and Appendix~\ref{sec:appendix_sweeps}).

\item We further tested the DREAMS-trained model on galaxies selected from the IllustrisTNG and \Fire simulations (Section~\ref{sec:domain_adaptation}). 
When applied to Milky Way-mass galaxies from TNG50, the model achieves performance comparable to that on the DREAMS test set up to $\sim20$\,kpc from the galactic centers, despite the difference in simulation resolution. 
When applied to \Fire galaxies, the model recovers the broad dark matter distribution but systematically over-predicts the central dark matter density. 
This behavior highlights the physical limitation of using stellar density alone: the central dark matter profile can depend on feedback histories that are not uniquely encoded in the present-day stellar density field.

\end{itemize}

These results establish a baseline for simulation-based, uncertainty-aware inference of dark matter structure from baryonic observables. 
They show that stellar density contains substantial information about the dark matter distribution of Milky Way-mass galaxies, while also identifying the central regions and feedback-sensitive regimes where additional information is required. 
Future work will extend this framework to forward-modeled images with realistic instrumental effects, broader halo-mass ranges, and richer conditioning fields such as velocity moments, gas morphology, and multi-band photometry. 
Together, these extensions will not only allow the model to more effectively probe dark matter distribution in extragalactic galaxies, but also move the method toward a general experimental-design framework for determining which galaxy observations are most informative for constraining dark matter density profiles.

\begin{acknowledgments}

XO, PT, AMG, and NK acknowledge support from the NSF-Simons AI Institute for Cosmic Origins which is supported by the National Science Foundation under Cooperative Agreement 2421782 and the Simons Foundation award MPS-AI-00010515.
LN is supported by the Sloan Fellowship, the NSF CAREER award 2337864, NSF award 2307788, and by the NSF award PHY2019786 (The NSF AI Institute for Artificial Intelligence and Fundamental Interactions, \href{http://iaifi.org/}{http://iaifi.org/}). LN also gratefully acknowledges the continued support of the Adam J. Burgasser Endowed Chair of Astrophysics at MIT, which sustained many long hours of writing and revision of this manuscript.
The authors acknowledge Research Computing at The University of Virginia for providing computational resources and technical support that have contributed to the results reported within this publication. URL: \href{https://rc.virginia.edu}{https://rc.virginia.edu}
This work used Bridges-2 system at PSC through allocation PHY210118 from the Advanced Cyberinfrastructure Coordination Ecosystem: Services \& Support (ACCESS) program, which is supported by U.S. National Science Foundation grants \#2138259, \#2138286, \#2138307, \#2137603, and \#2138296.
AI tools were used to assist with refinement of the public release of code used in the research process, to allow clean reproduction of the model training and analysis, as well as potential follow-up work. The GAI tool used was Claude Opus 5.

\end{acknowledgments}

%

\software{%
BaryonBridge \citep{horowitz25},
PyTorch \citep{paszke19},
Lightning \citep{falcon24},
matplotlib \citep{hunter07},
numpy \citep{vanderwalt11},
scipy \citep{jones01}, 
astropy \citep{astropy:2013,astropy:2018}.}

\clearpage

\bibliography{xou,xou_add}{}
\bibliographystyle{aasjournalv7}



\renewcommand{\thefigure}{\thesection\arabic{figure}}
\renewcommand{\thetable}{\thesection\arabic{table}}

\appendix

\section{Model behavior across the DREAMS test population}
\label{sec:appendix_stamps}
\setcounter{figure}{0}
\setcounter{table}{0}

Figures~\ref{fig:dreams_map}-\ref{fig:dreams_profile_comp} illustrate the model's diagnostics for a single representative held-out DREAMS test galaxy.
To demonstrate that this behavior is representative of the held-out population, we show the same diagnostics computed independently for 16 randomly selected held-out DREAMS test galaxies.

\begin{figure*}
    \centering
    \includegraphics[width=0.95\linewidth]{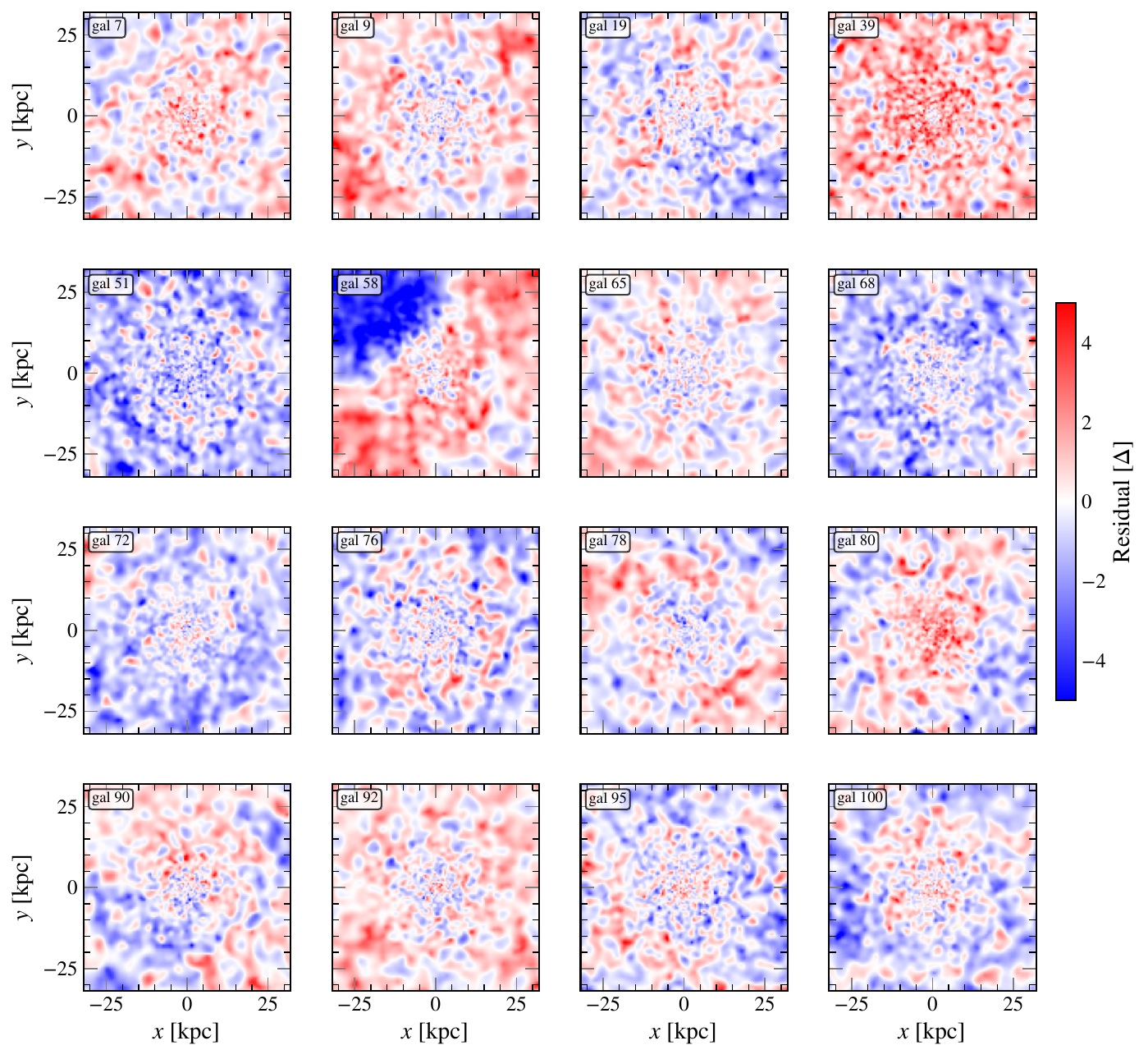}
    \caption{
    $\Delta$-residual maps for 16 randomly selected held-out DREAMS test galaxies, shown in the same format and color scale as the residual map in Figure~\ref{fig:dreams_map}.
    Most galaxies show a spatially incoherent, approximately zero-mean residual pattern. 
    A minority show coherent structure spanning all four residual-morphology modes discussed in Section~\ref{sec:ind_gal}.
    }
    \label{fig:appendix_stamp_deviation}
\end{figure*}

Figure~\ref{fig:appendix_stamp_deviation} shows the $\Delta$-residual map for each of the 16 galaxies.
Most galaxies show the same broadly unbiased, spatially incoherent residual pattern as the example galaxy in the main text (for instance galaxies 7, 65, 72, 90, and 100).
A minority show coherent structure, and between them the panel illustrates all four ``individual-galaxy residual morphology'' modes introduced in Section~\ref{sec:ind_gal}.
Galaxy 39 is globally over-predicted and galaxy 68 globally under-predicted; galaxy 80 shows a coherent radial trend, with a central excess surrounded by an outer deficit; and galaxy 58 shows pronounced azimuthal structure, a large-scale gradient running across the field of view.
These cases illustrate the range of individual-galaxy behavior rather than a systematic population-level failure mode.

\begin{figure*}
    \centering
    \includegraphics[width=0.95\linewidth]{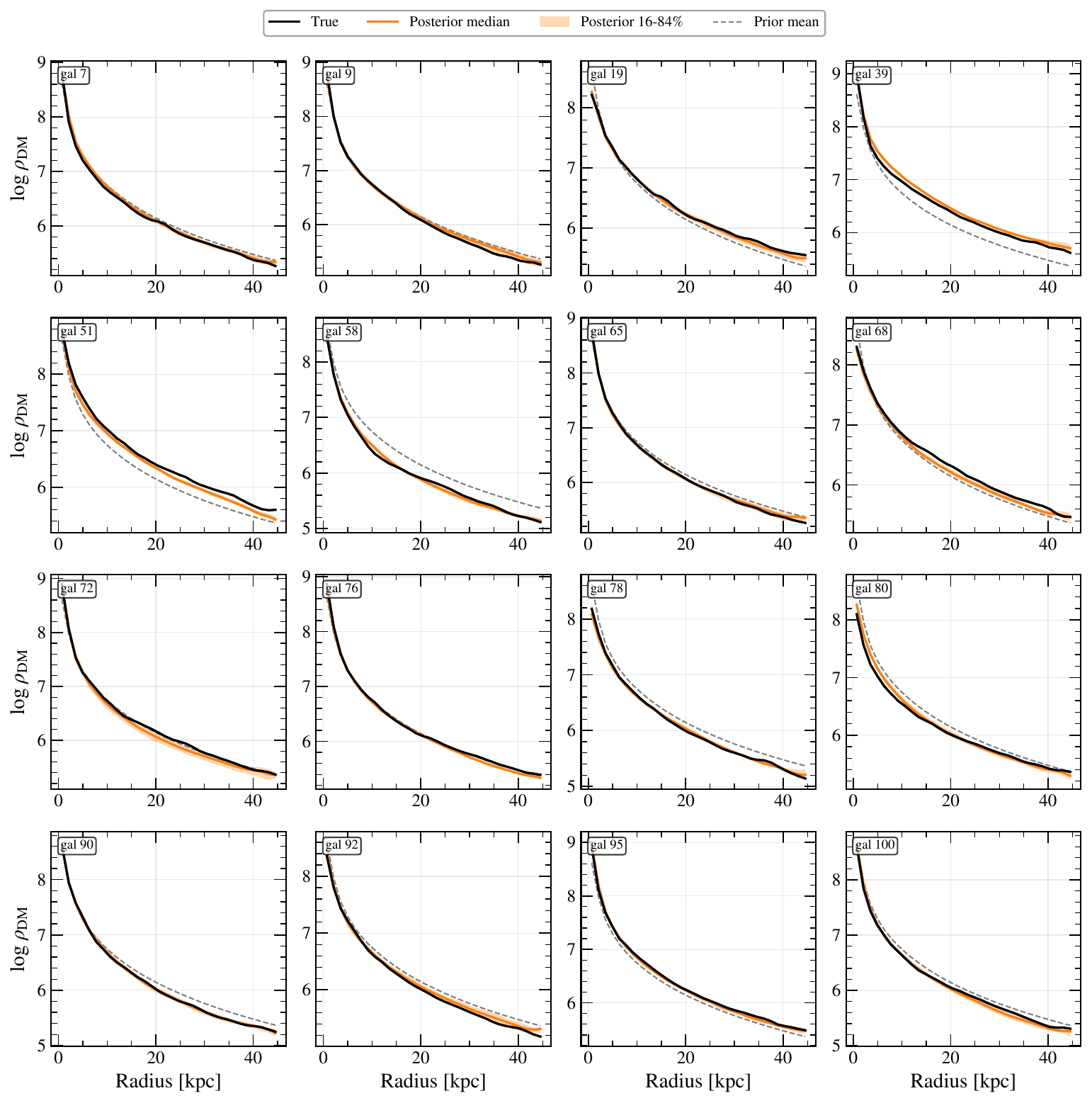}
    \caption{
    Prior/posterior radial dark matter density profile comparison for the same 16 held-out DREAMS test galaxies shown in Figure~\ref{fig:appendix_stamp_deviation}, in the same format as Figure~\ref{fig:dreams_profile_comp}.
    We omit the individual training-set profiles for readability.
    In most cases the posterior is centered close to the true profile.
    }
    \label{fig:appendix_stamp_prior_posterior}
\end{figure*}

Figure~\ref{fig:appendix_stamp_prior_posterior} shows the corresponding prior/posterior radial density profile comparison for the same 16 galaxies, in the same format as Figure~\ref{fig:dreams_profile_comp}.
In all 16 cases, the posterior mean tracks the true profile closely, reproducing across the sample the behavior illustrated for the single example galaxy in the main text.

\begin{figure*}
    \centering
    \includegraphics[width=0.95\linewidth]{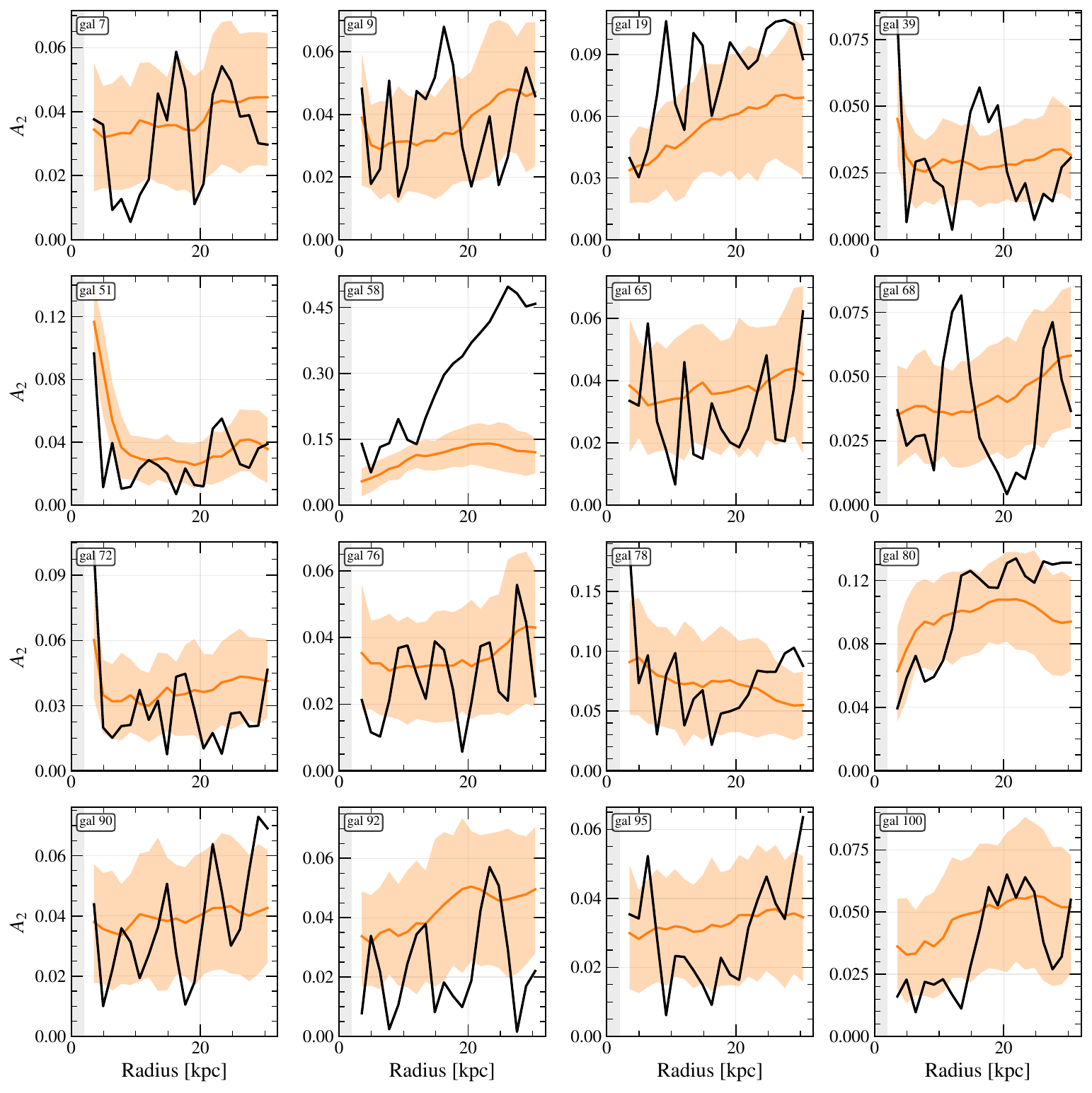}
    \caption{
    $m=2$ amplitude $A_2$ against projected radius for the same 16 held-out DREAMS test galaxies shown in Figure~\ref{fig:appendix_stamp_deviation}, in the same format as the top panel of Figure~\ref{fig:dreams_anisotropy}.
    Black is the truth, orange the mean over the $100$ stochastic draws, and the shaded band the $16$th-$84$th percentile range of the ensemble.
    The shaded strip at small radius marks the annuli excluded from the metric.
    }
    \label{fig:appendix_stamp_anisotropy_A2}
\end{figure*}

\begin{figure*}
    \centering
    \includegraphics[width=0.95\linewidth]{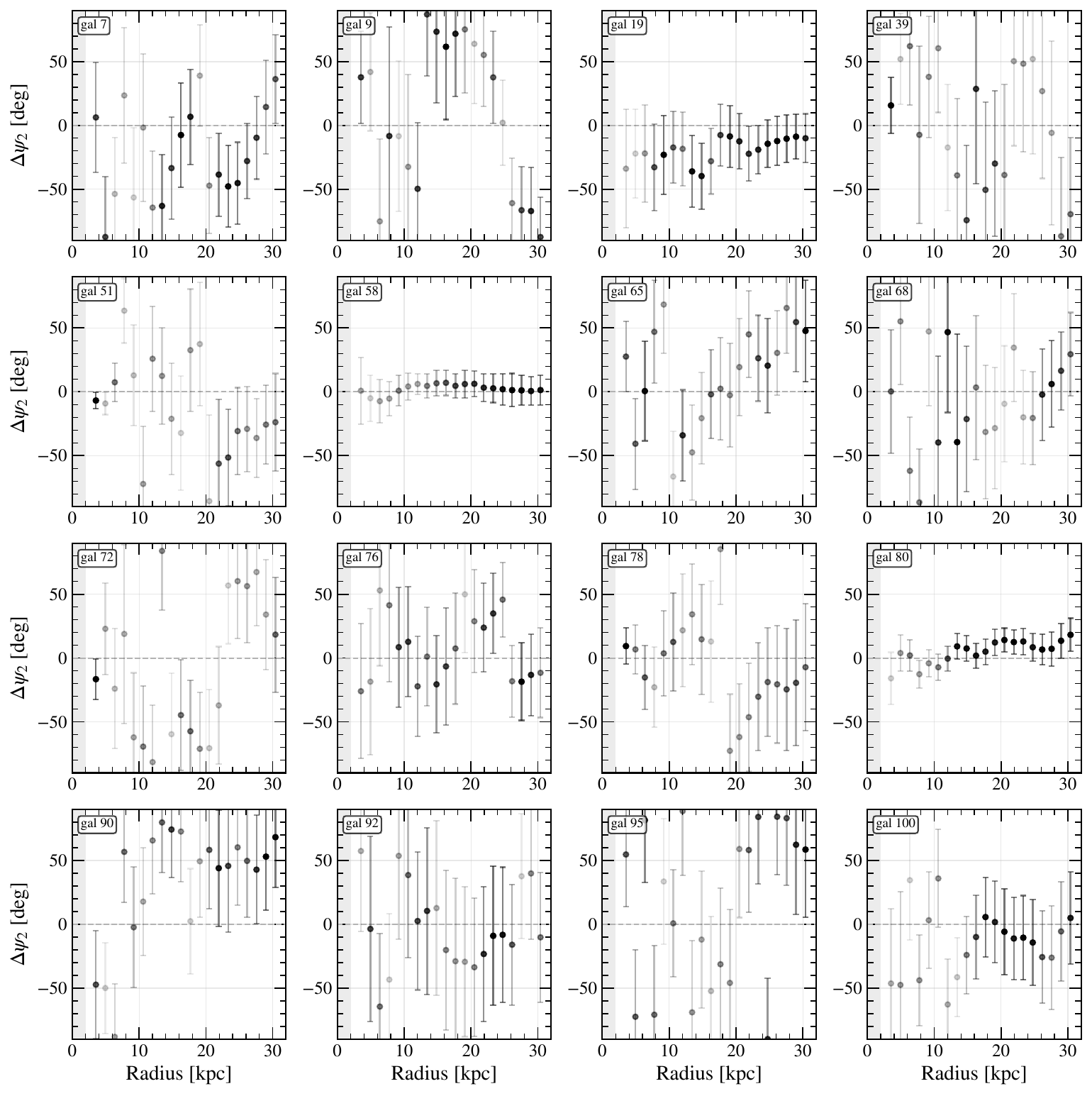}
    \caption{
    Offset $\Delta\psi_2$ between the predicted and true major-axis position angles for the same 16 galaxies, in the same format as the bottom panel of Figure~\ref{fig:dreams_anisotropy}.
    Error bars are the circular standard deviation of $\psi_2$ across the $100$ draws, and marker opacity is scaled by the true $A_2$ within each panel, so annuli whose position angle is ill-defined because the galaxy is nearly round appear faded.
    Every panel spans the full range of distinguishable offsets, $\pm90^{\circ}$, so the vertical scale is shared across the grid.
    }
    \label{fig:appendix_stamp_anisotropy_phase}
\end{figure*}

Figures~\ref{fig:appendix_stamp_anisotropy_A2} and~\ref{fig:appendix_stamp_anisotropy_phase} repeat the azimuthal $m=2$ diagnostics of Figure~\ref{fig:dreams_anisotropy} for the same 16 galaxies.
The two panels together show that the amplitude and the orientation of the anisotropy are recovered with accuracy depending on how anisotropic the galaxy is.
Galaxies 58 and 80, the two most elongated in the sample, have their major axis recovered in every examined radial annuli to better than $20^{\circ}$, with median offsets of $4^{\circ}$ and $8^{\circ}$ and $\mathcal{A}=0.99$ and $0.94$.
For the four nearly axisymmetric galaxies with $\mathcal{A}<0$ (galaxies 9, 72, 90 and 95), the true $A_2$ never exceeds $0.04$, so there is no well-defined major axis to recover and the measured offsets scatter over the full $\pm90^{\circ}$ range.

\begin{figure*}
    \centering
    \includegraphics[width=0.95\linewidth]{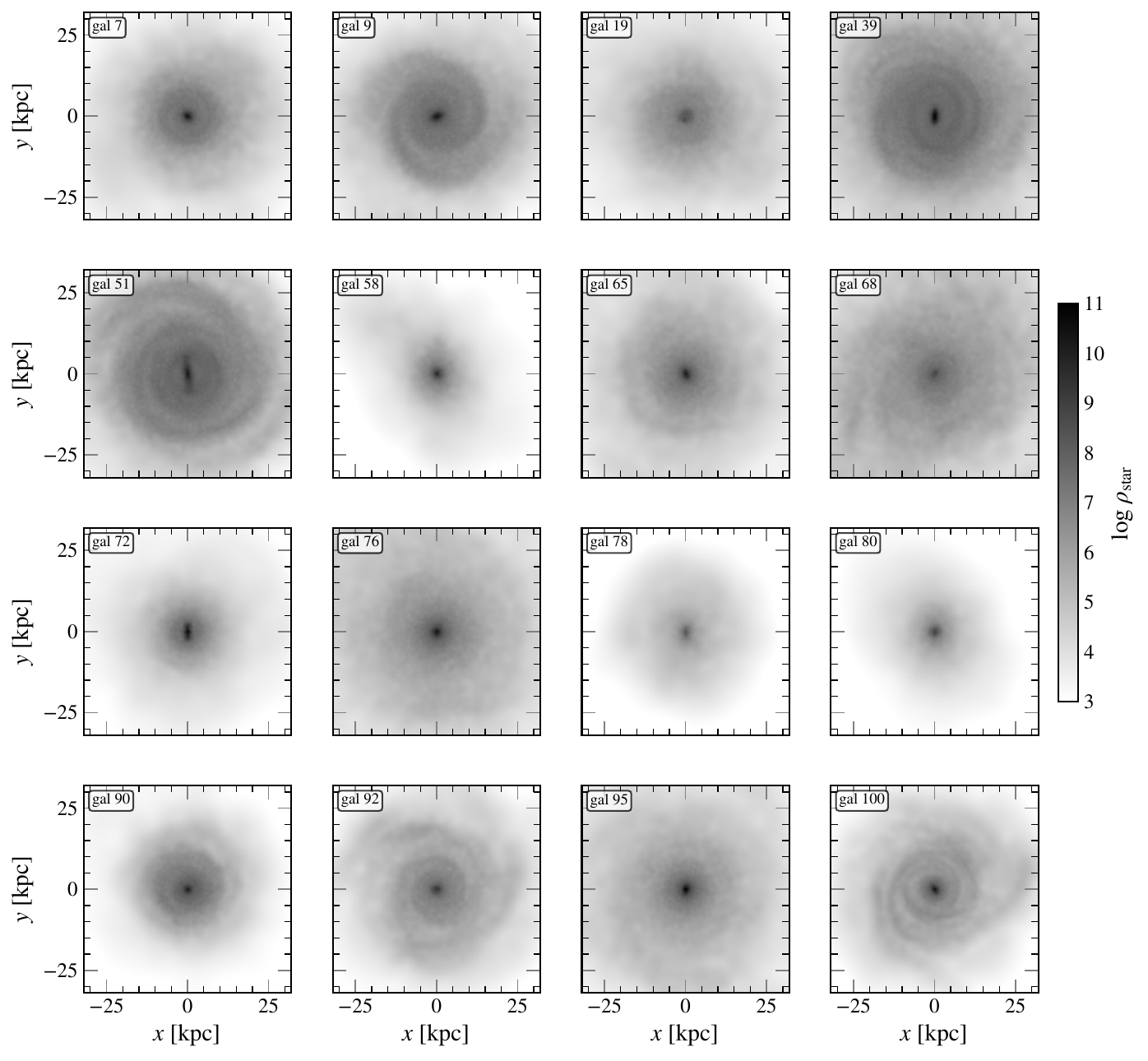}
    \caption{
    Input stellar density conditioning maps for the same 16 held-out DREAMS test galaxies shown in Figure~\ref{fig:appendix_stamp_deviation}, on the same color scale as the condition panel of Figure~\ref{fig:dreams_map}.
    These are the only galaxy-specific inputs the model receives, alongside the five scalar simulation parameters.
    }
    \label{fig:appendix_stamp_stellar}
\end{figure*}

\begin{figure*}
    \centering
    \includegraphics[width=0.95\linewidth]{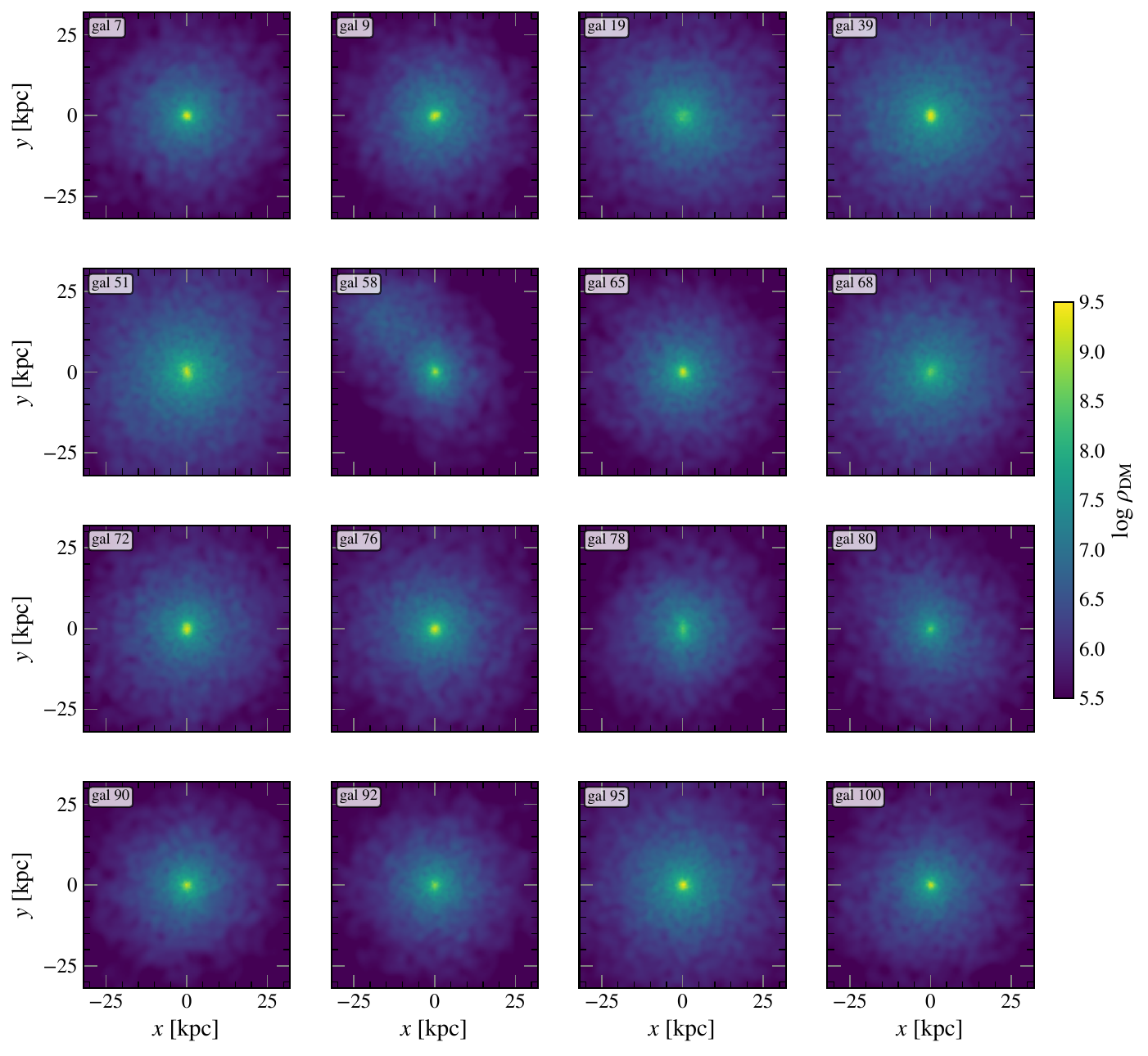}
    \caption{
    True dark matter density maps for the same 16 galaxies, on the same color scale as the corresponding panel of Figure~\ref{fig:dreams_map}.
    }
    \label{fig:appendix_stamp_truth}
\end{figure*}

\begin{figure*}
    \centering
    \includegraphics[width=0.95\linewidth]{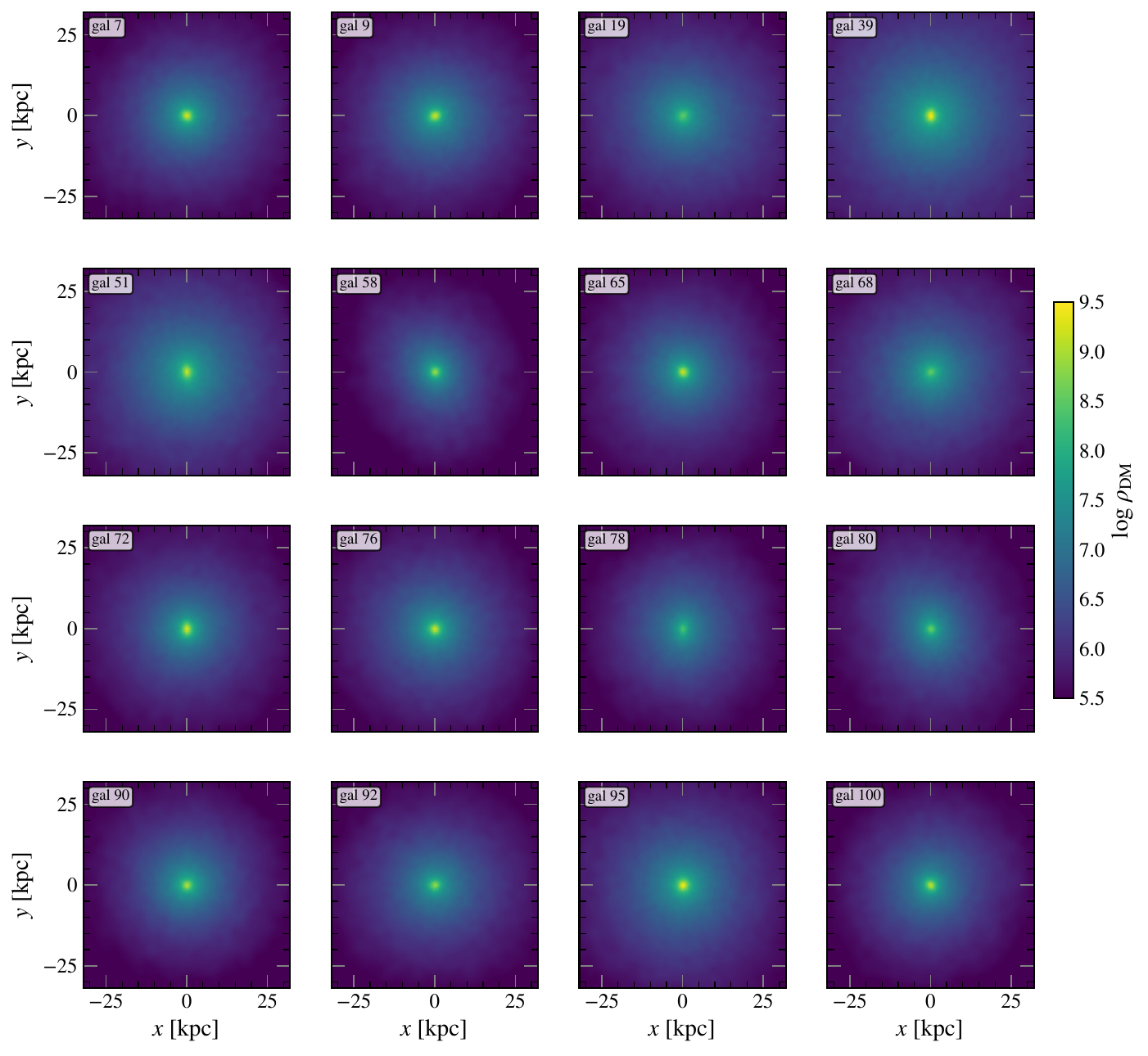}
    \caption{
    Mean predicted dark matter density maps for the same 16 galaxies, on the same color scale as Figure~\ref{fig:appendix_stamp_truth}, and to be compared with it panel by panel.
    The residuals between the two are the maps shown in Figure~\ref{fig:appendix_stamp_deviation}.
    }
    \label{fig:appendix_stamp_pred}
\end{figure*}

\begin{figure*}
    \centering
    \includegraphics[width=0.95\linewidth]{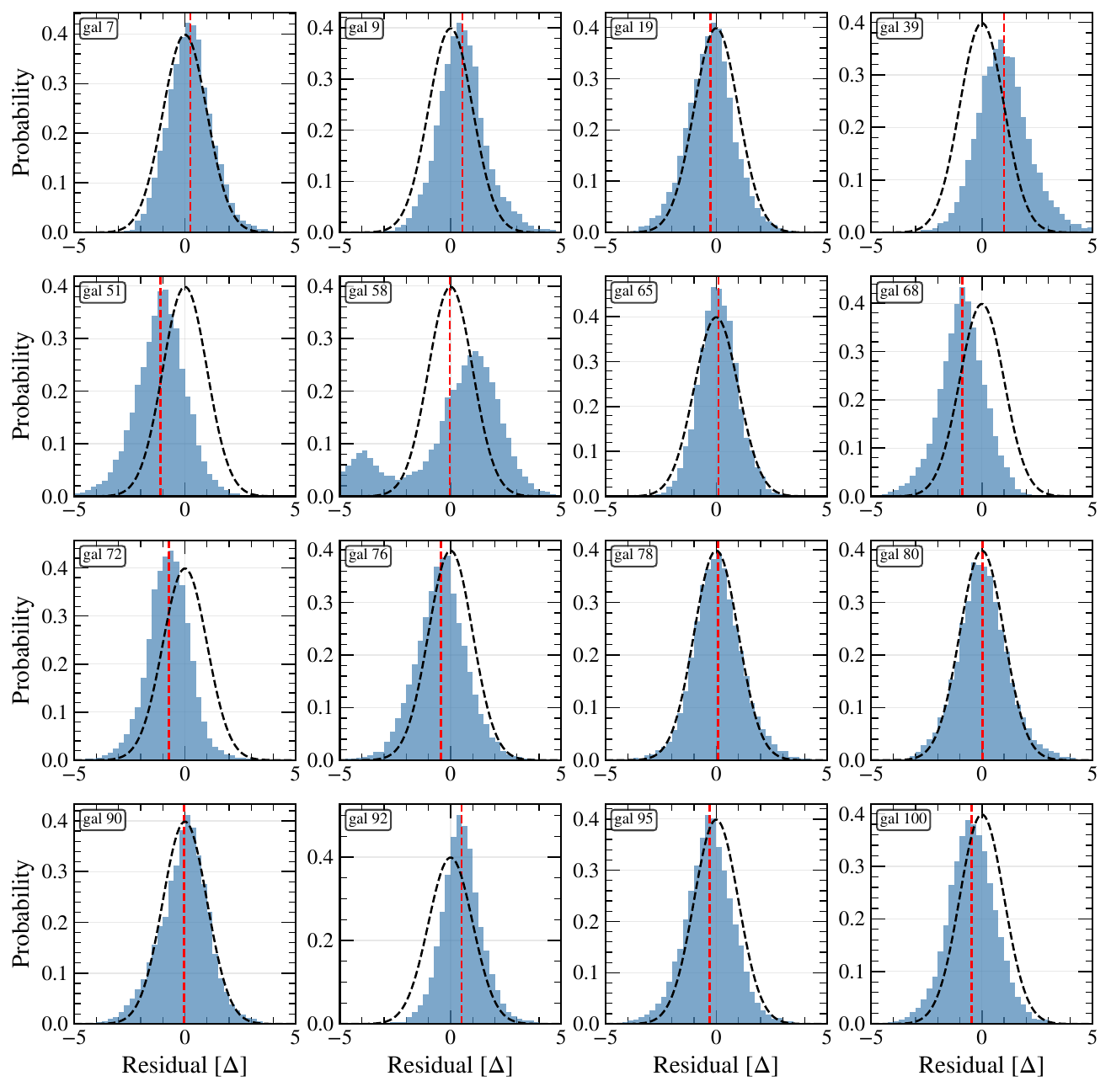}
    \caption{
    Pixel-level $\Delta$-residual distributions for the same 16 galaxies, in the same format as the top panel of Figure~\ref{fig:dreams_residual}.
    The black dashed curve is a unit Gaussian, shown as a reference rather than as a fit, and the red dashed line marks the mean $\Delta$ of each galaxy.
    }
    \label{fig:appendix_stamp_hist}
\end{figure*}

\begin{figure*}
    \centering
    \includegraphics[width=0.95\linewidth]{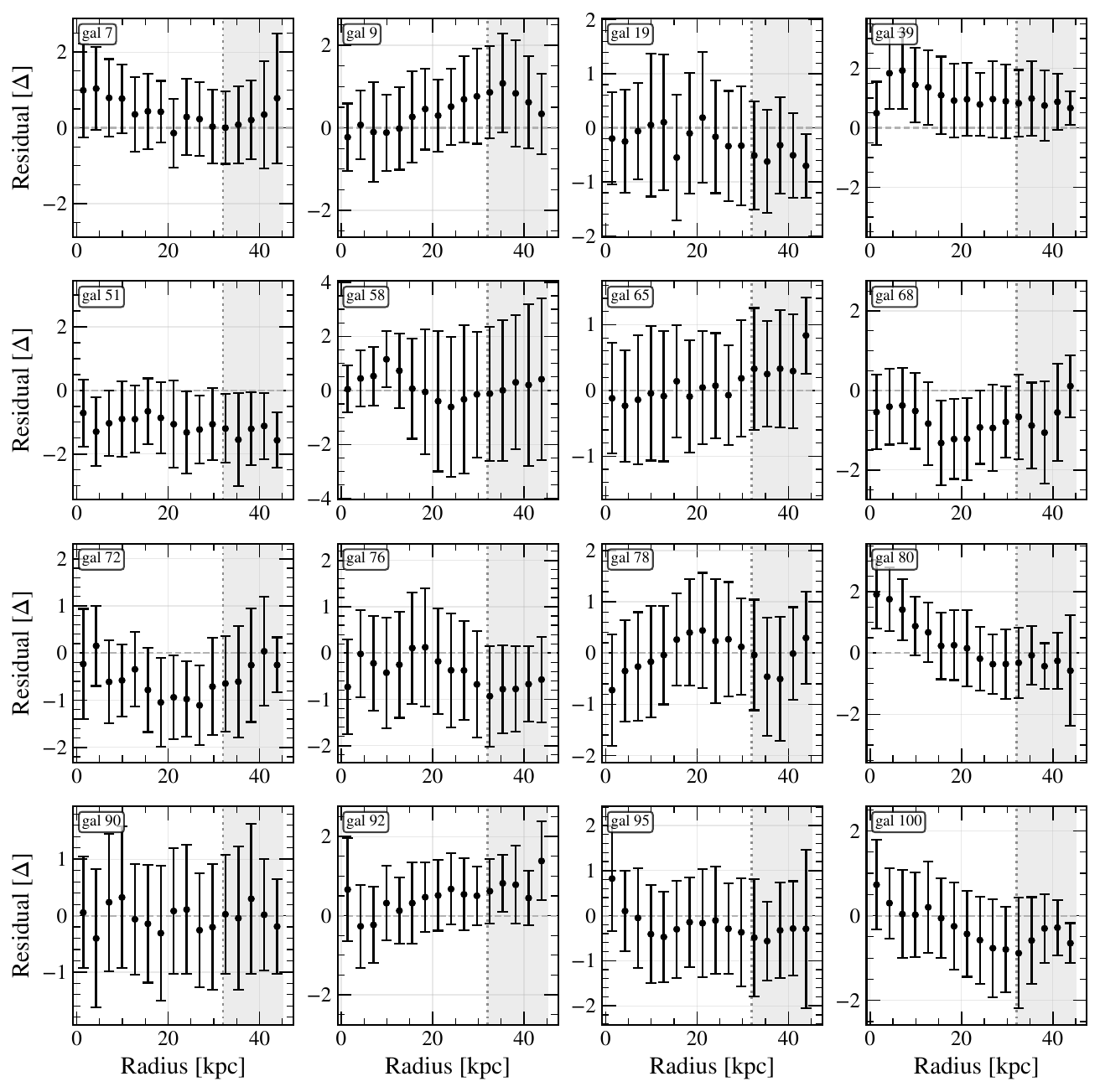}
    \caption{
    Radial dependence of the binned $\Delta$-residuals for the same 16 galaxies, in the same format as the middle panel of Figure~\ref{fig:dreams_residual}: points and error bars are the mean and standard deviation within each of 16 radial bins, and the dashed line marks $\Delta = 0$.
    The shaded region beyond $r=32$\,kpc marks the radii sampled only by the corners of the field of view.
    Note that the vertical scale is set independently for each galaxy.
    }
    \label{fig:appendix_stamp_radial}
\end{figure*}

\begin{figure*}
    \centering
    \includegraphics[width=0.95\linewidth]{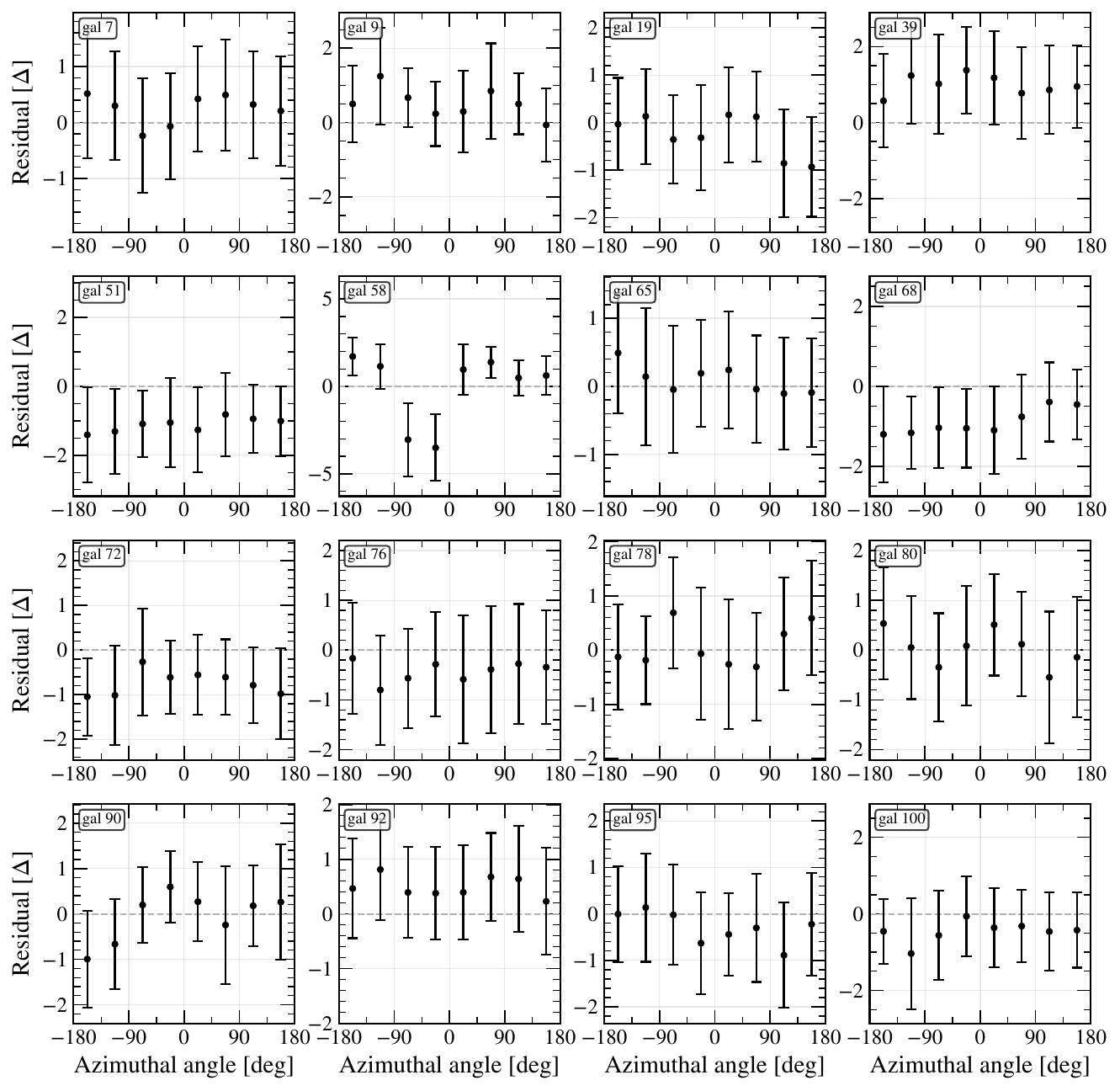}
    \caption{
    As Figure~\ref{fig:appendix_stamp_radial}, but binned in eight azimuthal angle bins, matching the bottom panel of Figure~\ref{fig:dreams_residual}.
    Galaxy 58 is the clearest case of coherent azimuthal structure in the sample.
    }
    \label{fig:appendix_stamp_azimuthal}
\end{figure*}

The two figures above show the residual map and the prior/posterior profile comparison, the diagnostics most directly complementary to Figures~\ref{fig:dreams_map}-\ref{fig:dreams_profile_comp}.
For completeness we collect the remaining six panel types for the same 16 galaxies.
Figures~\ref{fig:appendix_stamp_stellar}-\ref{fig:appendix_stamp_pred} show the quantities that enter and leave the model: the input stellar density map, the true dark matter map, and the mean predicted dark matter map on the same color scales as Figure~\ref{fig:dreams_map}, so that each prediction can be compared with its own target.
Figures~\ref{fig:appendix_stamp_hist}-\ref{fig:appendix_stamp_azimuthal} show the remaining residual diagnostics: the pixel-level $\Delta$ distribution and its radial and azimuthal dependence, which are the stamp-grid counterparts of the three panels of Figure~\ref{fig:dreams_residual}.

These reinforce the picture described above, and make the individual failure modes easier to attribute to a specific diagnostic.
Most $\Delta$ distributions are centered near zero and comparable in width to the unit Gaussian reference, while the global offsets of galaxies 39, 51, and 68 appear as bodily shifts of the entire distribution rather than as changes in its shape.
The radial panels isolate the coherent radial trend of galaxy 80, which declines monotonically from $\Delta \approx +2$ in the innermost bin to slightly below zero beyond $25$\,kpc, and the azimuthal panels isolate the large-scale gradient of galaxy 58, where two adjacent angular bins sit near $\Delta \approx -3$ while the other six lie between $0$ and $+2$.
Galaxy 58 is also the one clearly bimodal $\Delta$ distribution in Figure~\ref{fig:appendix_stamp_hist}, which is the same structure seen through a different projection.

\section{Radially resolved residual distributions}
\label{sec:appendix_radial}
\setcounter{figure}{0}
\setcounter{table}{0}

\begin{figure*}
    \centering
    \includegraphics[width=0.95\linewidth]{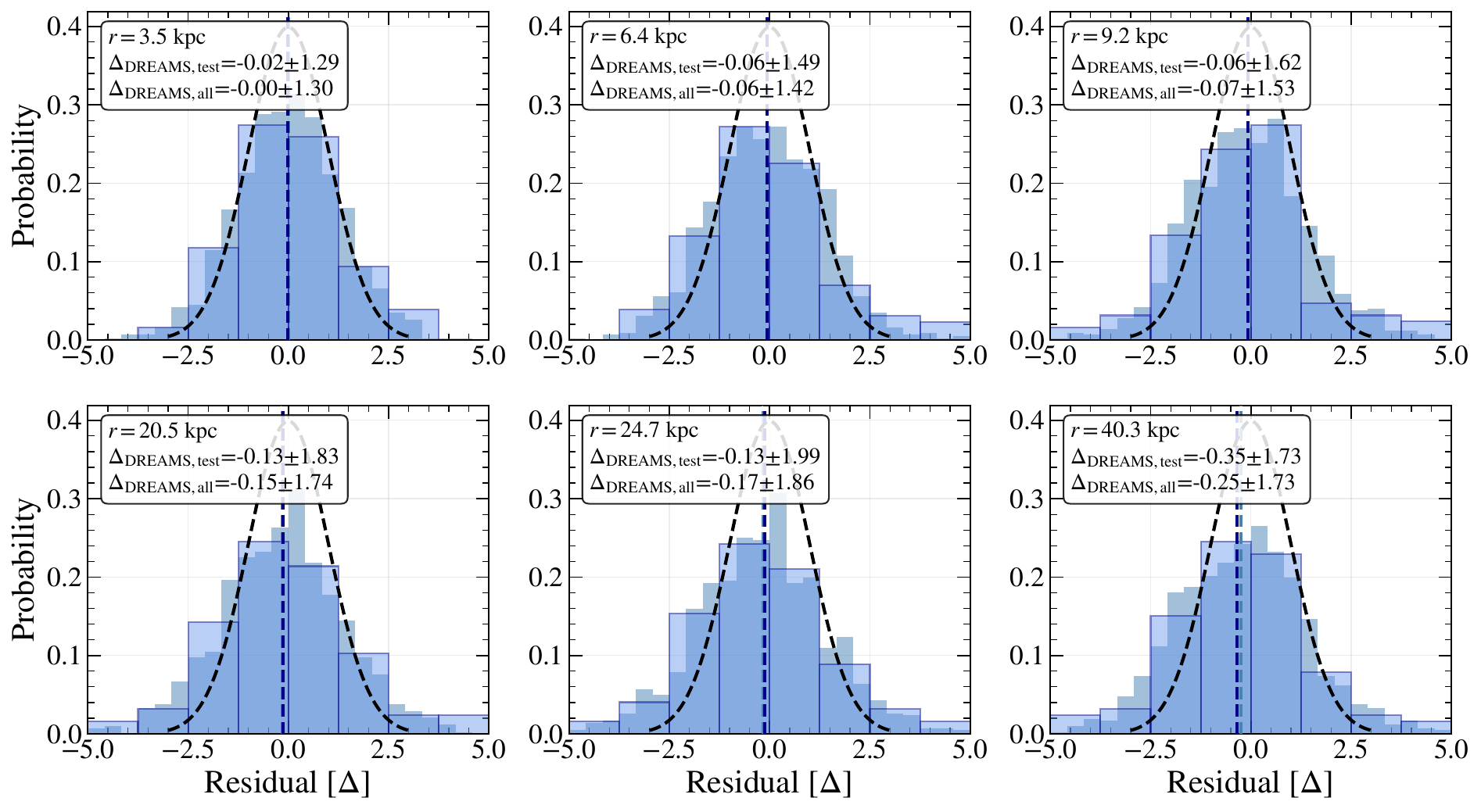}
    \caption{
    Uncalibrated profile-level $\Delta$-residual distributions for the DREAMS sample at six radial locations.
    The 102 held-out test galaxies are the darker histogram.
    The $R^2$ statistics are quoted separately for the held-out split and for all 1024 galaxies.
    The standard deviation rises from $1.29$ at $r=3.5$\,kpc to $1.99$ at $r=24.7$\,kpc, which is the radial dependence of the calibration factor applied in Figure~\ref{fig:calibration}.
    }
    \label{fig:appendix_dreams_sixpanel}
\end{figure*}

\begin{figure*}
    \centering
    \includegraphics[width=0.95\linewidth]{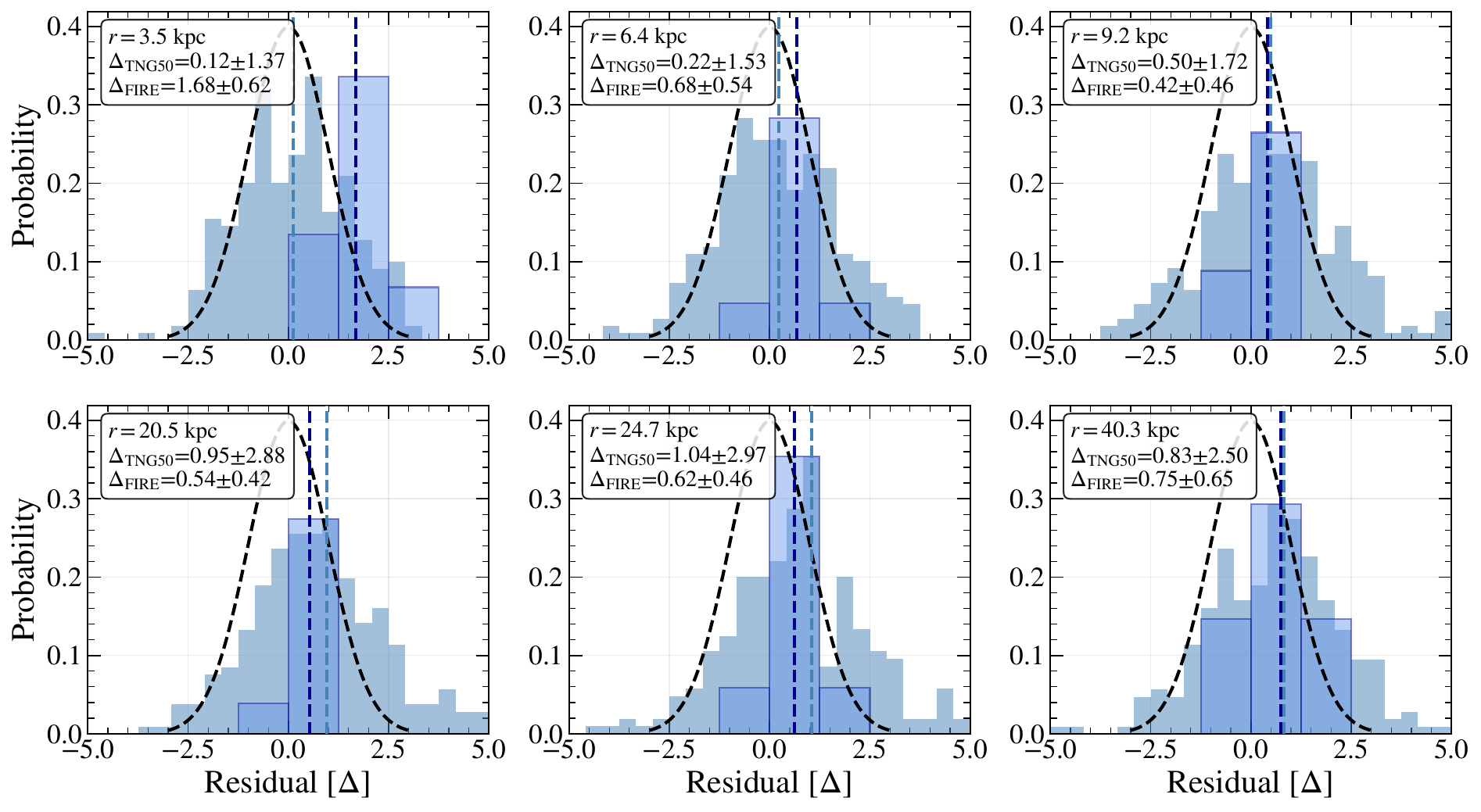}
    \caption{
    As Figure~\ref{fig:appendix_dreams_sixpanel}, but for the out-of-domain TNG50 (light blue) and \Fire (dark blue) samples.
    }
    \label{fig:appendix_ood_sixpanel}
\end{figure*}

Figure~\ref{fig:calibration} shows the profile-level residual distribution at a single radius.
Here we show the same diagnostic at six radial locations, without calibration, for the DREAMS sample (Figure~\ref{fig:appendix_dreams_sixpanel}) and for the out-of-domain samples (Figure~\ref{fig:appendix_ood_sixpanel}).
These are the distributions from which the calibration factors quoted in Section~\ref{sec:all_gal} are measured, and they show that the required correction grows with radius.

\section{Simulation-parameter sweep tests}
\label{sec:appendix_sweeps}
\setcounter{figure}{0}
\setcounter{table}{0}

Section~\ref{sec:all_gal} shows that deliberately setting the supernova-feedback energy parameter ($\bar{e}_w$) to values away from a galaxy's true value shifts the population-level $\Delta$-residual distribution in a physically consistent direction, supporting the interpretation that the model has learned to use the scalar simulation-parameter conditioning rather than relying on the stellar density image alone.
Here we extend that test to all five DREAMS simulation parameters used for conditioning: the two cosmological parameters ($\Omega_m$, $\sigma_8$) and the three baryonic feedback parameters (supernova energy $\bar{e}_w$, supernova wind velocity normalization $\kappa_w$, and AGN coupling efficiency $\epsilon_{f,\,{\rm high}}$).
For each parameter, we fix the stellar density input and hold the other four conditioning parameters at their true values, while sweeping the parameter of interest over five values spanning its DREAMS training range, and generate predictions for all 1024 DREAMS galaxies at each grid point.

\begin{figure*}
    \centering
    \includegraphics[width=0.95\linewidth]{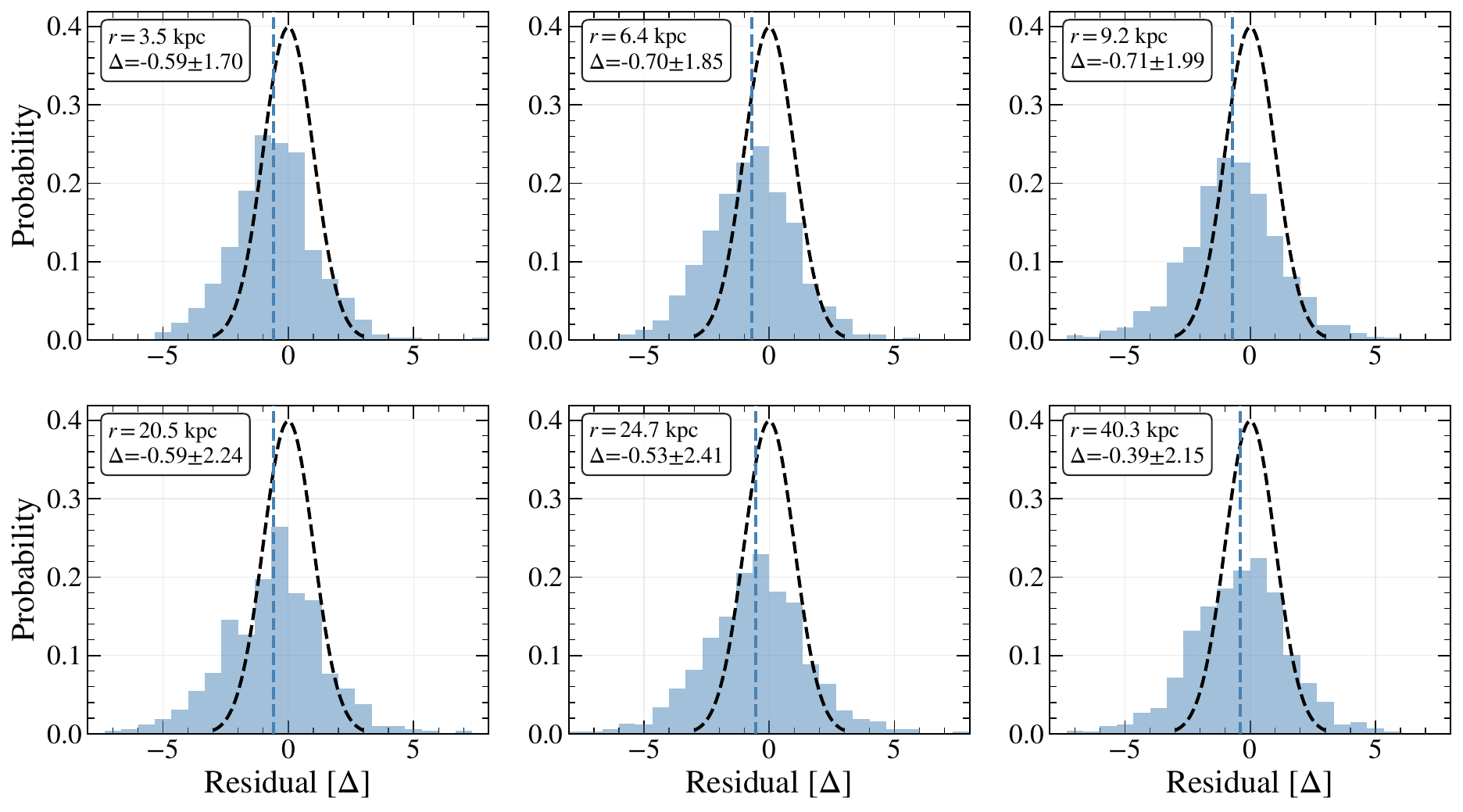} \\
    \includegraphics[width=0.95\linewidth]{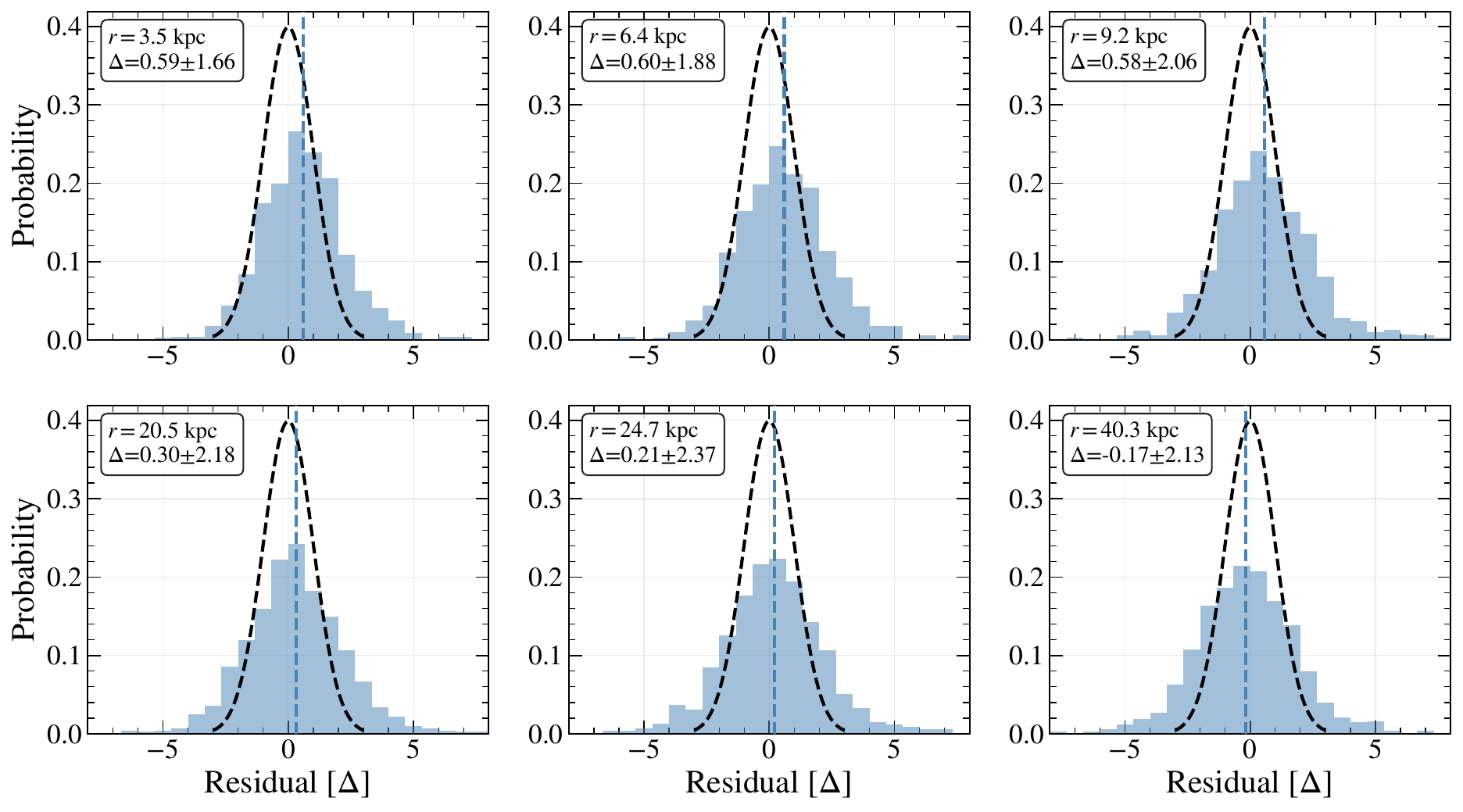}
    \caption{
    Population-level $\Delta$-residual distributions with $\Omega_m$ set to the minimum (top, $\Omega_m=0.274$) and maximum (bottom, $\Omega_m=0.354$) values of the DREAMS training range during inference, with the stellar density input and the other four conditioning parameters held fixed.
    Same format as Figure~\ref{fig:sim_par_test}, except that the residuals here are \textit{uncalibrated}, so that all five parameters are compared on a common footing.
    This figure and Figures~\ref{fig:appendix_sweep_s8} and \ref{fig:appendix_sweep_bhff} share an axis range of $[-8,8]$. 
    Figures~\ref{fig:appendix_sweep_sn1} and~\ref{fig:appendix_sweep_sn2} use a wider range because the two supernova responses run much further from zero.
    }
    \label{fig:appendix_sweep_om}
\end{figure*}

\begin{figure*}
    \centering
    \includegraphics[width=0.95\linewidth]{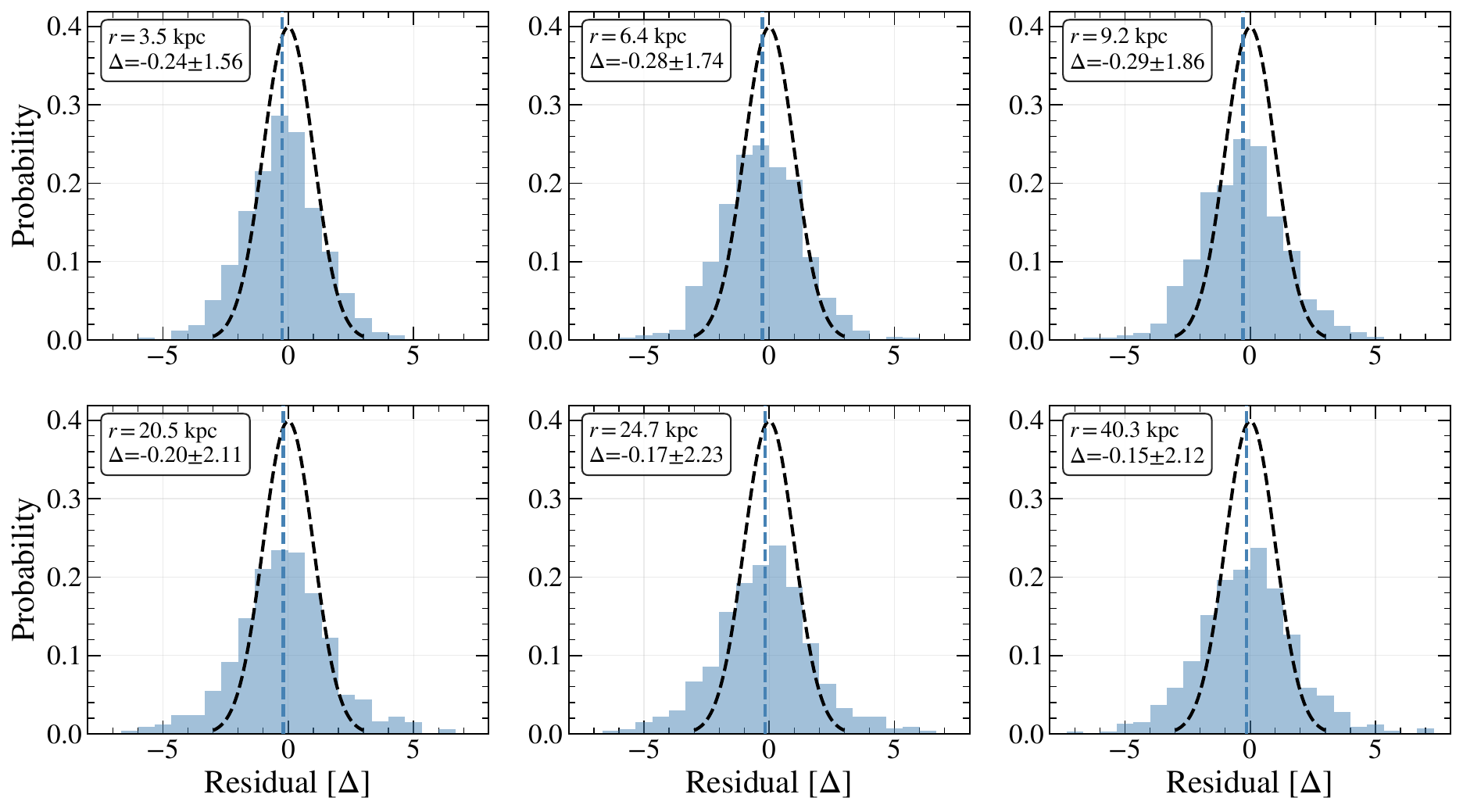} \\
    \includegraphics[width=0.95\linewidth]{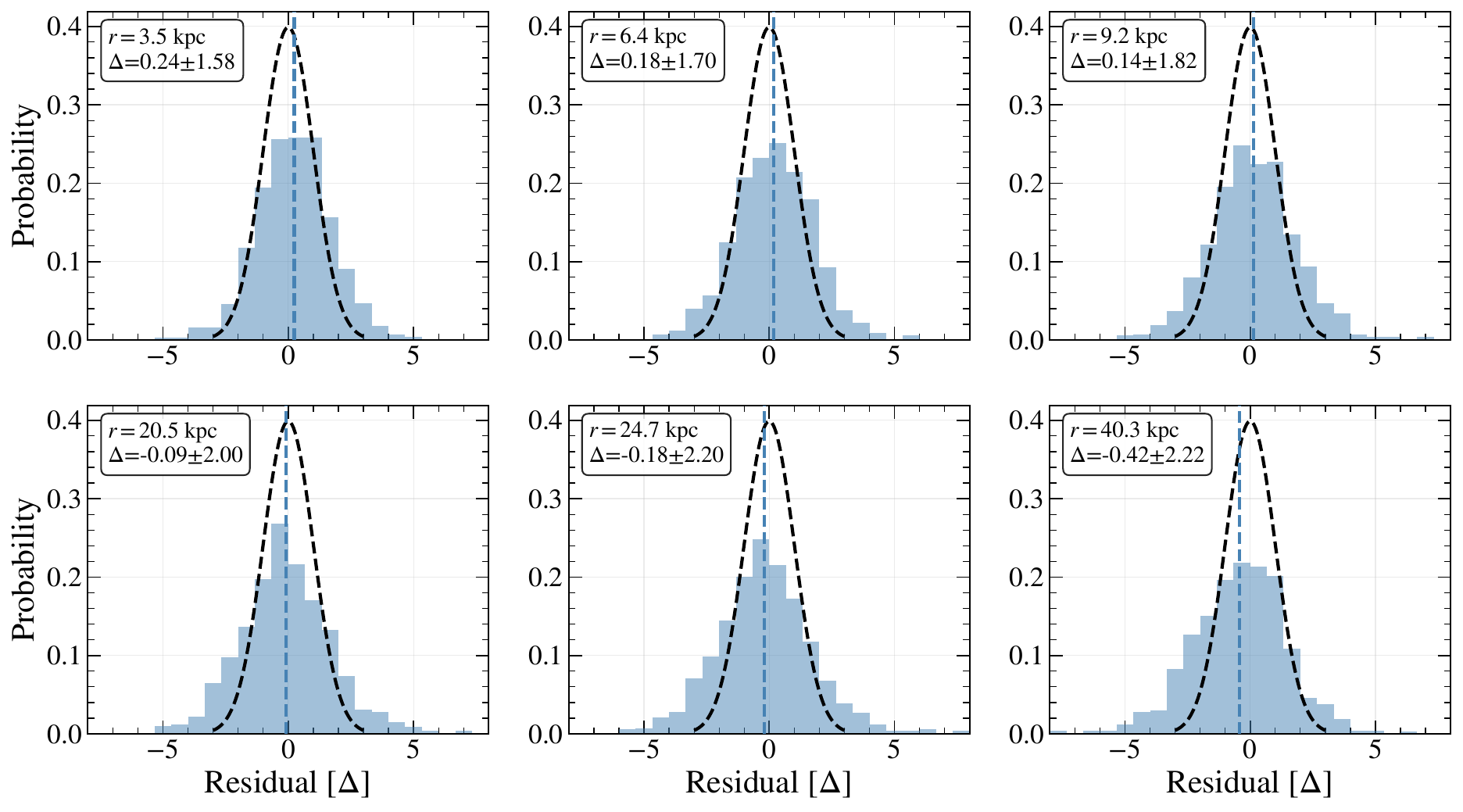}
    \caption{
    As Figure~\ref{fig:appendix_sweep_om}, but sweeping $\sigma_8$ from $0.780$ (top) to $0.888$ (bottom).
    This is the weakest response of the five parameters.
    }
    \label{fig:appendix_sweep_s8}
\end{figure*}

\begin{figure*}
    \centering
    \includegraphics[width=0.95\linewidth]{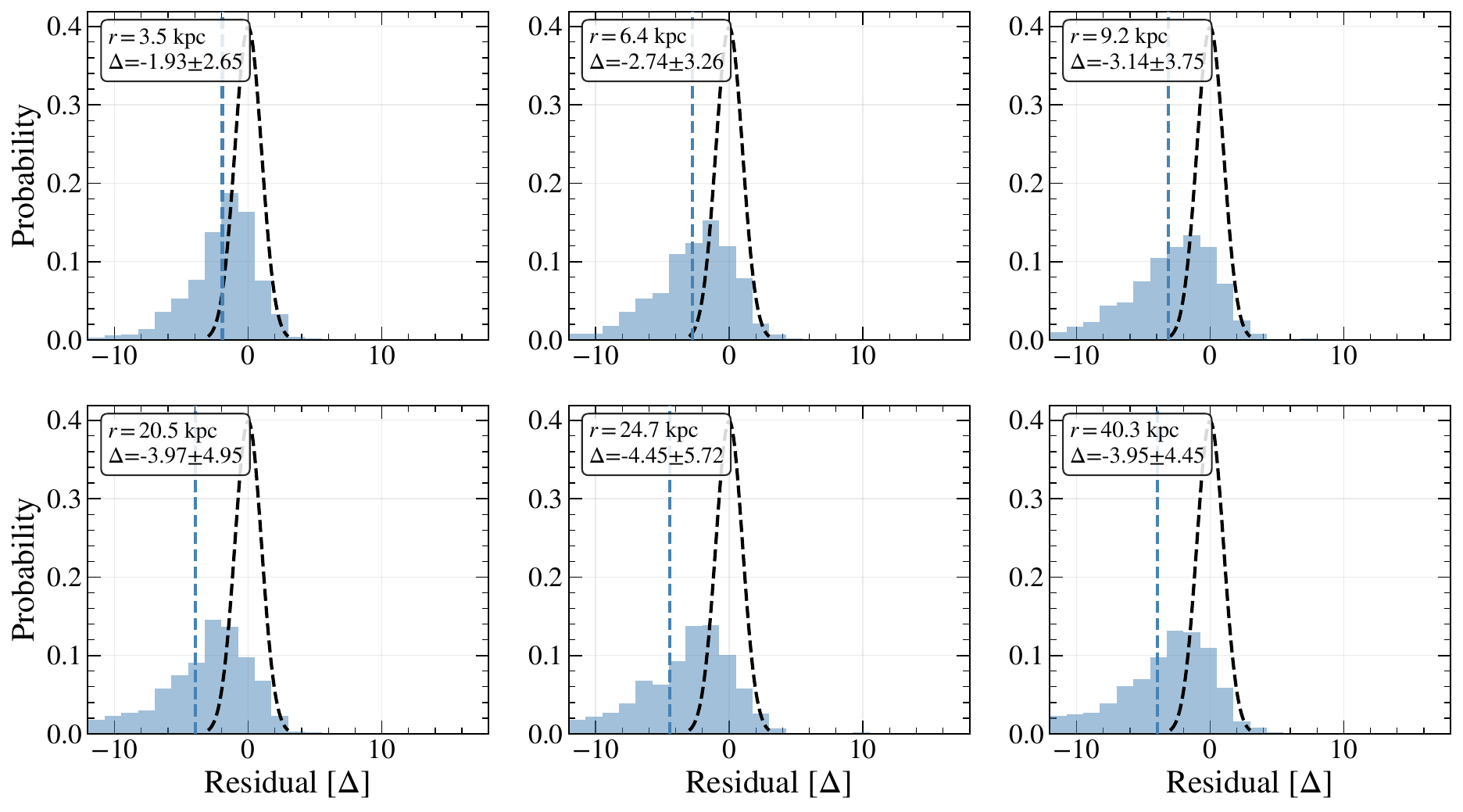} \\
    \includegraphics[width=0.95\linewidth]{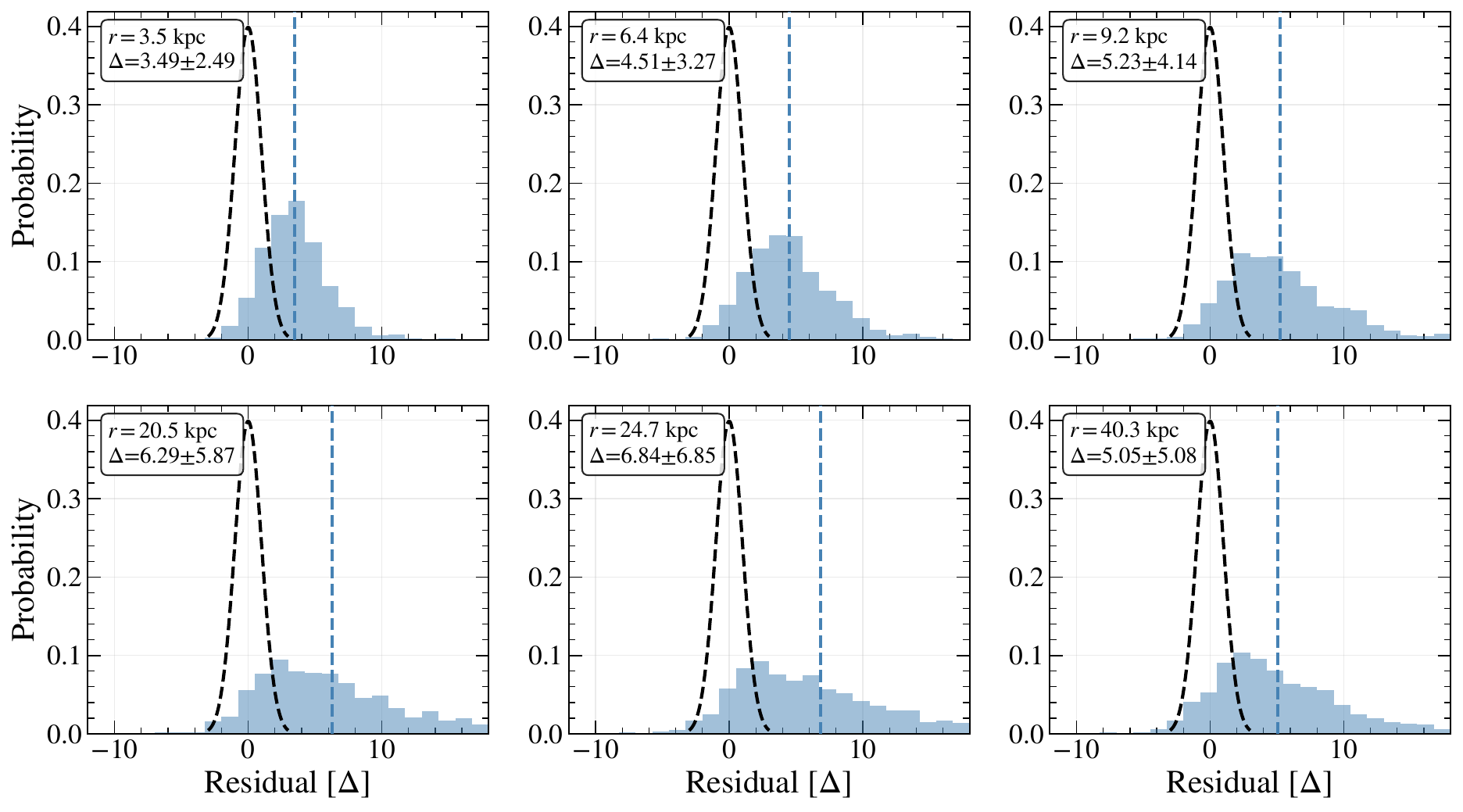}
    \caption{
    As Figure~\ref{fig:appendix_sweep_om}, but sweeping the supernova wind energy $\bar{e}_w$ from $0.25$ (top) to $4.0$ (bottom), and on the wider $[-12,18]$ axis range.
    This is the largest response of the five parameters.
    Figure~\ref{fig:sim_par_test} shows the same sweep after calibration and at two radii. 
    Here it is uncalibrated and at all six, so that it can be compared directly with the other four parameters.
    }
    \label{fig:appendix_sweep_sn1}
\end{figure*}

\begin{figure*}
    \centering
    \includegraphics[width=0.95\linewidth]{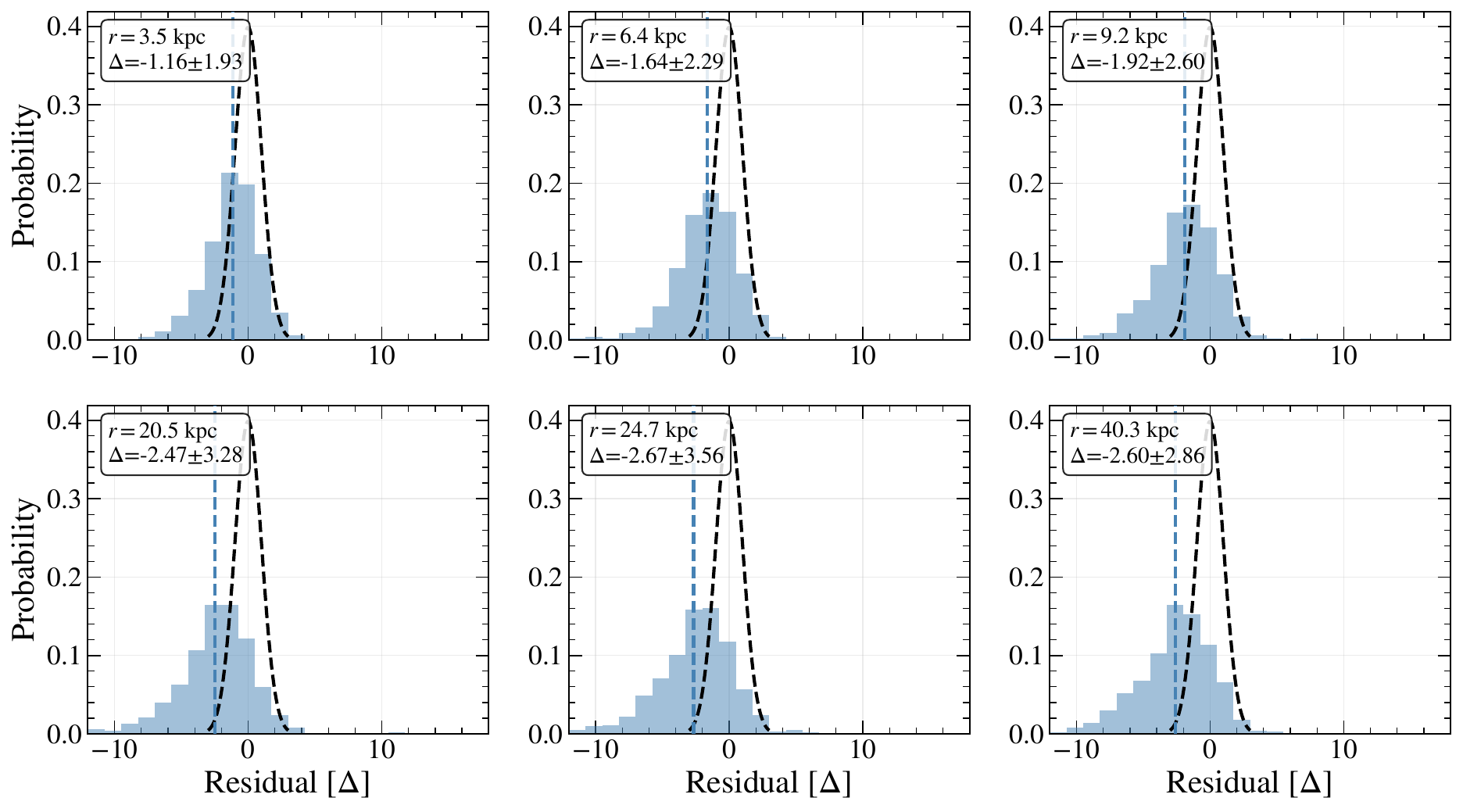} \\
    \includegraphics[width=0.95\linewidth]{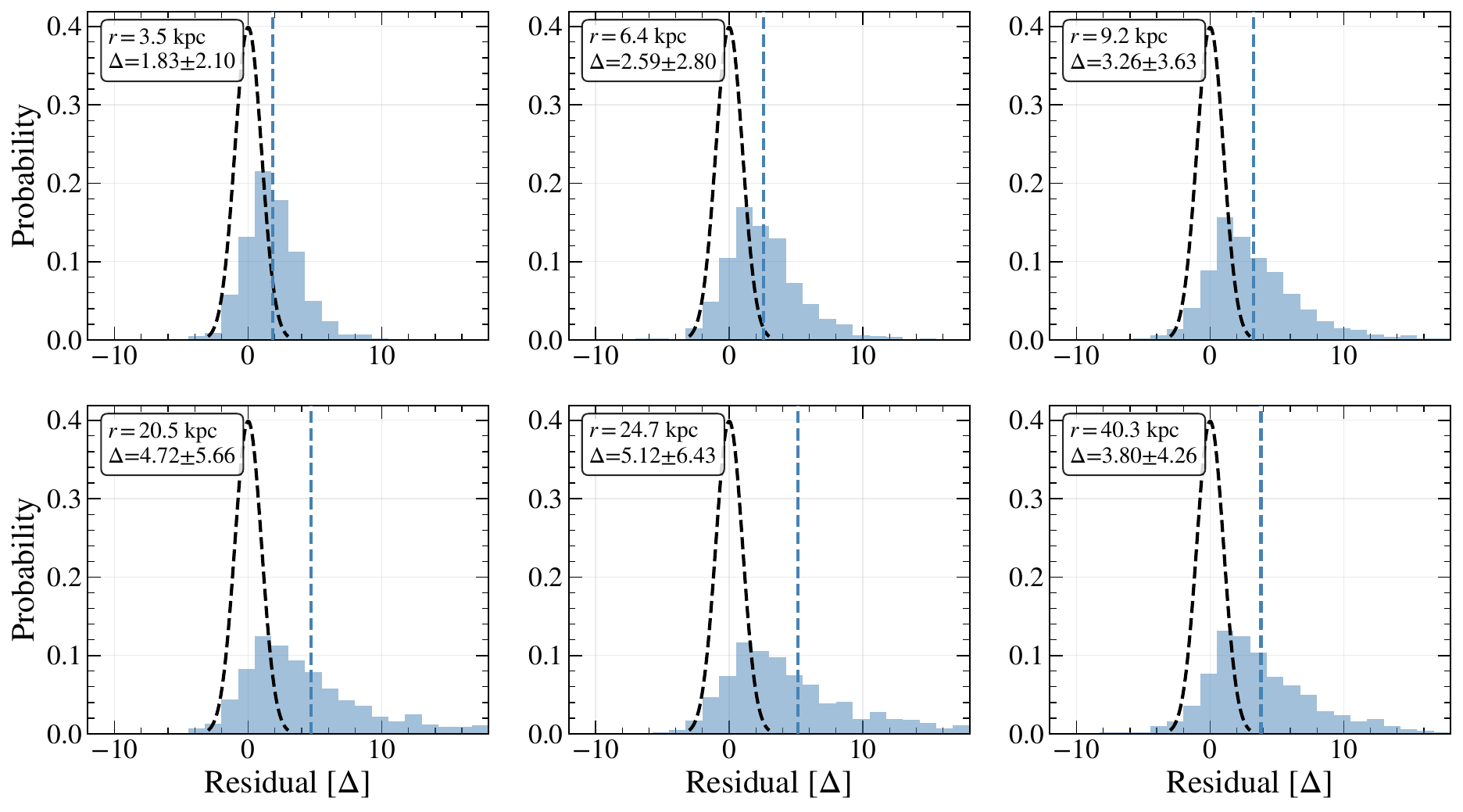}
    \caption{
    As Figure~\ref{fig:appendix_sweep_om}, but sweeping the supernova wind velocity normalization $\kappa_w$ from $0.5$ (top) to $2.0$ (bottom).
    The shift runs in the same sense as for $\bar{e}_w$ in Figure~\ref{fig:appendix_sweep_sn1}, and is the second largest of the five parameters.
    Note the wider horizontal range, $[-12,18]$, needed to contain this distribution.
    }
    \label{fig:appendix_sweep_sn2}
\end{figure*}

\begin{figure*}
    \centering
    \includegraphics[width=0.95\linewidth]{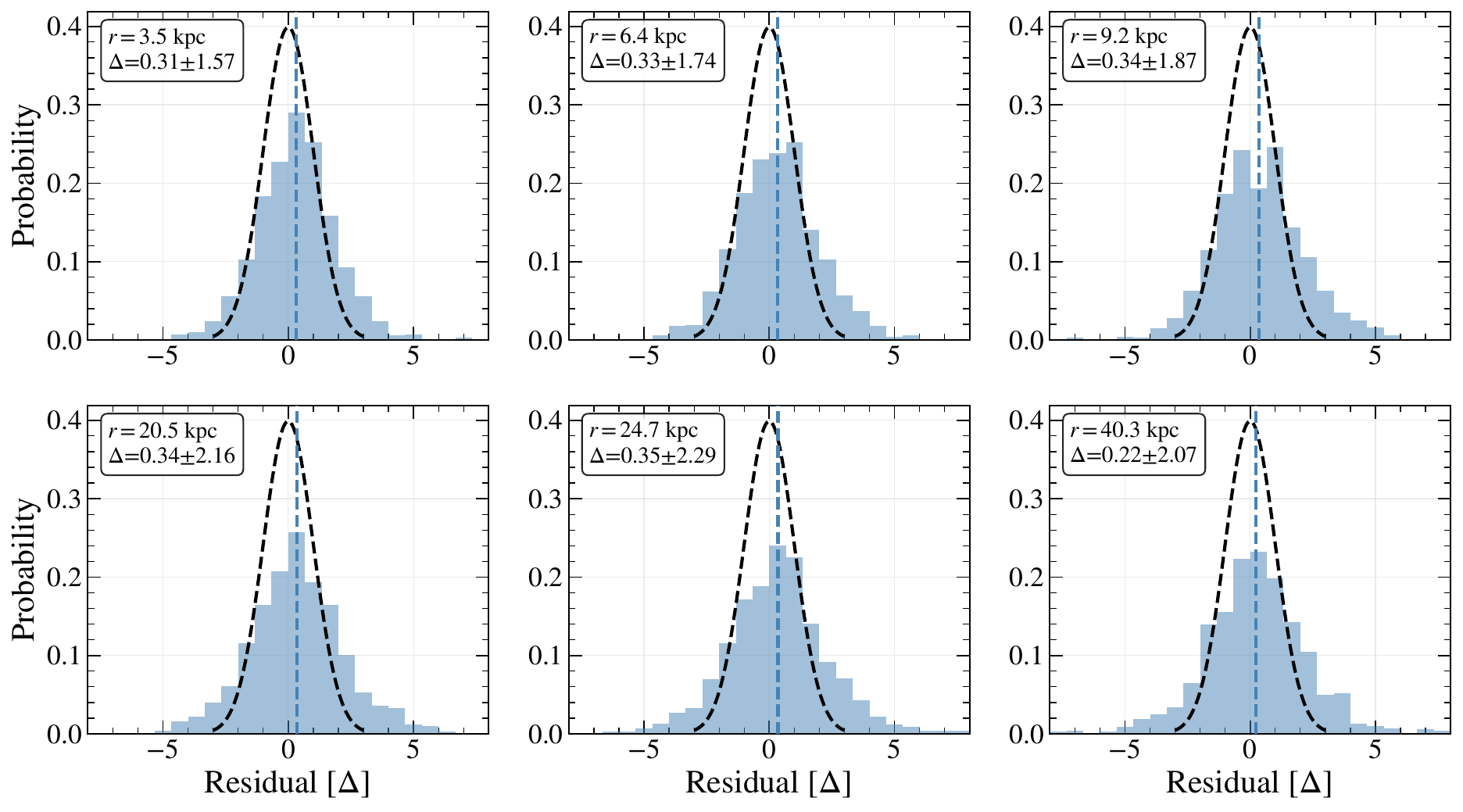} \\
    \includegraphics[width=0.95\linewidth]{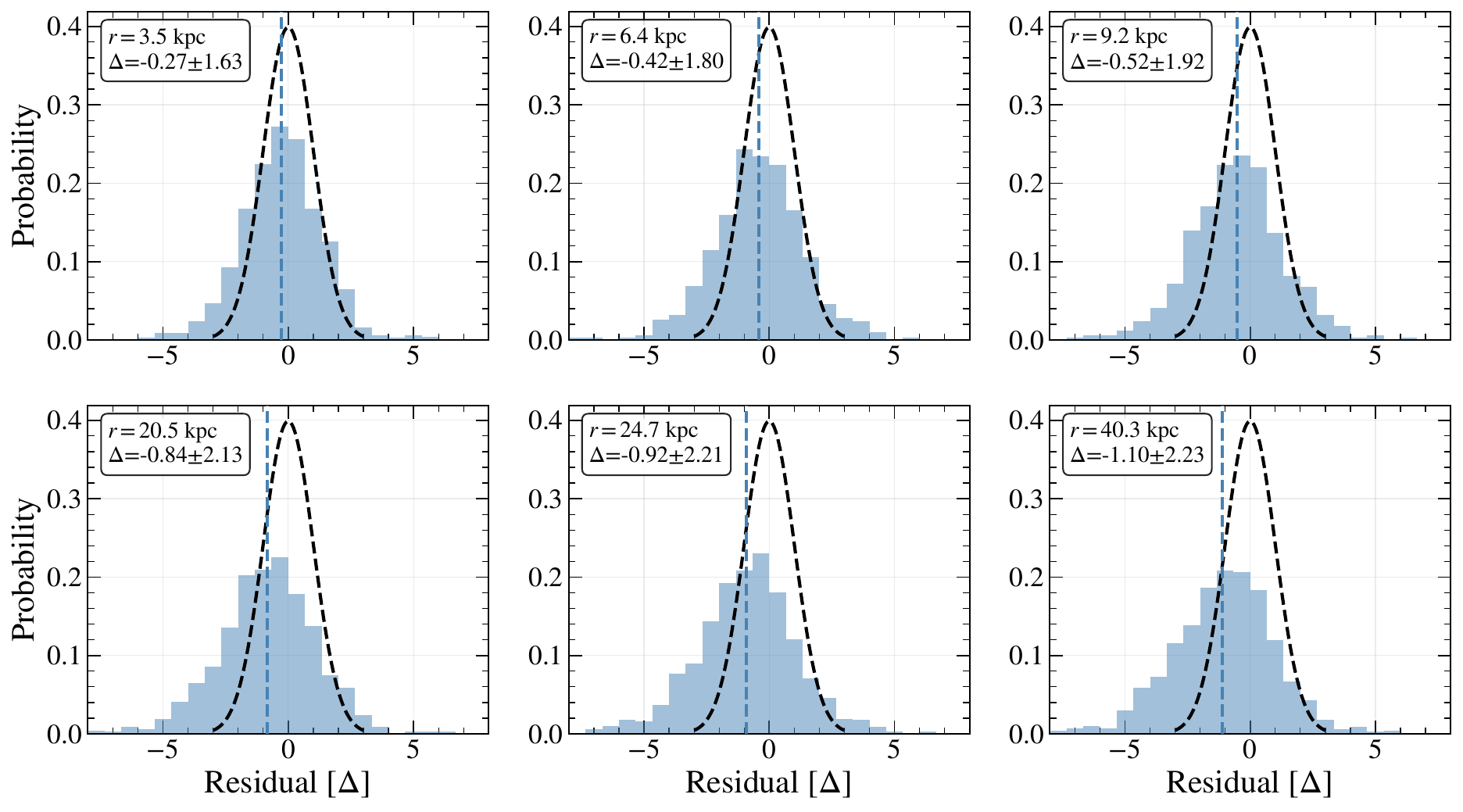}
    \caption{
    As Figure~\ref{fig:appendix_sweep_om}, but sweeping the AGN coupling efficiency $\epsilon_{f,\,{\rm high}}$ from $0.25$ (top) to $4.0$ (bottom).
    Note that the shift runs in the opposite sense to the supernova parameters, and that its amplitude is comparable to that of $\Omega_m$ rather than to the other baryonic parameters.
    }
    \label{fig:appendix_sweep_bhff}
\end{figure*}

Figures~\ref{fig:appendix_sweep_om}-\ref{fig:appendix_sweep_bhff} show the resulting population-level $\Delta$-residual distributions for the minimum and maximum grid value of each parameter, in the same format as Figure~\ref{fig:sim_par_test}.
The model responds monotonically to all five conditioning parameters, but the amplitudes differ by more than an order of magnitude.
Measured as the shift in the population mean $\Delta$ between the extreme grid values, the two supernova-feedback parameters dominate: the wind energy $\bar{e}_w$ produces the largest response, with the population mean residual running from $\Delta\simeq-3.1$ to $\sim5.2$ at $r=9.2$\,kpc, and the wind velocity normalization $\kappa_w$ the second largest, reaching $\Delta\simeq5.1$ at its maximum grid value.
In both cases larger values of the parameter drive the prediction toward higher dark matter density.
The values quoted in this appendix are \textit{uncalibrated}, so that all five parameters are compared on the same footing. 
Figure~\ref{fig:sim_par_test} shows the $\bar{e}_w$ response after the calibration of Section~\ref{sec:all_gal}, which divides these shifts by roughly a factor of two.
By contrast, the AGN coupling efficiency $\epsilon_{f,\,{\rm high}}$ and the two cosmological parameters all produce shifts of $|\Delta|\lesssim1$, with $\sigma_8$ the weakest of the five.
Taken together, these tests confirm that the simulation parameter conditioning is used when available, and that its effect is concentrated in the parameters that most directly govern the baryonic modification of the halo.

\section{Training without simulation-parameter conditioning}
\label{sec:appendix_noh}
\setcounter{figure}{0}
\setcounter{table}{0}

The sweep tests of Appendix~\ref{sec:appendix_sweeps} show that the model does use the scalar simulation parameters, but they do not say how much of its performance depends on them.
To measure that directly, we retrained the model with simulation-parameter conditioning removed entirely, so that the projected stellar density map is the only information the network receives.
Every other ingredient of the model is held fixed.

The figures in this appendix are generated from $10$-draw ensembles, except for the \Fire test (see below), rather than the $100$ used in the main text, for reasons of computational cost. 
We have verified on the held-out split that this does not affect any conclusion drawn here.

\begin{figure*}
    \centering
    \includegraphics[width=0.95\linewidth]{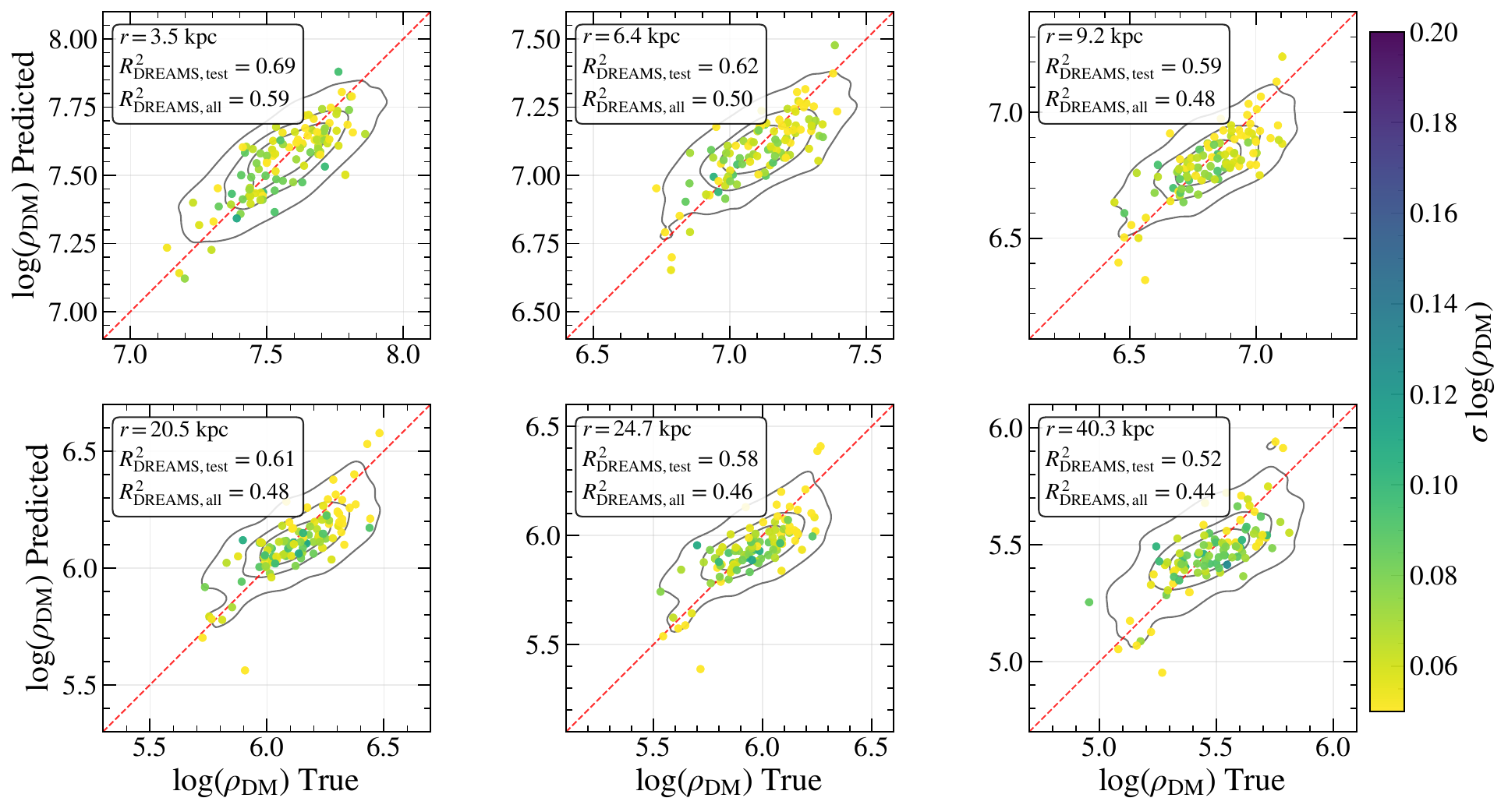}
    \caption{
    Predicted versus true azimuthally averaged dark matter density for the model trained \textit{without} simulation-parameter conditioning, in the same format as Figure~\ref{fig:dreams_profiles}. 
    Compared with the conditioned model, $R^2$ on the held-out split falls from $0.69$-$0.79$ to $0.52$-$0.69$, so most of the recoverable signal is still obtained from the stellar density map alone.
    }
    \label{fig:appendix_noh_scatter}
\end{figure*}

Figure~\ref{fig:appendix_noh_scatter} repeats the predicted-versus-true comparison of Figure~\ref{fig:dreams_profiles} for the unconditioned model.
Removing all five simulation parameters costs roughly $0.10$-$0.19$ in $R^2$ on the held-out split, yet the unconditioned model still recovers most of the galaxy-to-galaxy variance in the dark matter profile from the stellar density image alone.
The information is therefore carried predominantly by the conditioning image itself.
This is ideal as the stellar density map can be more readily connected to an observational counterpart. 

\begin{figure*}
    \centering
    \includegraphics[width=0.95\linewidth]{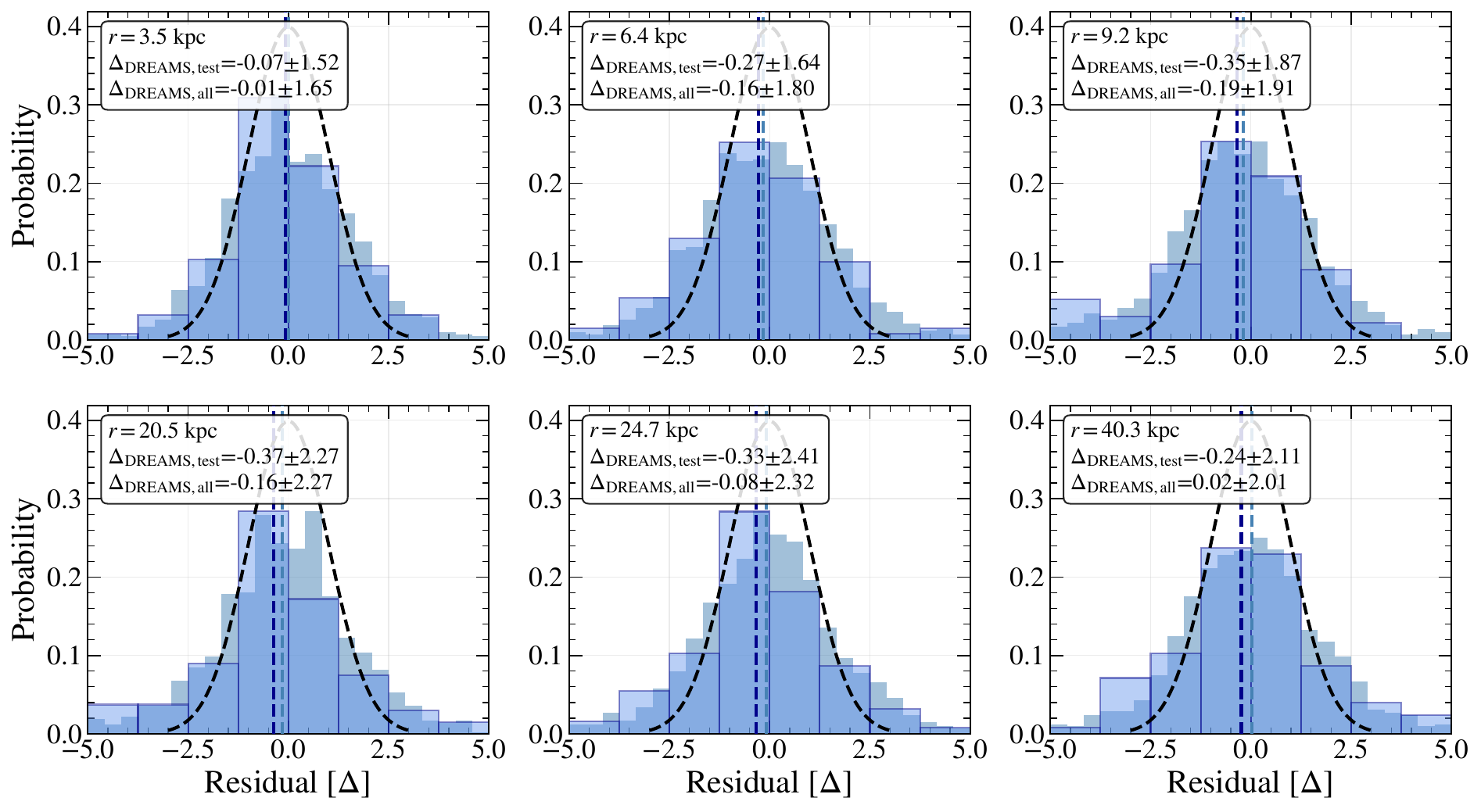}
    \caption{
    Uncalibrated profile-level $\Delta$-residual distributions for the model trained without simulation-parameter conditioning, in the same format as Figure~\ref{fig:appendix_dreams_sixpanel}.
    The distributions remain centered near zero, indicating that removing the conditioning degrades precision without introducing a systematic bias.
    }
    \label{fig:appendix_noh_deviation}
\end{figure*}

Figure~\ref{fig:appendix_noh_deviation} shows the corresponding profile-level $\Delta$-residual distributions.
The unconditioned model remains close to unbiased, and its residual widths are statistically indistinguishable from those of the conditioned model.
In other words, while dropping the parameter conditioning costs accuracy in the per-galaxy prediction, it does not introduce any systematic bias.

\begin{figure}
    \centering
    \includegraphics[width=0.95\linewidth]{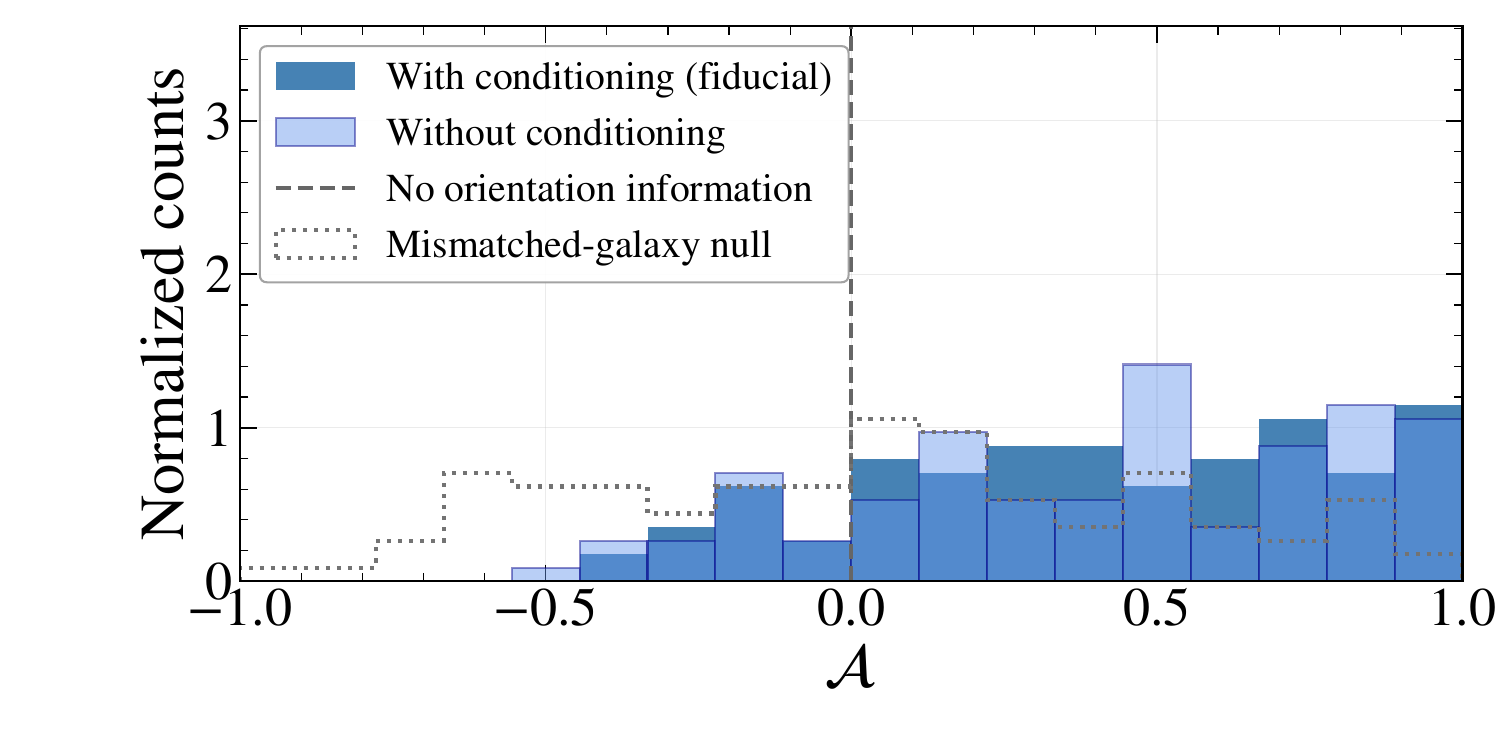}
    \caption{
    Distribution of the alignment statistic $\mathcal{A}$ of Equation~\ref{eq:alignment} over the $102$ held-out DREAMS galaxies, with and without the scalar simulation-parameter conditioning $h$.
    The dashed vertical line marks the value expected of a predictor carrying no azimuthal information, and the dotted histogram is the null obtained by scoring each galaxy's prediction against a different galaxy's truth.
    Both models pile up at $\mathcal{A}>0$ and are statistically indistinguishable from one another.
    }
    \label{fig:appendix_noh_alignment}
\end{figure}

For the azimuthal diagnostics, Figure~\ref{fig:appendix_noh_alignment} compares the distribution of the alignment statistic $\mathcal{A}$ over the held-out split for the two models.
The unconditioned model gives $\mathcal{A}=0.42\pm0.04$ with $82\%$ of galaxies scoring positively, against $0.41\pm0.04$ and $84\%$ for the conditioned model, and a paired comparison over the same galaxies finds no significant difference.  
Removing the scalar parameters therefore costs essentially nothing in the recovery of the orientation of the dark matter distribution, as expected.

\begin{figure*}
    \centering
    \includegraphics[width=0.95\linewidth]{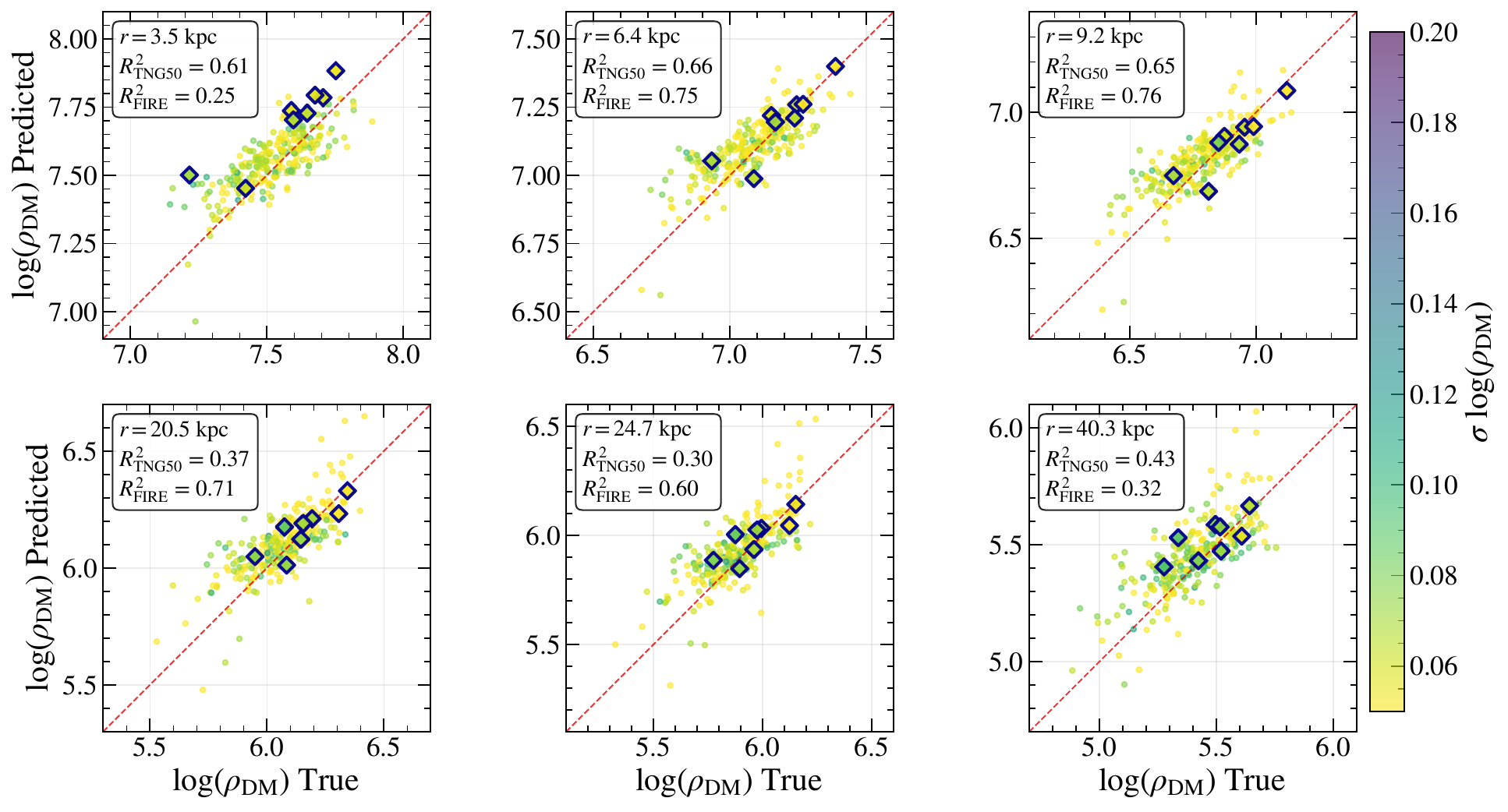}
    \caption{
    Predicted versus true azimuthally averaged dark matter density for the model trained \textit{without} simulation-parameter conditioning, evaluated out of domain on TNG50 (circles) and \Fire (diamonds), in the same format as Figure~\ref{fig:tng50_fire_profiles}.
    TNG50 uses $10$-draw ensembles and \Fire the $100$-draw set described above.
    }
    \label{fig:appendix_noh_ood_scatter}
\end{figure*}

\begin{figure*}
    \centering
    \includegraphics[width=0.95\linewidth]{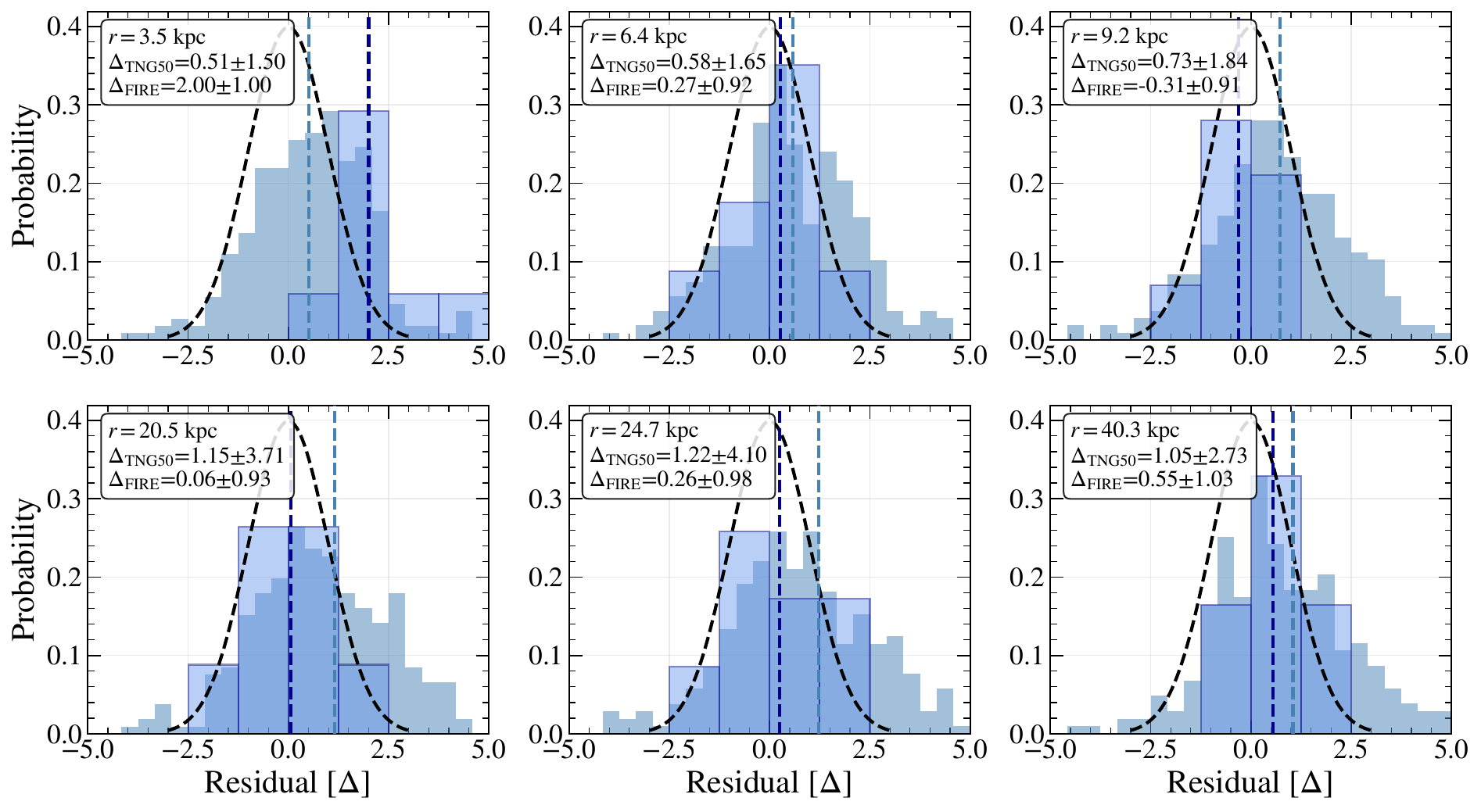}
    \caption{
    Uncalibrated profile-level $\Delta$-residual distributions for the unconditioned model on the out-of-domain TNG50 (light blue) and \Fire (dark blue) samples, in the same format as Figure~\ref{fig:appendix_ood_sixpanel}.
    As noted at the start of this appendix, TNG50 here is a $10$-draw ensemble and \Fire a $100$-draw one, so \Fire's $\Delta$ carries a denominator about $8$ per cent larger than TNG50's.
    }
    \label{fig:appendix_noh_ood_deviation}
\end{figure*}

Figures~\ref{fig:appendix_noh_ood_scatter} and~\ref{fig:appendix_noh_ood_deviation} repeat the two diagnostics above for TNG50 and \Fire, and are the ablation counterparts of Figure~\ref{fig:tng50_fire_profiles} and of Figure~\ref{fig:appendix_ood_sixpanel}.
Because \Fire has no counterpart to the DREAMS feedback parametrization, the conditioned model's \Fire predictions marginalize over a flat prior in $\bar{e}_w$, $\kappa_w$, and $\epsilon_{f,\,{\rm high}}$ (Section~\ref{sec:validation}), whereas the unconditioned model requires no such marginalization. 
We generate \Fire predictions at $100$ draws with the unconditioned model to match the ensemble size from the conditioned model. 
We find that beyond $5$\,kpc, marginalization over the three baryon feedback parameters accounts for roughly two thirds of the \Fire predictive variance, inflating the plotted $1\,\sigma$ band by a factor of $\sim1.7$.
Removing it leaves $\sim0.093$\,dex, matching both the $0.097$\,dex the conditioned model achieves in domain and the $0.106$\,dex width of the unconditioned model on the same galaxies.
The central over-prediction discussed in Section~\ref{sec:domain_adaptation} remain significant. 
Inside 5 kpc, the flat prior contributes least of any radial range (an inflation of $\sim1.3$), and the unconditioned model still over-predicts the same pixels by $\sim0.1$ dex.

\end{CJK*}
\end{document}